%% file: main.tex
\documentclass[lettersize,journal]{IEEEtran}

\usepackage{amsmath,amsthm,amssymb}
\usepackage{graphicx}
\usepackage{booktabs} 
\usepackage{multirow}
\usepackage{siunitx} 
\usepackage{hyperref}
\usepackage{float}
\usepackage[caption=false]{subfig}
\usepackage[linesnumbered,ruled]{algorithm2e} 
\usepackage{cite}
\usepackage[capitalize]{cleveref}
\usepackage{url}
\usepackage{balance}
\usepackage{etoolbox} 
\makeatletter
\patchcmd{\algocf@makecaption@ruled}{\hsize}{\textwidth}{}{} 
\patchcmd{\@algocf@start}{-1.5em}{0em}{}{}
\patchcmd\algocf@Vline{\vrule}{\vrule \kern-0.4pt}{}{}
\patchcmd\algocf@Vsline{\vrule}{\vrule \kern-0.4pt}{}{}
\makeatother
\AtBeginEnvironment{algorithm}{\linespread{0.93}\selectfont}
\AtEndEnvironment{algorithm}{\linespread{1.0}\selectfont}

\begin{document}

\title{Exact, Efficient, and Reliable Multi-Objective and Multi-Constrained IoT Workflow Scheduling in Edge-Hub-Cloud \\ Cyber-Physical Systems
}

\author{Andreas Kouloumpris, Georgios L. Stavrinides, \IEEEmembership{Member, IEEE}, Maria K. Michael, \IEEEmembership{Member, IEEE}, and Theocharis Theocharides, \IEEEmembership{Senior Member, IEEE} 
\thanks{Received 27 October 2025; revised 11 January 2026; accepted 12 February 2026.
This work has been supported by the European Union’s Horizon 2020 research and innovation programme under grant agreement No. 739551 (KIOS CoE) and from the Government of the Republic of Cyprus through the Cyprus Deputy Ministry of Research, Innovation and Digital Policy.
\emph{(Andreas Kouloumpris and Georgios L. Stavrinides are co-first authors.)}
\emph{(Corresponding author: Andreas Kouloumpris.)}
}
\thanks{Andreas Kouloumpris, Maria K. Michael, and Theocharis Theocharides are with the KIOS Research and Innovation Center of Excellence and the Department of Electrical and Computer Engineering, University of Cyprus, 1678 Nicosia, Cyprus (e-mail: kouloumpris.andreas@ucy.ac.cy; mmichael@ucy.ac.cy; ttheocharides@ucy.ac.cy).}
\thanks{Georgios L. Stavrinides is with the KIOS Research and Innovation Center of Excellence, University of Cyprus, 1678 Nicosia, Cyprus (e-mail: stavrinides.georgios@ucy.ac.cy).}
\thanks{This article has supplementary downloadable material available at https://doi.org/10.1109/JIOT.2026.3665298, provided by the authors.}
\thanks{Digital Object Identifier 10.1109/JIOT.2026.3665298}
%
}



\maketitle

\input{0_abstract.tex}


\input{1_introduction}
\input{2_related}
\input{3_framework}
\input{4_heuristic}

\input{5_results}

\input{6_conclusion}

\bibliographystyle{IEEEtran}
\bibliography{references.bib}


\vspace{-20pt}
\begin{IEEEbiography}[{\includegraphics[width=1in,height=1.25in,clip,keepaspectratio]{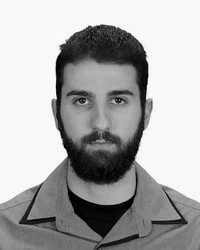}}]{Andreas Kouloumpris}
graduated top of his class receiving the BSc degree in Computer Engineering from the Department of Electrical and Computer Engineering of the University of Cyprus in 2016. He received the PhD degree in Computer Engineering from the same university in 2024. He joined the KIOS Research and Innovation Center of Excellence at the University of Cyprus as a Research Assistant in 2017. He is a member of the Cyprus Scientific and Technical Chamber and a member of the high IQ society MENSA. His research interests include task allocation and scheduling, mathematical optimization, graph theory, edge and cloud computing.
\end{IEEEbiography}

\vspace{-20pt}
\begin{IEEEbiography}[{\includegraphics[width=1in,height=1.25in,clip,keepaspectratio]{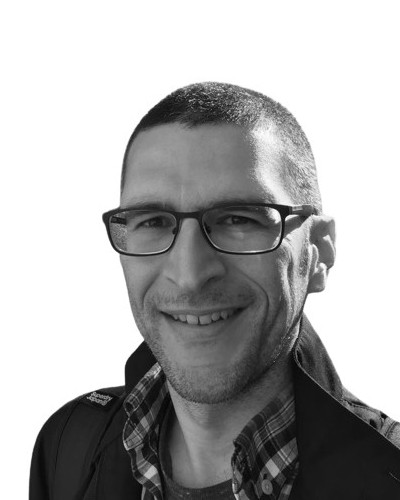}}]{Georgios L. Stavrinides} 
(Member, IEEE) received the BSc degree in Informatics from Aristotle University of Thessaloniki, Greece in 2006 and the MSc degree in Advanced Computing from Imperial College London, UK in 2007. He received the PhD degree in Informatics from Aristotle University of Thessaloniki, Greece in 2014. He is currently a Postdoctoral Fellow at the KIOS Research and Innovation Center of Excellence at the University of Cyprus. He serves on the editorial board of Simulation Modelling Practice and Theory. His research interests include task scheduling in distributed environments and performance optimization of cyber-physical systems in the edge-cloud continuum.
\end{IEEEbiography}

\vspace{-20pt}
\begin{IEEEbiography}[{\includegraphics[width=1in,height=1.25in,clip,keepaspectratio]{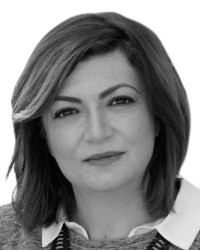}}]{Maria K. Michael}
(Member, IEEE) received the PhD degree in Computer Engineering from Southern Illinois University, USA. She is currently an Associate Professor at the Department of Electrical and Computer Engineering, and a co-founding faculty member of the KIOS Research and Innovation Center of Excellence, at the University of Cyprus. Her current research focuses on IC/embedded test and reliability, hardware-inspired cyber-security, and resource allocation and reliability in cyber-physical and edge/embedded systems. Her research has been funded by local and international organizations and the industry. She is an associate editor of ACM Computing Surveys and serves on the steering, organizing, and program committees of several IEEE/ACM conferences in the areas of test and reliability.
\end{IEEEbiography}

\vspace{-20pt}
\begin{IEEEbiography}[{\includegraphics[width=1in,height=1.25in,clip,keepaspectratio]{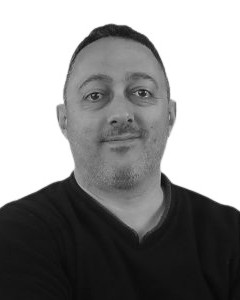}}]{Theocharis Theocharides}
(Senior Member, IEEE) received the PhD degree in Computer Engineering from Pennsylvania State University, USA. He is currently an Associate Professor at the Department of Electrical and Computer Engineering, and the Director of Research at the KIOS Research and Innovation Center of Excellence, at the University of Cyprus. His research focuses on the design, development and deployment of low-power and reliable on-chip application-specific architectures, VLSI design, and real-time embedded/edge computing systems. His research has been funded by national/European agencies and the industry. He is an associate editor of IEEE Transactions on Computer-Aided Design of Integrated Circuits and Systems, ACM Computing Surveys, and ACM Journal on Emerging Technologies in Computing Systems.
\end{IEEEbiography}

\end{document}




\title{Exact, Efficient, and Reliable Multi-Objective and Multi-Constrained IoT Workflow Scheduling in Edge-Hub-Cloud\\ Cyber-Physical Systems\\ 
\vspace{6pt} \hspace{3pt} Supplementary Material
\vspace{6pt}}

\author{Andreas Kouloumpris, Georgios L. Stavrinides, \IEEEmembership{Member, IEEE}, Maria K. Michael, \IEEEmembership{Member, IEEE}, and Theocharis Theocharides, \IEEEmembership{Senior Member, IEEE}
\vspace{20pt} 
}

\maketitle

\appendices

\counterwithin{table}{section} 

\counterwithin{figure}{section}

\renewcommand{\thealgocf}{\thesection.\arabic{algocf}}
\numberwithin{algocf}{section}

\numberwithin{equation}{section}


\section{}
The main notations used in the manuscript are summarized in \cref{table:notations}.

\begin{table}[!h]
\centering
\caption{Main Notations}
\label{table:notations}
\footnotesize
    \begin{tabular}{ll} 
        \toprule
        Notation & Definition\\
        \hline
        $\mathcal{U}$ & Set of devices in the considered edge-hub-cloud cyber-physical system (CPS).\\
        
        $u_{\mu k}$ & Device in $\mathcal{U}$, where $\mu$ indicates the type of the device ($\mathrm{e}$=edge device, $\mathrm{h}$=hub device, and $\mathrm{c}$=cloud server) and $k$ is the\\
            &  device index for the specific type (e.g., $u_{\mathrm{e} 1}$ denotes edge device 1).\\ 
        
        $\mathcal{P}_{\mu k}$ & Set of reserved cores on device $u_{\mu k}$.\\
        
        $p_{\mu k\hspace{-0.5pt}.\hspace{-0.5pt}q}$ & Reserved core on device $u_{\mu k}$ (e.g., $p_{\mathrm{e} 1.1}$ denotes reserved core 1 on edge device 1).\\
        
        $\lambda_{\mu k\hspace{-0.5pt}.\hspace{-0.5pt}q}$ & Failure rate of core $p_{\mu k\hspace{-0.5pt}.\hspace{-0.5pt}q}$.\\
        
        $\beta_{\mu k}$ & Number of reserved cores on device $u_{\mu k}$ (i.e., $\beta_{\mu k} = |\mathcal{P}_{\mu k}|$).\\
        
        $\mathcal{P}$ & Set of all reserved cores on all system devices in $\mathcal{U}$.\\ 
        
        $M_{\mu k}^{\mathrm{bgt}}$ & Memory budget for device $u_{\mu k}$.\vspace{1.25pt}\\
        
        $S_{\mu k}^{\mathrm{bgt}}$ & Storage budget for device $u_{\mu k}$.\vspace{1.25pt}\\
        
        $E_{\mu k}^{\mathrm{bgt}}$ & Energy budget for device $u_{\mu k}$.\\
        
        $\psi_{\mu k\hspace{-0.5pt}, \nu l}$ & Communication channel between devices $u_{\mu k}$ and $u_{\nu l}$.\\
        
        $\epsilon_{\mu k\hspace{-0.5pt}, \nu l}$ & Bandwidth of communication channel $\psi_{\mu k\hspace{-0.5pt}, \nu l}$.\\
        
        $\pi_{\mu k\hspace{-0.5pt}, \nu l}$ & Energy required to transmit a unit of data over communication channel $\psi_{\mu k\hspace{-0.5pt}, \nu l}$.\\
        
        $\sigma_{\mu k\hspace{-0.5pt}, \nu l}$ & Energy required to receive a unit of data over communication channel $\psi_{\mu k\hspace{-0.5pt}, \nu l}$.\\
        
        $\mathcal{I}_{\mu k\hspace{-0.5pt}, \nu l}$ & Set containing the device used for the communication between devices $u_{\mu k}$ and $u_{\nu l}$ when they cannot communicate directly.\\
        
        $\mathcal{C}$ & Set of sensing, actuating, or other device capabilities in the considered CPS.\\
        
        $c_a$ & Device capability in $\mathcal{C}$.\\
        
        $\mathcal{C}_{\mu k}$ & Set of capabilities of device $u_{\mu k}$ $( \mathcal{C}_{\mu k} \subseteq \mathcal{C} )$.\vspace{1.25pt}\\
        
        $y_{\mu k}^{c_a}$ & Binary parameter denoting whether device $u_{\mu k}$ has capability $c_a \in  \mathcal{C}$.\vspace{1.25pt}\\

        $\mathcal{T}$ & Set of tasks of the considered IoT workflow application.\\
        
        $\tau_i$ & Task in $\mathcal{T}$.\\
        
        $G=(\mathcal{N}, \mathcal{A})$ & Initial task graph (TG) of the application, where $\mathcal{N}$ and $\mathcal{A}$ are the sets of its nodes and arcs, respectively.\\
        
        $N_{i}$ & Node in $\mathcal{N}$ representing task $\tau_i \in \mathcal{T}$.\\
        
        $A_{i \rightarrow\hspace{-0.5pt}j}$ & Arc in $\mathcal{A}$ representing the communication and precedence relationship between tasks $\tau_i, \tau_j \in \mathcal{T}$.\\
        
        $z_i^{c_a}$ & Binary parameter denoting whether task $\tau_i$ requires capability $c_a \in \mathcal{C}$.\\
        
        $L_{\mathrm{thr}}$ & Predefined deadline before which the application should be completed (latency threshold).\\

        $G^{\prime}=\left(\mathcal{N}^{\prime}, \mathcal{A}^{\prime}\right)$ & Intermediate task allocation graph (TAG), where $\mathcal{N}^{\prime}$ and $\mathcal{A}^{\prime}$ are the sets of its composite nodes and arcs, respectively.\\
        
        $N^{\prime}_i$ & Composite node in $\mathcal{N}^{\prime}$ comprising a set of $N_{i\hspace{-0.5pt}, \mu k\hspace{-0.5pt}.\hspace{-0.5pt}q}$ nodes.\\
        
        $N_{i\hspace{-0.5pt}, \mu k\hspace{-0.5pt}.\hspace{-0.5pt}q}$ & Node in $N^{\prime}_i$ representing the possible allocation of task $\tau_i$ on core $p_{\mu k\hspace{-0.5pt}.\hspace{-0.5pt}q}$.\\

        $A^{\prime}_{i \rightarrow\hspace{-0.5pt}j}$ & Composite arc in $\mathcal{A}^{\prime}$ comprising a set of $A_{i\hspace{-0.5pt}, \mu k\hspace{-0.5pt}.\hspace{-0.5pt}q \rightarrow\hspace{-0.5pt}j\hspace{-0.5pt}, \nu l\hspace{-0.5pt}.\hspace{-0.5pt}r}$ arcs.\\

        $A_{i\hspace{-0.5pt}, \mu k\hspace{-0.5pt}.\hspace{-0.5pt}q \rightarrow\hspace{-0.5pt}j\hspace{-0.5pt}, \nu l\hspace{-0.5pt}.\hspace{-0.5pt}r}$ & Arc in $A^{\prime}_{i \rightarrow\hspace{-0.5pt}j}$ representing the transfer of data from node $N_{i\hspace{-0.5pt}, \mu k\hspace{-0.5pt}.\hspace{-0.5pt}q}$ to $N_{j\hspace{-0.5pt}, \nu l\hspace{-0.5pt}.\hspace{-0.5pt}r}$.\\

        \bottomrule
        \multicolumn{2}{r}{\scriptsize \emph{Continues on next page.}}\\
    \end{tabular}
\end{table}

\begin{table}[!h]
\ContinuedFloat 
\centering
\caption{Main Notations (Cont.)}
\label{table:notations2}
\footnotesize
    \begin{tabular}{ll} 
        \toprule
        Notation & Definition\\
        \hline 
        $M_i$ & Main memory required by task $\tau_i$.\\

        $S_i$ & Storage required by task $\tau_i$.\\

        $D_i$ & Output data size of task $\tau_i$.\\
        
        $\mathcal{Q}_i$ & Set of child tasks of task $\tau_i$.\\

        $\phi_i$ & Binary parameter denoting whether task $\tau_i$ is an exit task.\\

        $L_{i\hspace{-0.5pt}, \mu k\hspace{-0.5pt}.\hspace{-0.5pt}q}$ & Execution time of task $\tau_i$ on core $p_{\mu k\hspace{-0.5pt}.\hspace{-0.5pt}q}$.\\
   
        $P_{i\hspace{-0.5pt}, \mu k\hspace{-0.5pt}.\hspace{-0.5pt}q}$ & Power required to execute task $\tau_i$ on core $p_{\mu k\hspace{-0.5pt}.\hspace{-0.5pt}q}$.\\ 
    
        $E_{i\hspace{-0.5pt}, \mu k\hspace{-0.5pt}.\hspace{-0.5pt}q}$ & Energy required to execute task $\tau_i$ on core $p_{\mu k\hspace{-0.5pt}.\hspace{-0.5pt}q}$.\\

        $R_{i\hspace{-0.5pt}, \mu k\hspace{-0.5pt}.\hspace{-0.5pt}q}$ & Reliability of task $\tau_i$ on core $p_{\mu k\hspace{-0.5pt}.\hspace{-0.5pt}q}$.\vspace{1.25pt}\\

        $R_i^{\mathrm{thr}}$ &  Reliability threshold (lowest required reliability) of task $\tau_i$.\\

        $\zeta_{i\hspace{-0.5pt}, \mu k\hspace{-0.5pt}.\hspace{-0.5pt}q}$ & Binary parameter denoting whether task $\tau_{i}$ when allocated on core $p_{\mu k\hspace{-0.5pt}.\hspace{-0.5pt}q}$ requires duplication.\vspace{1.25pt}\\

        $\theta_{i\hspace{-0.5pt}, \mu k\hspace{-0.5pt}.\hspace{-0.5pt}q \rightarrow\hspace{-0.5pt}j\hspace{-0.5pt}, \nu l\hspace{-0.5pt}.\hspace{-0.5pt}r}^{\xi m}$ & Binary parameter denoting whether arc $A_{i\hspace{-0.5pt}, \mu k\hspace{-0.5pt}.\hspace{-0.5pt}q \rightarrow\hspace{-0.5pt}j\hspace{-0.5pt}, \nu l\hspace{-0.5pt}.\hspace{-0.5pt}r}$ involves indirect communication between devices $u_{\mu k}$ and $u_{\nu l}$\\
            & through device $u_{\xi m}$.\\

        $CL_{i\hspace{-0.5pt}, \mu k\hspace{-0.5pt}.\hspace{-0.5pt}q \rightarrow\hspace{-0.5pt}j\hspace{-0.5pt}, \nu l\hspace{-0.5pt}.\hspace{-0.5pt}r}$ & Time required to transfer output data $D_{i}$ from task $\tau_i$ (allocated on $p_{\mu k\hspace{-0.5pt}.\hspace{-0.5pt}q}$) to task $\tau_j$ (allocated on $p_{\nu l\hspace{-0.5pt}.\hspace{-0.5pt}r}$).\\
        
        $CE_{i\hspace{-0.5pt}, \mu k\hspace{-0.5pt}.\hspace{-0.5pt}q \rightarrow\hspace{-0.5pt}j\hspace{-0.5pt}, \nu l\hspace{-0.5pt}.\hspace{-0.5pt}r}$ & Energy required to transfer output data $D_{i}$ from task $\tau_i$ (allocated on $p_{\mu k\hspace{-0.5pt}.\hspace{-0.5pt}q}$) to task $\tau_j$ (allocated on $p_{\nu l\hspace{-0.5pt}.\hspace{-0.5pt}r}$).\vspace{1.25pt}\\

        $G^{\prime \prime}=\left(\mathcal{N}^{\prime \prime}, \mathcal{A}^{\prime \prime}\right)$ & Final extended task allocation graph (ETAG), where $\mathcal{N}^{\prime \prime}$ and $\mathcal{A}^{\prime \prime}$ are the sets of its composite nodes and arcs, respectively.\\

        $\tau_{i\hspace{-0.5pt}.\hspace{-0.5pt}n}$ & Primary (if $n=1$) or replica (if $n=2$) task of task $\tau_i$.\vspace{1.25pt}\\ 
        
        $N_i^{\prime \prime}$ & Composite node in $\mathcal{N}^{\prime \prime}$ consisting of subsets $N_{i\hspace{-0.5pt}.\hspace{-0.5pt}1}^{\prime \prime}$ and $N_{i\hspace{-0.5pt}.\hspace{-0.5pt}2}^{\prime \prime}$.\vspace{1.25pt}\\

        $N_{i\hspace{-0.5pt}.\hspace{-0.5pt}1}^{\prime \prime}$ & Set of primary candidate nodes $N_{i\hspace{-0.5pt}.\hspace{-0.5pt}1, \mu k\hspace{-0.5pt}.\hspace{-0.5pt}q}$.\vspace{1.25pt}\\

        $N_{i\hspace{-0.5pt}.\hspace{-0.5pt}2}^{\prime \prime}$ & Set of replica candidate nodes $N_{i\hspace{-0.5pt}.\hspace{-0.5pt}2, \mu k\hspace{-0.5pt}.\hspace{-0.5pt}q}$.\\
        
        $N_{i\hspace{-0.5pt}.\hspace{-0.5pt}n\hspace{-0.5pt}, \mu k\hspace{-0.5pt}.\hspace{-0.5pt}q}$ & Primary (if $n=1$) or replica (if $n=2$) candidate node in $N_i^{\prime \prime}$ representing the possible allocation of $\tau_{i\hspace{-0.5pt}.\hspace{-0.5pt}1}$ or $\tau_{i\hspace{-0.5pt}.\hspace{-0.5pt}2}$ on\\
            & core $p_{\mu k\hspace{-0.5pt}.\hspace{-0.5pt}q}$, respectively.\\

        $A^{\prime \prime}_{i \rightarrow\hspace{-0.5pt}j}$ & Composite arc in $\mathcal{A}^{\prime \prime}$ comprising a set of $A_{i\hspace{-0.5pt}.\hspace{-0.5pt}n\hspace{-0.5pt}, \mu k\hspace{-0.5pt}.\hspace{-0.5pt}q \rightarrow\hspace{-0.5pt}j\hspace{-0.5pt}.\hspace{-0.5pt}o\hspace{-0.5pt}, \nu l\hspace{-0.5pt}.\hspace{-0.5pt}r}$ arcs.\\

        $A_{i\hspace{-0.5pt}.\hspace{-0.5pt}n\hspace{-0.5pt}, \mu k\hspace{-0.5pt}.\hspace{-0.5pt}q \rightarrow\hspace{-0.5pt}j\hspace{-0.5pt}.\hspace{-0.5pt}o\hspace{-0.5pt}, \nu l\hspace{-0.5pt}.\hspace{-0.5pt}r}$ & Arc in  $A^{\prime \prime}_{i \rightarrow\hspace{-0.5pt}j}$ representing the transfer of data from candidate node $N_{i\hspace{-0.5pt}.\hspace{-0.5pt}n\hspace{-0.5pt}, \mu k\hspace{-0.5pt}.\hspace{-0.5pt}q}$ to $N_{j\hspace{-0.5pt}.\hspace{-0.5pt}o\hspace{-0.5pt}, \nu l\hspace{-0.5pt}.\hspace{-0.5pt}r}$.\\

        $\hat{R}_{i\hspace{-0.5pt}, \mu k\hspace{-0.5pt}.\hspace{-0.5pt}q}$ & Total reliability of task $\tau_{i}$ when its primary task $\tau_{i\hspace{-0.5pt}.\hspace{-0.5pt}1}$ is allocated on core $p_{\mu k\hspace{-0.5pt}.\hspace{-0.5pt}q}$, considering whether task\\
            & duplication is required ($\hat{R}_{i\hspace{-0.5pt}, \mu k\hspace{-0.5pt}.\hspace{-0.5pt}q} = R_{i\hspace{-0.5pt}, \mu k\hspace{-0.5pt}.\hspace{-0.5pt}q\hspace{-0.5pt}, \nu l\hspace{-0.5pt}.\hspace{-0.5pt}r}$) or not ($\hat{R}_{i\hspace{-0.5pt}, \mu k\hspace{-0.5pt}.\hspace{-0.5pt}q}=R_{i\hspace{-0.5pt}, \mu k\hspace{-0.5pt}.\hspace{-0.5pt}q}$).\\

        $R_{i\hspace{-0.5pt}, \mu k\hspace{-0.5pt}.\hspace{-0.5pt}q\hspace{-0.5pt}, \nu l\hspace{-0.5pt}.\hspace{-0.5pt}r}$ & Total reliability of primary task  $\tau_{i\hspace{-0.5pt}.\hspace{-0.5pt}1}$ and its replica $\tau_{i\hspace{-0.5pt}.\hspace{-0.5pt}2}$ when allocated on cores $p_{\mu k\hspace{-0.5pt}.\hspace{-0.5pt}q}$ and $p_{\nu l\hspace{-0.5pt}.\hspace{-0.5pt}r}$, respectively.\\

        $x_{i\hspace{-0.5pt}.\hspace{-0.5pt}n\hspace{-0.5pt}, \mu k\hspace{-0.5pt}.\hspace{-0.5pt}q}$ & Binary decision variable corresponding to candidate node $N_{i\hspace{-0.5pt}.\hspace{-0.5pt}n\hspace{-0.5pt}, \mu k\hspace{-0.5pt}.\hspace{-0.5pt}q} \hspace{-0.5pt} \in \hspace{-0.5pt} G^{\prime \prime}$.\\
   
        $x_{i\hspace{-0.5pt}.\hspace{-0.5pt}n\hspace{-0.5pt}, \mu k\hspace{-0.5pt}.\hspace{-0.5pt}q \rightarrow\hspace{-0.5pt}j\hspace{-0.5pt}.\hspace{-0.5pt}o\hspace{-0.5pt}, \nu l\hspace{-0.5pt}.\hspace{-0.5pt}r}$ & Binary decision variable corresponding to arc $A_{i\hspace{-0.5pt}.\hspace{-0.5pt}n\hspace{-0.5pt}, \mu k\hspace{-0.5pt}.\hspace{-0.5pt}q \rightarrow\hspace{-0.5pt}j\hspace{-0.5pt}.\hspace{-0.5pt}o\hspace{-0.5pt}, \nu l\hspace{-0.5pt}.\hspace{-0.5pt}r} \hspace{-1pt} \in \hspace{-1pt} G^{\prime \prime}$.\\

        $t_{i\hspace{-0.5pt}.\hspace{-0.5pt}n}$ & Continuous decision variable denoting the start time of task $\tau_{i\hspace{-0.5pt}.\hspace{-0.5pt}n}$ (corresponding to set $N_{i\hspace{-0.5pt}.\hspace{-0.5pt}n}^{\prime \prime} \hspace{-2pt} \in \hspace{-1pt} G^{\prime \prime}$).\\

        $T$ & Continuous decision variable denoting the completion time of the application.\\

        $x_{i\hspace{-0.5pt}, \mu k\hspace{-0.5pt}.\hspace{-0.5pt}q\hspace{-0.5pt}, \nu l\hspace{-0.5pt}.\hspace{-0.5pt}r}$ & Auxiliary binary decision variable corresponding to candidate nodes $N_{i\hspace{-0.5pt}.\hspace{-0.5pt}1\hspace{-0.5pt}, \mu k\hspace{-0.5pt}.\hspace{-0.5pt}q}$ and $N_{i\hspace{-0.5pt}.\hspace{-0.5pt}2\hspace{-0.5pt}, \nu l\hspace{-0.5pt}.\hspace{-0.5pt}r}$ of task $\tau_i$.\\ 

        $x_{i\hspace{-0.5pt}.\hspace{-0.5pt}n\hspace{-0.5pt}, j\hspace{-0.5pt}.\hspace{-0.5pt}o}$ & Auxiliary binary decision variable denoting whether task $\tau_{i\hspace{-0.5pt}.\hspace{-0.5pt}n}$ will be executed before task $\tau_{j\hspace{-0.5pt}.\hspace{-0.5pt}o}$.\vspace{1.25pt}\\

        $x_h^{i\hspace{-0.5pt}.\hspace{-0.5pt}n\hspace{-0.5pt}, \mu k\hspace{-0.5pt}.\hspace{-0.5pt}q}$ & Auxiliary binary decision variable denoting whether task $\tau_{i\hspace{-0.5pt}.\hspace{-0.5pt}n}$ will be in execution on core $p_{\mu k\hspace{-0.5pt}.\hspace{-0.5pt}q}$ at time $s_h  \hspace{-1.5pt} \in \hspace{-1.5pt} \mathcal{S}$.\vspace{1.25pt}\\ 

        $\hat{x}_h^{i\hspace{-0.5pt}.\hspace{-0.5pt}n\hspace{-0.5pt}, \mu k\hspace{-0.5pt}.\hspace{-0.5pt}q}$ & Auxiliary binary decision variable used in conjunction with $x_h^{i\hspace{-0.5pt}.\hspace{-0.5pt}n\hspace{-0.5pt}, \mu k\hspace{-0.5pt}.\hspace{-0.5pt}q}$ to preserve the linearity of the model.\vspace{1.25pt}\\

        $\mathcal{S}$ & Set of start times of all primary and replica tasks of the application.\\

        $s_h$ & Time instant in $\mathcal{S}$.\\

        $\dot{f}_{\mathrm{lat}}$ & Normalized objective function of overall latency.\\
        
        $\dot{f}_{\mathrm{en}}$ & Normalized objective function of overall energy.\\
        
        $\ddot{f}_{\mathrm{rel}}$ & Negated normalized objective function of overall reliability.\\

        $g$ & Multi-objective function comprising the weighted sum of $\dot{f}_{\mathrm{lat}}$,  $\dot{f}_{\mathrm{en}}$, and $\ddot{f}_{\mathrm{rel}}$.\\
        
        $w_{\mathrm{lat}}$ & Weight reflecting the relative importance of overall latency in $g$.\\
        
        $w_{\mathrm{en}}$ & Weight reflecting the relative importance of overall energy in $g$.\\
        
        $w_{\mathrm{rel}}$ & Weight reflecting the relative importance of overall reliability in $g$.\\

        $\rho_{i\hspace{-0.6pt}.\hspace{-0.6pt}n\hspace{-0.6pt}, \mu k\hspace{-0.6pt}.\hspace{-0.6pt}q}$ & Upward rank of candidate node $N_{i\hspace{-0.5pt}.\hspace{-0.5pt}n\hspace{-0.5pt}, \mu k\hspace{-0.5pt}.\hspace{-0.5pt}q} \hspace{-1pt} \in \hspace{-1pt} G^{\prime \prime}\hspace{-1.25pt}$.${}^{1}$\\

        $EFT_{i\hspace{-0.5pt}.\hspace{-0.5pt}n\hspace{-0.5pt}, \mu k\hspace{-0.5pt}.\hspace{-0.5pt}q}$ & Earliest finish time of candidate node $N_{i\hspace{-0.5pt}.\hspace{-0.5pt}n\hspace{-0.5pt}, \mu k\hspace{-0.5pt}.\hspace{-0.5pt}q} \hspace{-1pt} \in \hspace{-1pt} G^{\prime \prime}\hspace{-1.25pt}$.${}^{1}$\\

        \bottomrule
        \multicolumn{2}{l}{\scriptsize{${}^{1}$Used in extended HEFT.}}\\
    \end{tabular}
\end{table}

\clearpage
\section{}
\begin{figure}[!h]
    \centering
    \subfloat[]{\includegraphics[width=0.33\linewidth]{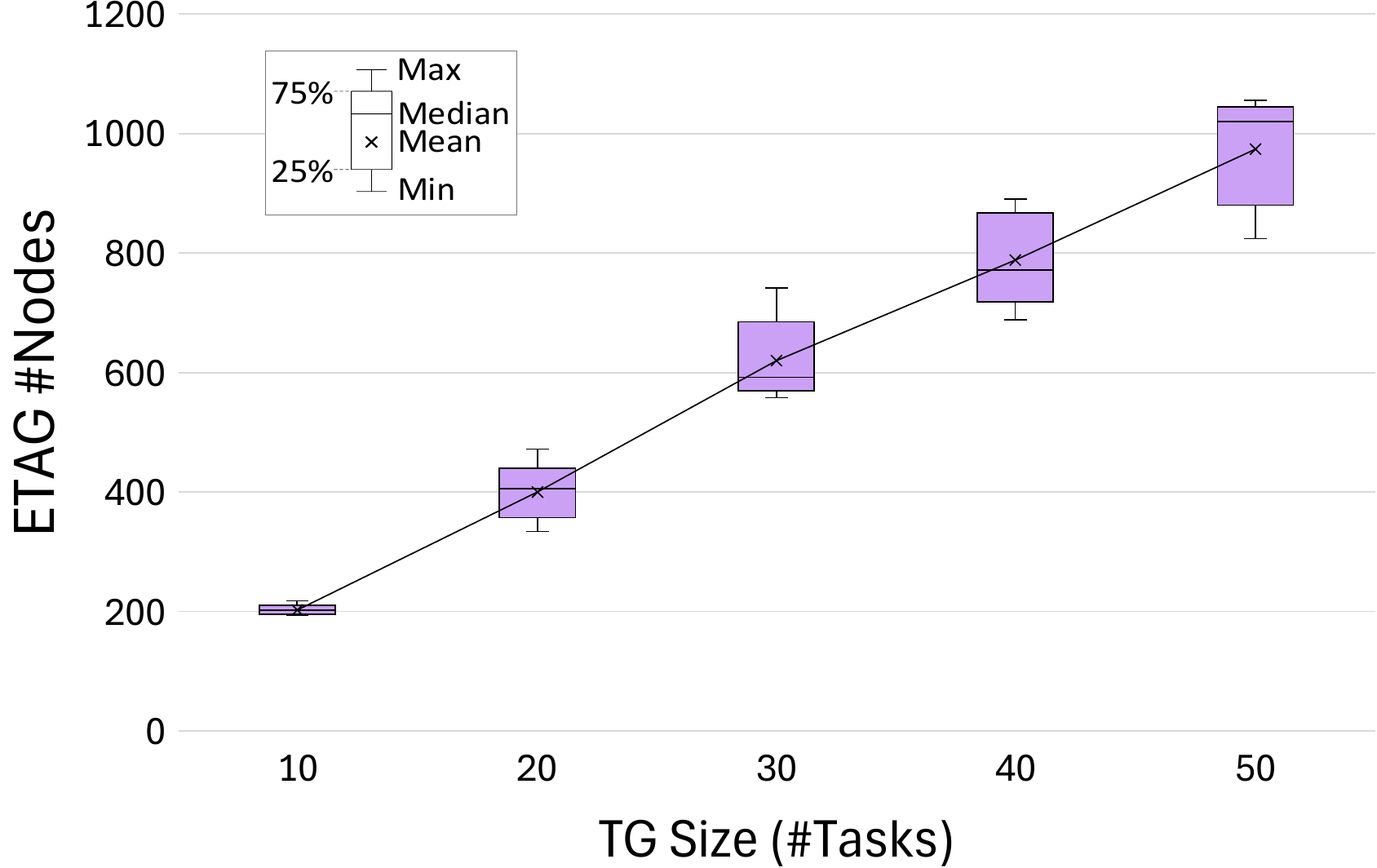}
        \label{fig:ETAGNodes}}
    \hspace{50pt}
    \subfloat[]{\includegraphics[width=0.33\linewidth]{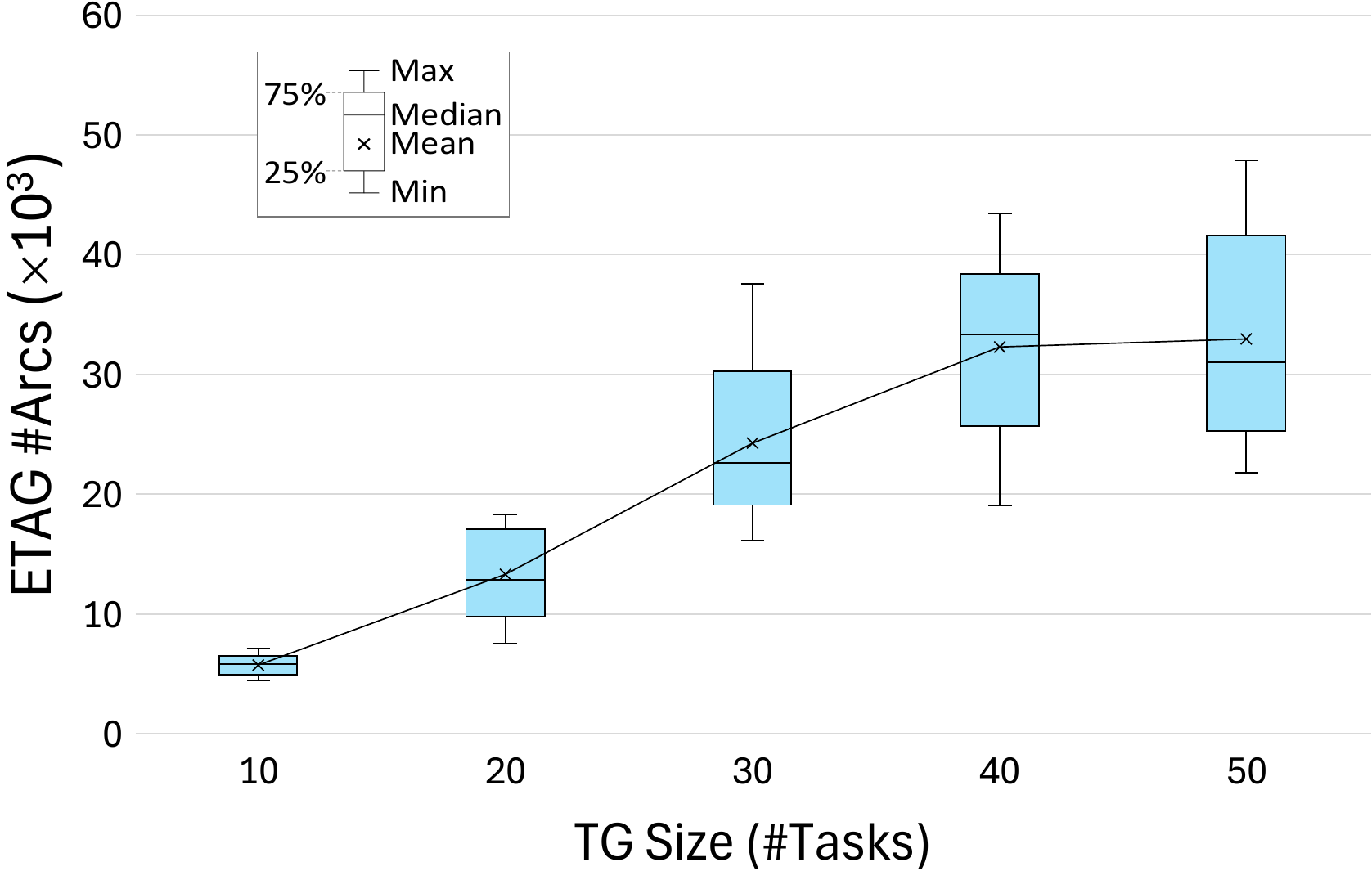}
        \label{fig:ETAGArcs}}
    \caption{ETAG (a) number of nodes and (b) number of arcs under increasing TG size. Box plots show the distribution of the number of nodes and arcs across ETAGs for each TG size, while the solid line connects the mean values.}
    \label{fig:ETAG}
\end{figure}

\begin{figure}[!h]
    \centering
    \subfloat[]{\includegraphics[width=0.33\linewidth]{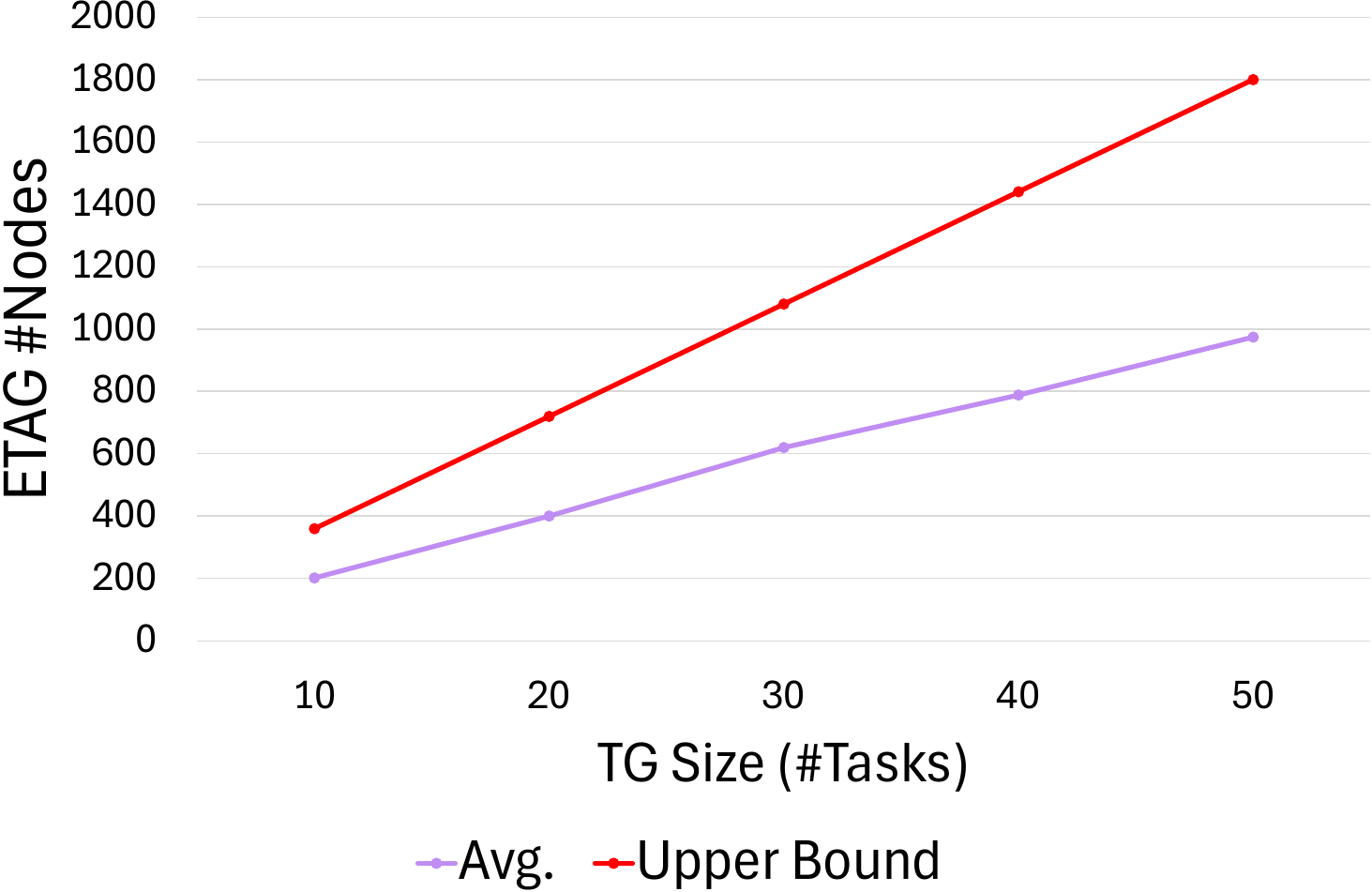}
        \label{fig:ETAGNodes2}}
    \hspace{50pt}
    \subfloat[]{\includegraphics[width=0.33\linewidth]{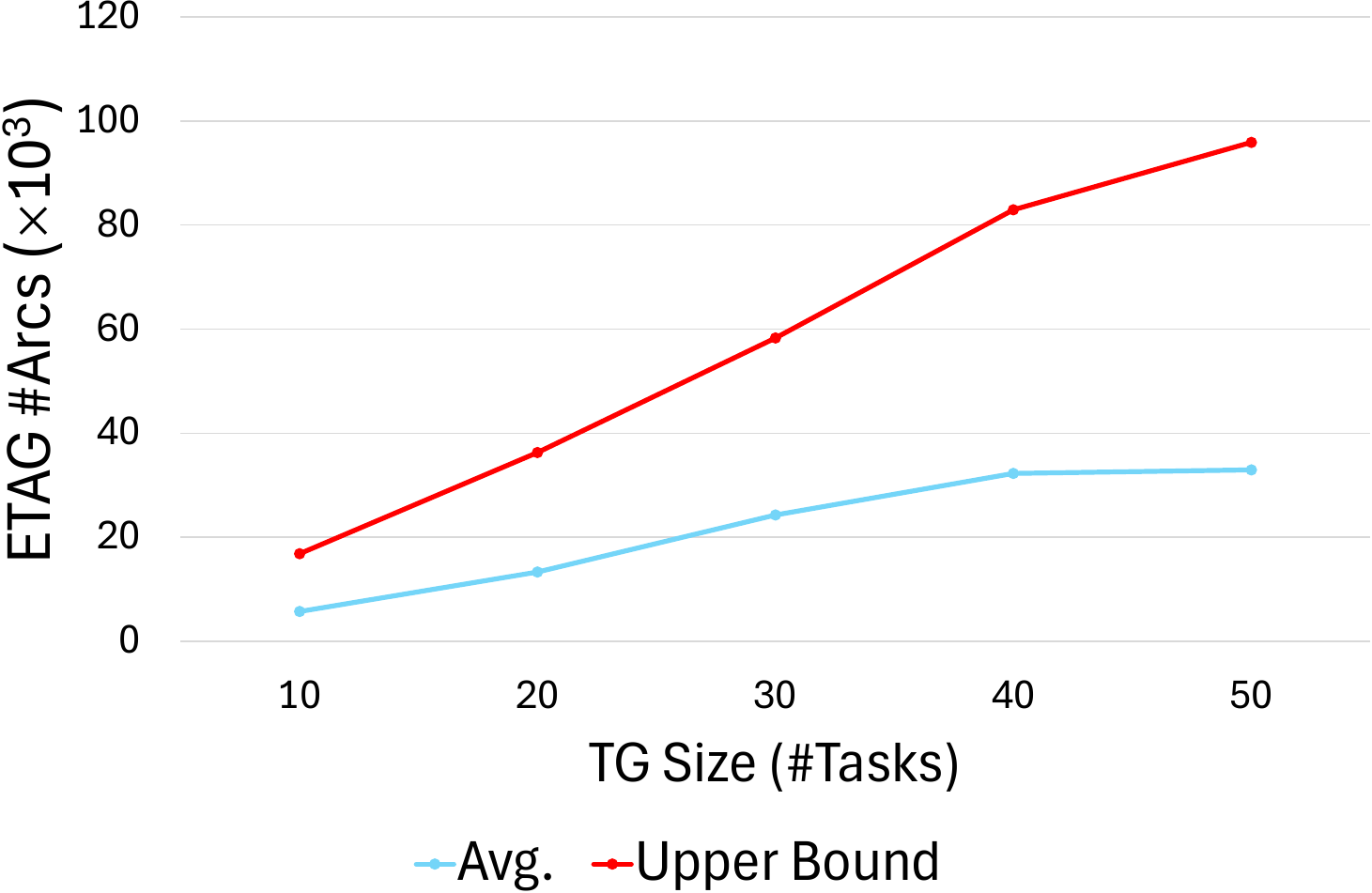}
        \label{fig:ETAGArcs2}}
    \caption{ETAG (a) average number of nodes and (b) average number of arcs compared against their corresponding theoretical upper bounds under increasing TG size.}
    \label{fig:ETAGcomparison}
\end{figure}

Figs. \ref{fig:ETAGNodes} and \ref{fig:ETAGArcs} demonstrate the growth of ETAG nodes and arcs, respectively, under increasing TG size for synthetic IoT workflows. Figs. \ref{fig:ETAGNodes2} and \ref{fig:ETAGArcs2} illustrate the comparison between the average number of ETAG nodes and arcs, respectively, against their corresponding theoretical upper bounds under increasing TG size. The analysis of these results is provided in Section V-C2 of the manuscript.

%% file: 0_abstract.tex
\begin{abstract}

Emerging Internet of Things (IoT)-enabled cyber-physical applications, such as autonomous critical infrastructure inspection, demand low-latency, energy-efficient, and reliable execution across resource-constrained edge devices with heterogeneous multicore processors and diverse sensing and actuating capabilities, in collaboration with a hub device and a cloud server. 
These workflow-based applications comprise interdependent tasks that must be executed under stringent deadline, reliability, capability, memory, storage, and energy constraints.
Given their critical nature, exact optimization is necessary to obtain optimal schedules that ensure dependable operation.
Existing scheduling approaches, both exact and heuristic, fail to jointly address all these objectives and constraints.
To this end, we propose an exact multi-objective and multi-constrained workflow scheduling approach for edge-hub-cloud cyber-physical systems, based on continuous-time mixed integer linear programming.
The proposed formulation jointly optimizes latency, energy, and reliability, while holistically addressing timing and resource constraints. To enhance reliability while avoiding the overhead of unnecessary task replicas, it selectively employs task duplication.
We evaluate our approach against a widely used heuristic, which we extend to ensure a fair and meaningful comparison, using a real-world IoT workflow and synthetic task graphs of varying sizes, across different system configurations and objective trade-offs.
The proposed method consistently outperforms the heuristic, achieving up to 29.83\%, 33.96\%, and 28.49\% average improvements in latency, energy, and reliability, respectively, while attaining practical runtimes. 
Overall, the experimental results demonstrate the effectiveness of our approach under various system configurations and objective trade-offs, and show its practical scalability to task graphs of sizes relevant to the targeted applications and system architecture.
\end{abstract}

%% file: 1_introduction.tex
\vspace{20pt}

\section{Introduction}
\label{sec:intro}

\IEEEPARstart{T}{he} rapid growth of the Internet of Things (IoT) has driven the development of critical cyber-physical applications that demand real-time, energy-efficient, and reliable execution at the network edge \cite{Qu2025, Kant2024}. Such IoT-enabled applications often follow the edge-hub-cloud paradigm, which deviates from the conventional edge computing model that comprises three distinct layers (edge devices, edge servers, and cloud servers). 
In this paradigm, edge and hub devices may be battery-powered and form the bottom layer, while cloud servers form the top layer \cite{Kouloumpris2024}.
Edge devices equipped with sensors collect data and process them locally or transmit them to a hub device for aggregation and further processing. The hub device may also interact with a cloud server to execute more demanding tasks, while the final results often drive actuators on edge devices that interact with the physical environment.
Hub devices, such as smartphones or laptops, offer higher computational capacity than edge devices, such as wearables or single-board computers. Although less capable than edge servers, hub devices are physically closer to edge devices, facilitating their coordination and communication with remote cloud servers.

\begin{figure}[t]
    \centering
    \subfloat[]{\includegraphics[width=0.45\columnwidth]{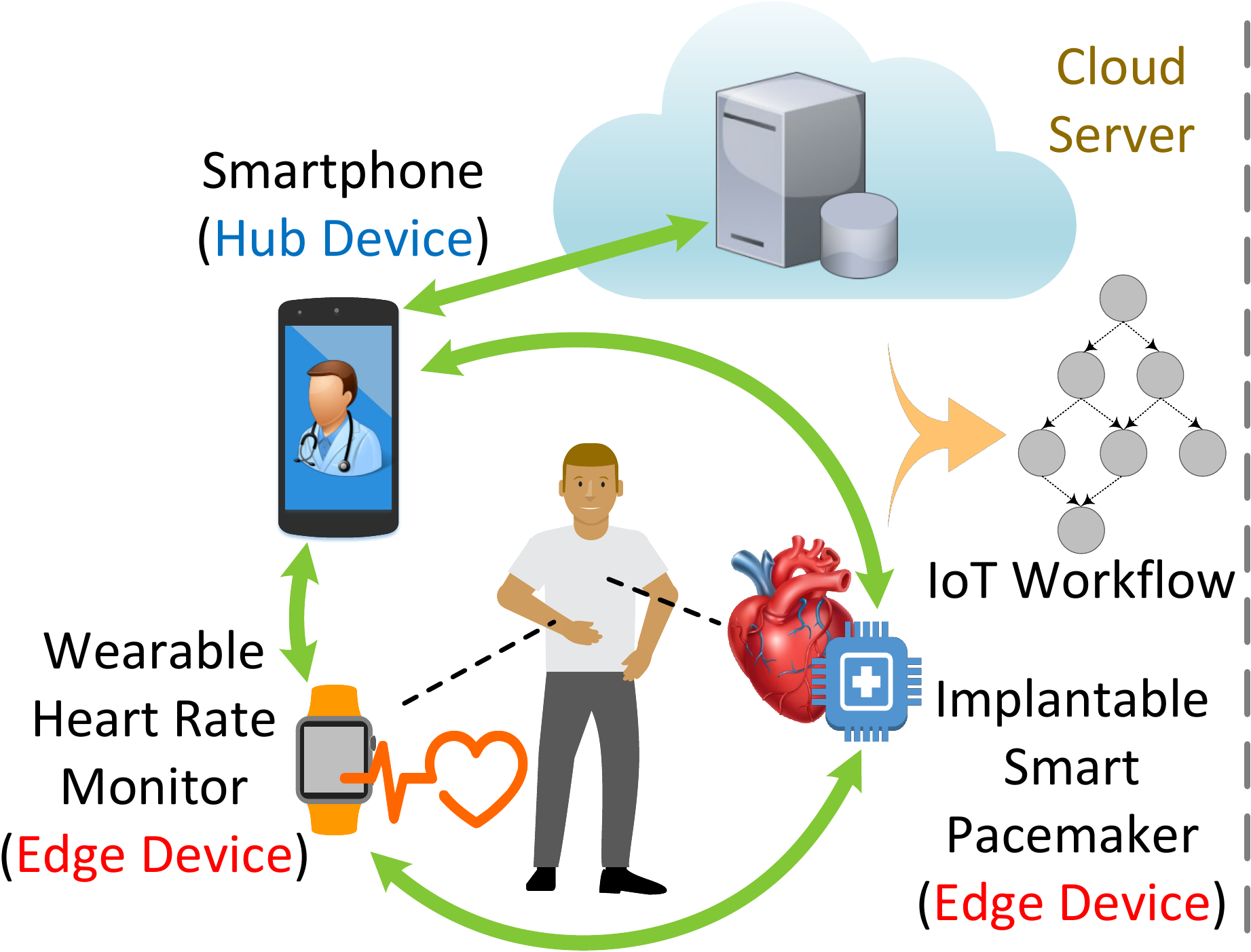}
        \label{fig:exampleBM}}
    \subfloat[]{\includegraphics[width=0.51\columnwidth]{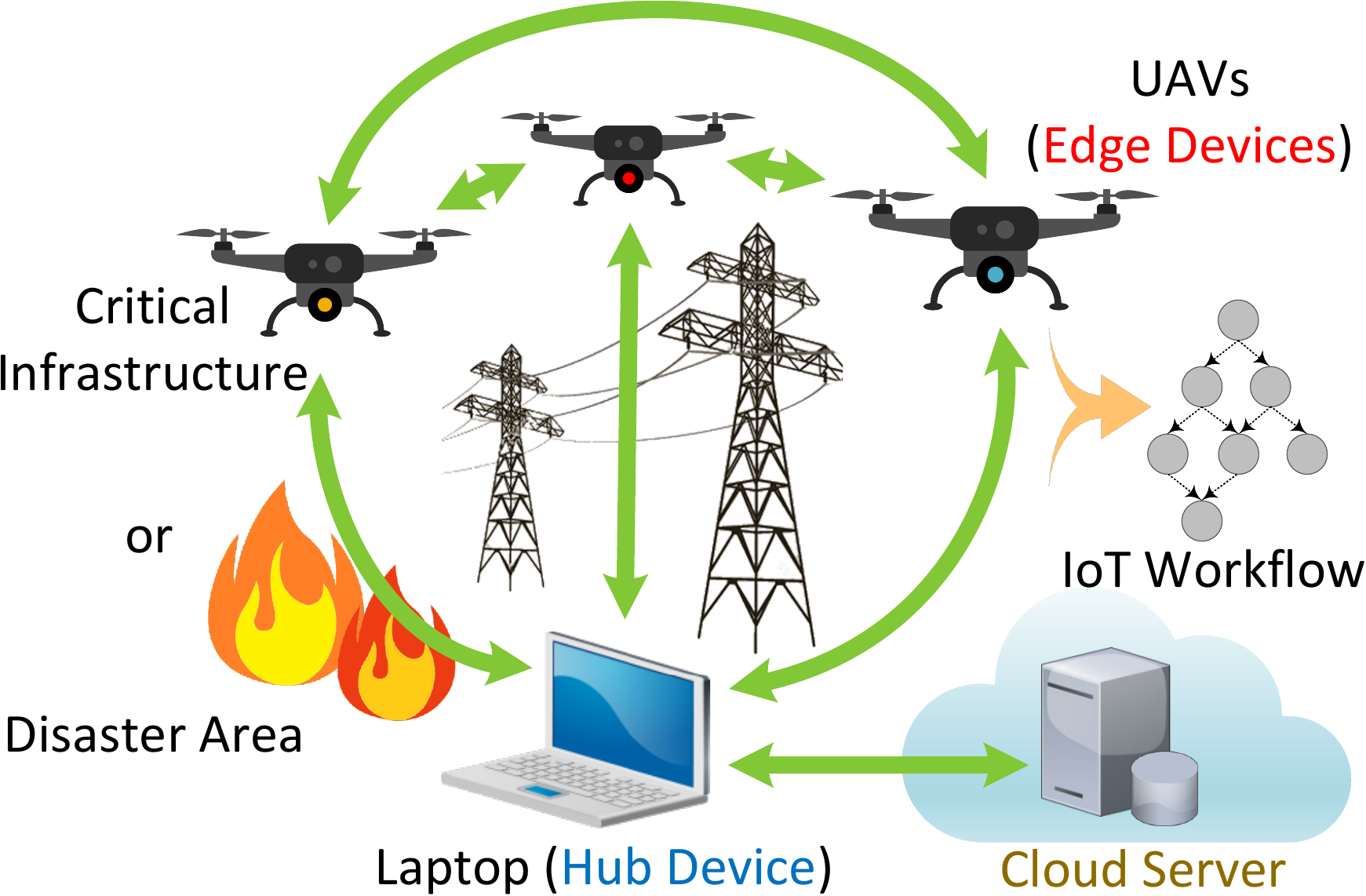}
        \label{fig:exampleUAV}}
    \caption{Examples of critical IoT-enabled cyber-physical applications following the edge-hub-cloud paradigm: (a) remote patient monitoring and support, and (b) autonomous critical infrastructure inspection or search-and-rescue missions. Edge devices equipped with various sensors (e.g., heart rate sensors or thermal cameras) and actuators (e.g., heart stimulators or payload release mechanisms) collaborate with each other and a hub device, which also interacts with a cloud server to jointly execute the IoT workflow.}
    \label{fig:examples}
\end{figure}

Cyber-physical IoT applications based on this paradigm often involve deadline- and precedence-constrained tasks that require specialized device capabilities, such as sensors, actuators, or specific software/hardware modules.
These applications, commonly referred to as workflows \cite{Mohammadzadeh2021}, encompass a wide range of latency-critical, energy-restricted, and reliability-sensitive scenarios.
Prominent examples include the use of biomedical edge devices for remote patient monitoring and support (\cref{fig:exampleBM}) \cite{Zheng2021, Alam2017}, and the use of unmanned aerial vehicles (UAVs) for autonomous critical infrastructure inspection or search-and-rescue missions (\cref{fig:exampleUAV}) \cite{Kashino2019, Savva2021}.
In these use cases, multiple edge devices with diverse sensing and actuating capabilities cooperate with each other and a hub device, which in turn communicates with a cloud server, to execute the application tasks in a timely, energy-efficient, and reliable manner. The hub device and the cloud server may also feature specialized software or hardware, such as libraries or accelerators for machine learning inference.

Moving from the cloud towards the network edge, the heterogeneity of devices increases, as they encompass different multicore processors and varied sensing and actuating capabilities \cite{Rasouli2025, Hosseinzadeh2025, Khoshvaght2025}. 
Meanwhile, their computational, communication, memory, storage, and energy capacities become more limited, posing additional challenges. 
These challenges are further amplified by the operating setting, as environmental factors such as temperature, humidity, and cosmic rays can affect reliability, often leading to transient errors \cite{Ottavi2014, Biswas2024}.  
Edge devices, due to their higher integration and operating conditions, are typically more vulnerable to these threats than hub devices, which are in turn less reliable than cloud servers.
These reliability issues are often mitigated through task replication, where the same task is executed more than once to enhance reliability \cite{Maniatakos2014}.
However, task replication introduces additional overhead that can degrade performance and increase energy consumption. Therefore, it should be applied selectively.
Hence, minimizing latency and energy while maximizing reliability are inherently conflicting objectives.

Given the criticality of applications such as those in \cite{Zheng2021, Alam2017, Kashino2019, Savva2021}, an optimized balance between latency, energy, and reliability is essential. This can be achieved through an exact scheduling technique that optimally determines where to allocate and when to execute each task, based on the relative importance of each objective, ensuring dependable execution.
In contrast to heuristic approaches, which typically yield approximate solutions without guaranteeing global optimality or full constraint satisfaction, exact methods such as mixed integer linear programming (MILP) can provide globally optimal schedules that satisfy all formulated constraints \cite{Grodzevich2006}.
Although exact methods are computationally intensive, the pre-programmed and deterministic nature of the targeted applications allows their offline use, provided that solutions are obtained within a reasonable time frame. 
An additional challenge of exact methods is the representation of time, which can be modeled as discrete or continuous. 
Discrete-time models simplify problems by assuming that events occur only at predefined intervals. On the the other hand, continuous-time models provide higher accuracy by allowing events to occur at any time. However, this leads to additional complexity, especially when modeling time-dependent cumulative constraints for the concurrent use of limited resources by multiple tasks, such as main memory, storage, or specific device capabilities \cite{Floudas2005}.

Existing exact and heuristic scheduling approaches do not holistically address all objectives and constraints of the targeted applications and overlook the specific multi-tier architecture. Therefore, we propose a continuous-time MILP method for the multi-objective and multi-constrained scheduling of IoT workflows in an edge-hub-cloud cyber-physical system (CPS) comprising multiple edge devices, a hub device, and a cloud server.
The proposed exact approach jointly minimizes latency and energy while maximizing reliability under application and system constraints.
Building on our preliminary research in \cite{Kouloumpris2024b}, our main contributions are as follows:
\begin{itemize}
\item We propose a comprehensive multi-objective and multi-constrained continuous-time MILP formulation to optimally schedule an IoT workflow in an edge-hub-cloud CPS.
The formulation is enabled by a two-phase task graph transformation technique that incorporates selective task duplication to enhance reliability while avoiding the overhead of unnecessary task replicas.

\item We holistically address objectives and constraints often overlooked by existing approaches. Specifically, we jointly optimize latency, energy, and reliability under precedence, deadline, reliability, capability, memory, storage, and energy constraints. We further account for heterogeneous multicore processors and multiple capabilities per device, and explicitly model time-dependent cumulative constraints for the concurrent use of limited resources, which are often omitted due to their complexity.

\item As our approach is the first to provide an optimal schedule for the specific objectives, constraints, and CPS architecture, we evaluate and compare it against one of the most effective and well-established workflow scheduling heuristics, the heterogeneous earliest finish time (HEFT) algorithm \cite{Topcuoglu2002, Kuhbacher2019, Aldegheri2020}. To ensure a fair and meaningful comparison, we extend HEFT to integrate selective task duplication and the same objectives and constraints as our method.


\item Our experimental evaluation is based on a representative real-world IoT workflow for UAV-enabled autonomous power infrastructure inspection, considering different system configurations and objective trade-offs.
To further validate the proposed approach and assess its scalability across varying workflow sizes, we also employ synthetic IoT task graphs.
\end{itemize}

The rest of the paper is organized as follows. \cref{sec:related} provides an overview of related studies. \cref{sec:framework} presents the proposed MILP method. \cref{sec:heft} describes our extension to HEFT. \cref{sec:evaluation} presents the experimental setup and results. \cref{sec:conclusion} concludes the paper.

%% file: 2_related.tex
\section{Related Work}
\label{sec:related}

Scheduling workflow applications in multicore and distributed environments has been widely studied using a variety of techniques, both exact and heuristic \cite{Topcuoglu2002, Kuhbacher2019, Aldegheri2020, Genez2020, Zengen2020, Liu2019, Mo2022, Mo2023, Ye2025, Jiang2025, Stavrinides2019b, Bai2021, Maio2020, Tuli2022, Fard2012, Huang2020, Saeedi2020, Taghinezhad2023}.

\subsection{Exact Approaches}
\label{subsec:exactApproaches}
MILP-based approaches are proposed in \cite{Genez2020, Zengen2020, Liu2019, Mo2022, Mo2023}, with \cite{Zengen2020} and \cite{Mo2023} focusing on specific CPS architectures.
While these methods \cite{Genez2020, Zengen2020, Liu2019, Mo2022, Mo2023} yield optimal schedules for deadline-constrained workflows, they do not jointly optimize latency, energy, and reliability. Furthermore, they overlook the memory and storage limitations of devices and do not consider a multi-tier system.
Specifically, \cite{Genez2020} and \cite{Zengen2020} minimize the overall latency, whereas \cite{Liu2019, Mo2022, Mo2023} optimize energy consumption under reliability and energy constraints. 
\cite{Zengen2020} and \cite{Mo2023} focus on single-core processors, while \cite{Liu2019} and \cite{Mo2022} assume homogeneous multicore processors.
Although \cite{Mo2023} considers the sensing and actuating capabilities of CPS devices, it restricts each device to only one sensor or actuator.
On the other hand, the sensing and actuating capabilities of devices are not addressed in \cite{Zengen2020}, despite focusing on a CPS.
With respect to the problem formulation, \cite{Genez2020} and \cite{Zengen2020} adopt a discrete-time model, whereas \cite{Liu2019, Mo2022, Mo2023} employ a continuous-time approach.

\subsection{Heuristic Approaches}
\label{subsec:heuristicApproaches}
In addition to exact methods, various heuristics have been extensively explored for workflow scheduling in multicore and multi-tier architectures \cite{Topcuoglu2002, Kuhbacher2019, Aldegheri2020, Ye2025, Jiang2025, Stavrinides2019b, Bai2021, Maio2020, Tuli2022}. 
One of the most well-established and effective heuristics is HEFT \cite{Topcuoglu2002}, which remains widely used for heterogeneous multicore processors due to its ability to provide high-quality schedules.
For example, HEFT is employed in \cite{Kuhbacher2019} and \cite{Aldegheri2020} to improve latency in embedded workflow applications. 
However, HEFT does not address the energy and reliability objectives, nor the deadline, reliability, capability, memory, storage, and energy constraints of the considered CPS.
Among other relevant heuristics, \cite{Ye2025} proposes an ant colony optimization-based approach to minimize energy consumption under a reliability constraint. 
\cite{Jiang2025} jointly optimizes latency and energy consumption in a heterogeneous edge computing system, whereas \cite{Stavrinides2019b} enhances latency and energy efficiency in a cloud setting.
\cite{Bai2021} improves latency within a predefined deadline in a CPS with multiple sensors and actuators per device, while \cite{Maio2020} optimizes both latency and reliability in a multi-tier system with storage limitations.  
A multi-tier architecture is also examined in \cite{Tuli2022}, where latency and energy are jointly optimized under deadline, memory, and storage constraints.
Although these heuristics are applicable to systems with heterogeneous multicore processors, with \cite{Maio2020} and \cite{Tuli2022} also supporting a multi-tier setting, none of them addresses all the objectives and constraints of the considered CPS.

On the other hand, multi-objective workflow scheduling strategies that jointly optimize latency, energy, and reliability in heterogeneous multicore environments are introduced in \cite{Fard2012, Huang2020, Saeedi2020, Taghinezhad2023}.
Specifically, \cite{Fard2012} adapts HEFT for multi-objective scheduling in a cloud computing setting, whereas \cite{Huang2020} combines various problem-specific heuristics in a multiprocessor system with reliability and energy limitations. 
Similarly, \cite{Saeedi2020} and \cite{Taghinezhad2023} employ particle swarm optimization and task duplication, respectively, to improve all three objectives under reliability and deadline constraints.
However, these methods do not consider all the constraints and the multi-tier environment addressed by our approach. Moreover, they cannot guarantee an exact solution, due to their heuristic nature. 

\subsection{Summary of Research Gaps}
Overall, existing exact and heuristic workflow scheduling techniques fail to holistically address all the objectives (latency, energy, and reliability) and constraints (deadline, reliability, capability, memory, storage, and energy) considered in the examined problem. 
Moreover, they do not take into account the specific CPS architecture and do not provide an optimal solution based on the desired trade-off between these objectives.
\cref{table:comparison} highlights this gap by qualitatively comparing existing approaches with the proposed method, which jointly addresses all the considered aspects.

\begin{table}[t]
\setlength{\tabcolsep}{2pt}
\centering
\caption{Comparison With Existing Workflow Scheduling Approaches}
\label{table:comparison}
\resizebox{0.95\columnwidth}{!}{
    \begin{tabular}{@{\extracolsep{2pt}}
                    lcccccccccccc} 
        \toprule
        \multirow{2}{*}[-0.25ex]{Ref.} & \multicolumn{3}{c}{Objectives} & \multicolumn{6}{c}{Constraints} & \multirow{2}{*}[-1.75ex]{\shortstack{Exact \\ Solution \\ (d/c)${}^{2}$}} 
        & \multirow{2}{*}[-1.75ex]{\shortstack{Multi-tier \\ Environ.}}
        & \multirow{2}{*}[-1.75ex]{\shortstack{Multicore \\ Processors \\ (h/H)${}^{3}$}}\\
        \cline{2-4}
        \cline{5-10}
						& \rotatebox[origin=b]{90}{\shortstack{\hspace{1pt} Min. \\ \hspace{1pt} Latency}}	& \rotatebox[origin=b]{90}{\shortstack{\hspace{1pt} Min. \\ \hspace{1pt} Energy}}	& \rotatebox[origin=b]{90}{\shortstack{\hspace{1pt} Max. \\ \hspace{1pt} Reliability}}	& \rotatebox[origin=c]{90}{Deadline \hspace{1pt}}	& \rotatebox[origin=c]{90}{Reliability \hspace{1pt}}	& \rotatebox[origin=b]{90}{\shortstack{\hspace{1pt} Capability \\ \hspace{1pt} (s/m)${}^{1}$}}	& \rotatebox[origin=c]{90}{Memory \hspace{1pt}}	& \rotatebox[origin=c]{90}{Storage \hspace{1pt}}	& \rotatebox[origin=c]{90}{Energy \hspace{1pt}}	& & & \\
\hline                                                          

\cite{Topcuoglu2002} 	& \checkmark	& - 			& - 			& - 			& - 				& - 				& - 			& - 			& - 			& - 				& - 			& \checkmark (H)\\
\cite{Kuhbacher2019} 	& \checkmark	& - 			& - 			& - 			& - 				& - 				& - 			& - 			& - 			& - 				& - 			& \checkmark (H)\\
\cite{Aldegheri2020} 	& \checkmark	& - 			& - 			& - 			& - 				& - 				& - 			& - 			& - 			& - 				& - 			& \checkmark (H)\\

\cite{Genez2020} 		& \checkmark	& - 			& - 			& \checkmark	& - 			 	& - 				& - 			& - 			& - 			& \checkmark (d) 	& - 			& \checkmark (H)\\ 
\cite{Zengen2020} 		& \checkmark	& - 			& - 			& \checkmark	& - 			 	& - 				& - 			& - 			& - 			& \checkmark (d) 	& - 			& -\\
\cite{Liu2019} 			& - 			& \checkmark 	& - 			& \checkmark	& \checkmark 	 	& - 				& - 			& - 			& \checkmark 	& \checkmark (c) 	& - 			& \checkmark (h)\\
\cite{Mo2022} 			& - 			& \checkmark 	& - 			& \checkmark	& \checkmark 	 	& - 				& - 			& - 			& \checkmark 	& \checkmark (c) 	& - 			& \checkmark (h)\\ 
\cite{Mo2023} 			& - 			& \checkmark 	& - 			& \checkmark	& \checkmark 	 	& \checkmark (s)	& - 			& - 			& \checkmark 	& \checkmark (c)	& - 			& -\\

\cite{Ye2025} 	        & - 			& \checkmark 	& - 			& -            	& \checkmark 	 	& - 				& - 			& - 			& - 			& - 				& - 			& \checkmark (H)\\
\cite{Jiang2025} 		& \checkmark 	& \checkmark 	& - 	        & -	            & - 			 	& - 				& - 			& - 			& -           	& - 				& - 			& \checkmark (H)\\
\cite{Stavrinides2019b} & \checkmark   	& \checkmark    & -         	& \checkmark	& -            	   	& -            		& -         	& -         	& \checkmark   	& -             	& -         	& \checkmark (H)\\
\cite{Bai2021} 			& \checkmark 	& - 			& - 			& \checkmark	& - 			 	& \checkmark (m) 	& - 			& - 			& - 			& - 				& - 			& \checkmark (H)\\
\cite{Maio2020} 		& \checkmark 	& - 			& \checkmark 	& - 			& - 				& - 				& - 			& \checkmark 	& - 			& - 				& \checkmark 	& \checkmark (H)\\
\cite{Tuli2022} 		& \checkmark 	& \checkmark 	& - 			& \checkmark	& - 			 	& - 				& \checkmark 	& \checkmark 	& - 			& - 				& \checkmark 	& \checkmark (H)\\
																			
\cite{Fard2012} 		& \checkmark 	& \checkmark 	& \checkmark 	& - 			& - 				& - 				& - 			& - 			& - 			& - 				& - 			& \checkmark (H)\\
\cite{Huang2020} 		& \checkmark 	& \checkmark 	& \checkmark 	& - 			& \checkmark 		& - 				& - 			& - 			& \checkmark 	& - 				& - 			& \checkmark (H)\\
\cite{Saeedi2020} 		& \checkmark 	& \checkmark 	& \checkmark 	& \checkmark	& \checkmark 	 	& - 				& - 			& - 			& - 			& - 				& - 			& \checkmark (H)\\
\cite{Taghinezhad2023} 	& \checkmark 	& \checkmark 	& \checkmark 	& \checkmark	& \checkmark 	 	& - 				& - 			& - 			& - 			& - 				& - 			& \checkmark (H)\\
    
        Ours & \checkmark & \checkmark & \checkmark & \checkmark & \checkmark & \checkmark  (m) & \checkmark & \checkmark & \checkmark & \checkmark (c) & \checkmark & \checkmark (H)\\
        \bottomrule
        \multicolumn{13}{l}{${}^{1}$\underline{s}ingle/\underline{m}ultiple capabilities (e.g., sensors and/or actuators) per CPS device.}\\
        \multicolumn{13}{l}{${}^{2}$\underline{d}iscrete/\underline{c}ontinuous-time exact approach.}\\
        \multicolumn{13}{l}{${}^{3}$\underline{h}omogeneous/\underline{H}eterogeneous multicore processors.}\\
    \end{tabular}
}
\end{table}

\subsection{Differences from Preliminary Research}

This work builds upon our preliminary research in \cite{Kouloumpris2024b}. Although both works consider the same architectural paradigm and utilize a continuous-time MILP approach, the preliminary research is limited to a single-objective formulation targeting latency, without considering reliability.
In contrast, this work reformulates the problem into a multi-objective and multi-constrained optimization framework, jointly optimizing latency, energy, and reliability.
We employ a reliability model that utilizes reliability thresholds for each task and selective task duplication to enhance reliability, resulting in primary and replica tasks, which further complicate the examined scheduling problem.
We incorporate multiple reliability constraints and reformulate all other constraints (precedence, deadline, capability, memory, storage, and energy constraints) to account for both primary and replica tasks.
Furthermore, we propose a two-phase task graph transformation technique, in contrast to the single-step task graph extension used in \cite{Kouloumpris2024b}, to facilitate the new problem formulation through an extended task graph that encompasses the system, application, energy, and reliability models.
Since the proposed approach is the first to provide an exact solution for this expanded problem setting, we re-implement HEFT to operate on the new extended task graph and to incorporate the same objectives, constraints, and selective task duplication mechanism as our approach.
%
Moreover, we empirically investigate different trade-offs between latency, energy, and reliability, as well as various capability provisioning strategies for the edge devices, in contrast to \cite{Kouloumpris2024b}, which considers only latency and a fixed capability assignment.
\cref{tab:diffPreliminary} summarizes the key theoretical and empirical contributions of this work compared to our preliminary research in \cite{Kouloumpris2024b}.

\begin{table}[t]
    \setlength{\tabcolsep}{2pt}
    \centering
    \caption{Comparison With Preliminary Research}
    \resizebox{0.9\columnwidth}{!}{
    \begin{tabular}{lll}
        \toprule
        \multicolumn{1}{c}{Aspect} & \multicolumn{1}{c}{Preliminary Research \cite{Kouloumpris2024b}} & \multicolumn{1}{c}{This Work}\\
        \hline
        Optimization Type & Single-objective & \textbf{Multi-objective} \vspace{3.1pt}\\
        
        Objectives & Latency & Latency, \textbf{energy}, \textbf{reliability} \vspace{3.1pt}\\
        
        Constraints & Precedence, deadline,  & \textbf{Reliability} \& \textbf{reformulated} \\
        &  capability, memory, & (\textbf{replica-aware}) precedence,\\
        & storage, energy & deadline, capability, memory,\\
        &  constraints & storage, energy constraints \vspace{3.1pt}\\
        
        Reliability Model & Not considered & \textbf{Selective task duplication},\\
        & &  \textbf{reliability thresholds} \vspace{3.1pt}\\
        
        Task Representation & Primary tasks only & \textbf{Primary} \& \textbf{replica} tasks \vspace{3.1pt}\\
        
        Task Graph & Single-step transformation & \textbf{Two-phase} transformation\\
        Transformation & & \vspace{3.1pt}\\
        
        HEFT Extension & Single-objective,  & \textbf{Multi-objective} \& \textbf{reliability-} \\
         & no reliability support & \textbf{aware re-implementation} \vspace{3.1pt}\\
        
        Evaluation Focus & Latency-driven evaluation, & \textbf{Multi-objective-driven} evaluation,\\
          &  fixed capability assignment & various \textbf{capability provisioning}\\
          &  & strategies\\
      
        \bottomrule
    \end{tabular}
    }
\label{tab:diffPreliminary}
\end{table}

%% file: 3_framework.tex
\section{Proposed Exact Multi-Objective \& Multi-Constrained MILP Approach}
\label{sec:framework}

\subsection{High-Level Overview}
\label{subsec:overview}

To facilitate our problem formulation we utilize a two-phase task graph transformation technique.
In phase 1, we transform the initial task graph (TG) of the application into an intermediate task allocation graph (TAG), which represents all applicable task allocations on the processing cores of the system devices, taking into account the capability requirements of the tasks and the capabilities provided by each device.
Based on the resulting TAG, we subsequently define the adopted energy and reliability models.
In phase 2, we convert the intermediate TAG into the final extended task allocation graph (ETAG), driven by the reliability requirements of the tasks, using selective task duplication. 
We leverage the ETAG, which encapsulates the considered system, application, energy, and reliability models, to formulate and optimally solve the investigated problem, based on the desired trade-off between latency, energy, and reliability.

\subsection{System Model}
\label{subsec:sysModel}
The considered edge-hub-cloud CPS comprises a set of $\alpha$ devices $\mathcal{U} = \{ u_{\mu k} \,| \allowbreak \, \mu \in \{\mathrm{e, h, c}\}, \allowbreak \, 1 \leq k \leq \alpha_{\mu}, \allowbreak \, \sum_{\mu \in \{ \mathrm{e}, \mathrm{h}, \mathrm{c} \}} \alpha_{\mu} = \alpha \}$, where $\mu$ denotes the device type, with $\mathrm{e}$, $\mathrm{h}$, and $\mathrm{c}$ representing an edge device, a hub device, and a cloud server, respectively. $k$ is the device index and $\alpha_{\mu}$ is the number of devices of type $\mu$.
%
\paragraph{Processing resources}
Each device $u_{\mu k} \in \mathcal{U}$ has a set of $\beta_{\mu k}$ reserved processing cores (hereafter referred to as cores) $\mathcal{P}_{\mu k} = \{ p_{\mu k\hspace{-0.5pt}.\hspace{-0.5pt}q}\,| \allowbreak \, 1 \leq q \leq \beta_{\mu k} \}$ for the execution of the workflow application, where $q$ is the core index.
Each core can execute one task at a time.
Without loss of generality, we abstract the processor level (a device may comprise one or more multicore processors) and directly model cores as the computational resources, as in \cite{Genez2020}.
Each device $u_{\mu k}$ has memory $M_{\mu k}^{\mathrm{bgt}}$, storage $S_{\mu k}^{\mathrm{bgt}}$, and energy $E_{\mu k}^{\mathrm{bgt}}$ budgets, which are shared among its reserved cores.
The set of all reserved cores in the system is $\mathcal{P} = \bigcup_{u_{\mu k} \in \mathcal{U}} \mathcal{P}_{\mu k}$.
Cores are heterogeneous across the system and may also differ within the same device. Hence, the time and power required to execute a given task depend on its assigned core.
%
\paragraph{Communication}
The edge devices and the hub device are fully connected, while the hub device is also connected to the cloud server (see example in \cref{fig:example}a). This logical network topology enables close collaboration between the edge devices and the hub device, and facilitates the execution of more demanding tasks on the cloud server.
A communication channel between two devices $u_{\mu k}$ and $u_{\nu l}$ is defined as $\psi_{\mu k\hspace{-0.5pt}, \nu l} = \langle \epsilon_{\mu k\hspace{-0.5pt}, \nu l}, \allowbreak \pi_{\mu k\hspace{-0.5pt}, \nu l}, \allowbreak \sigma_{\mu k\hspace{-0.5pt}, \nu l}, \allowbreak \mathcal{I}_{\mu k\hspace{-0.5pt}, \nu l} \rangle$, where $\epsilon_{\mu k\hspace{-0.5pt}, \nu l}$ is its bandwidth, and $\pi_{\mu k\hspace{-0.5pt}, \nu l}$ and $\sigma_{\mu k\hspace{-0.5pt}, \nu l}$ indicate the energy required to transmit and receive a unit of data over the particular channel, respectively.
If devices $u_{\mu k}$ and $u_{\nu l}$ communicate directly, then $\mathcal{I}_{\mu k\hspace{-0.5pt}, \nu l}= \varnothing$. Otherwise, if communication occurs through an intermediate device $u_{\xi m}$ (i.e., the hub device), then $\mathcal{I}_{\mu k\hspace{-0.5pt}, \nu l}=\{ u_{\xi m} \}$. 
Intra-device data transfers are assumed to incur negligible latency and energy costs compared to inter-device communication \cite{Mo2023}.
Routing and protocol details are abstracted, as the focus of this work is on task scheduling rather than network-level optimization \cite{Ye2025}.
%
\paragraph{Device capabilities}
We consider a set of $\gamma$ device capabilities $\mathcal{C} = \{ c_a \,| \allowbreak \, 0 \leq a < \gamma \}$, where $c_0$ denotes the basic computational capability of a device, whereas $c_a$ with $a > 0$ represents a specialized capability (in addition to the basic computational capability), such as a specific sensor, actuator, or software/hardware module.
Each device $u_{\mu k}$ features a set of capabilities $\mathcal{C}_{\mu k} \subseteq \mathcal{C}$, such that  $c_0 \in \mathcal{C}_{\mu k}$.
The binary parameter $y_{\mu k}^{c_a}$ denotes whether device $u_{\mu k}$ has capability $c_a \in  \mathcal{C}$ ($y_{\mu k}^{c_a} = 1$) or not ($y_{\mu k}^{c_a}=0$).

\subsection{Application Model (TG)}
\label{subsec:appModel}
The considered workflow application comprises a set of $\delta$ tasks $\mathcal{T} = \{ \tau_i \,|\, 1 \leq i \leq \delta \}$.
A task is an indivisible unit of work.
The TG of the workflow is represented by a directed acyclic graph $G=( \mathcal{N, \mathcal{A}})$ \cite{Aldegheri2020}, where $\mathcal{N}=\{ N_i \,|\, \tau_i \in \mathcal{T} \}$ is the set of its nodes, and $\mathcal{A} = \{ A_{i \rightarrow\hspace{-0.5pt}j} \, | \, N_{i}, \allowbreak N_{j} \in \mathcal{N}, \, i \neq j, \, \exists \text{ a data dependency } N_{i} \hspace{-3pt}\rightarrow\hspace{-3pt} N_{j} \}$ is the set of its arcs.
A node $N_{i} \in \mathcal{N}$ represents a task $\tau_i \in \mathcal{T}$.
An arc $A_{i \rightarrow\hspace{-0.5pt}j} \in \mathcal{A}$ between two nodes $N_{i}$ and $N_{j}$ (corresponding to parent task $\tau_i$ and child task $\tau_j$, respectively) represents the communication and precedence relationship between the two tasks.
Each task requires a device capability \cite{Bai2021}.
The binary parameter $z_i^{c_a}$ denotes whether task $\tau_i$ requires capability $c_a \in  \mathcal{C}$  ($z_i^{c_a} = 1$) or not ($z_i^{c_a} = 0$).
All tasks of the application should be completed before a predefined deadline (latency threshold) $L_{\mathrm{thr}}$, which also indicates the time horizon of the scheduling problem.
For periodic applications, $L_{\mathrm{thr}}$ is equal to the application period \cite{Mo2023}.
Tasks are considered to be non-preemptive, as preemption in deadline-constrained applications may lead to performance degradation \cite{Stavrinides2019b}.

\subsection{TG Transformation Phase 1 (TAG)}
\label{subsec:TAG}
We first transform TG $G$ into TAG $G^{\prime}=\left(\mathcal{N}^{\prime}, \mathcal{A}^{\prime}\right)$.
Specifically, we transform each node  $N_i \in \mathcal{N}$ in $G$ into a composite node (i.e., set of nodes) $N_i^{\prime} \in \mathcal{N}^{\prime}$ in $G^{\prime}$, such that:
\begin{equation}
\label{eq:TAGnodes}
 N_i^{\prime} = \{ N_{i\hspace{-0.5pt}, \mu k\hspace{-0.5pt}.\hspace{-0.5pt}q} \, | \, p_{\mu k\hspace{-0.5pt}.\hspace{-0.5pt}q} \in \mathcal{P}, \, y_{\mu k}^{c_a}\,z_i^{c_a} = 1, \, c_a \in \mathcal{C} \}.
\end{equation}
An individual node $N_{i\hspace{-0.5pt}, \mu k\hspace{-0.5pt}.\hspace{-0.5pt}q} \! \in \! N_i^{\prime}$ represents the possible allocation of task $\tau_i$ on core $p_{\mu k\hspace{-0.5pt}.\hspace{-0.5pt}q}$.
Thus, $N_i^{\prime}$ encompasses all applicable allocations of $\tau_i$ on the reserved cores of the devices featuring its required capability.
Similarly, we transform each arc $A_{i \rightarrow\hspace{-0.5pt}j} \in \mathcal{A}$ in $G$ into a composite arc (i.e., set of arcs) $A^{\prime}_{i \rightarrow\hspace{-0.5pt}j} \in \mathcal{A}^{\prime}$ in $G^{\prime}$, such that:
\begin{equation}
 \label{eq:TAGarcs}
 \begin{split}
     A^{\prime}_{i \rightarrow\hspace{-0.5pt}j}  \hspace{-2pt} = &  \{ A_{i\hspace{-0.5pt}, \mu k\hspace{-0.5pt}.\hspace{-0.5pt}q \rightarrow\hspace{-0.5pt}j\hspace{-0.5pt}, \nu l\hspace{-0.5pt}.\hspace{-0.5pt}r} \hspace{-2pt} = \hspace{-2pt} N_{i\hspace{-0.5pt}, \mu k\hspace{-0.5pt}.\hspace{-0.5pt}q} \hspace{-3pt} \rightarrow \hspace{-3pt} N_{j\hspace{-0.5pt}, \nu l\hspace{-0.5pt}.\hspace{-0.5pt}r} \, | \, p_{\mu k\hspace{-0.5pt}.\hspace{-0.5pt}q}, p_{\nu l\hspace{-0.5pt}.\hspace{-0.5pt}r} \hspace{-2pt} \in \hspace{-2pt} \mathcal{P}, \\
     & \quad y_{\mu k}^{c_a}\,z_i^{c_a} \hspace{-2pt} = \hspace{-2pt} y_{\nu l}^{c_b}\,z_j^{c_b} \hspace{-2pt} = \hspace{-2pt} 1, \, c_a, c_b \hspace{-2pt} \in \hspace{-1pt} \mathcal{C} \}.
\end{split}
\end{equation}
An individual arc $A_{i\hspace{-0.5pt}, \mu k\hspace{-0.5pt}.\hspace{-0.5pt}q \rightarrow\hspace{-0.5pt}j\hspace{-0.5pt}, \nu l\hspace{-0.5pt}.\hspace{-0.5pt}r} \in A^{\prime}_{i \rightarrow\hspace{-0.5pt}j}$ denotes the transfer of data from node $N_{i\hspace{-0.5pt}, \mu k\hspace{-0.5pt}.\hspace{-0.5pt}q}$ to $N_{j\hspace{-0.5pt}, \nu l\hspace{-0.5pt}.\hspace{-0.5pt}r}$.

\subsubsection{TAG Node Parameters}
\label{subsubsec:TAGnodeParams}
A node $N_{i\hspace{-0.5pt}, \mu k\hspace{-0.5pt}.\hspace{-0.5pt}q} \in G^{\prime}$ has the following parameters, in addition to the binary parameters $y_{\mu k}^{c_a}$ and $z_i^{c_a}$ defined in \cref{subsec:sysModel,subsec:appModel}, respectively:
\begin{itemize}
    \item $M_i$: main memory required by task $\tau_i$.
    
    \item $S_i$: storage required by $\tau_i$.

    \item $D_i$: output data size of $\tau_i$.
     
    \item $\mathcal{Q}_i$: set of child tasks of $\tau_i$.

    \item $\phi_i$: binary parameter denoting whether $\tau_i$ is an exit task, i.e., without any child tasks:
    \begin{equation}
    \label{eq:isExitTask}
        \phi_i = 
        \begin{cases}
            1, & \text{if $|\mathcal{Q}_i| = 0$},\\
            0, & \text{if $|\mathcal{Q}_i| > 0$}.
        \end{cases}
    \end{equation}

    \item $L_{i\hspace{-0.5pt}, \mu k\hspace{-0.5pt}.\hspace{-0.5pt}q}$: execution time of $\tau_i$ on core $p_{\mu k\hspace{-0.5pt}.\hspace{-0.5pt}q}$.
   
    \item $P_{i\hspace{-0.5pt}, \mu k\hspace{-0.5pt}.\hspace{-0.5pt}q}$: power required to execute $\tau_i$ on $p_{\mu k\hspace{-0.5pt}.\hspace{-0.5pt}q}$. 
    
    \item $E_{i\hspace{-0.5pt}, \mu k\hspace{-0.5pt}.\hspace{-0.5pt}q}$: energy required to execute $\tau_i$ on $p_{\mu k\hspace{-0.5pt}.\hspace{-0.5pt}q}$. It is given by \eqref{eq:nodeEnergy}, based on the energy model in \cref{subsec:energyModel}.

    \item $R_{i\hspace{-0.5pt}, \mu k\hspace{-0.5pt}.\hspace{-0.5pt}q}$: reliability of $\tau_i$ on $p_{\mu k\hspace{-0.5pt}.\hspace{-0.5pt}q}$, i.e., probability that $\tau_i$ is executed on $p_{\mu k\hspace{-0.5pt}.\hspace{-0.5pt}q}$ without any failures. It is defined by \eqref{eq:nodeReliability}, based on the reliability model in \cref{subsec:reliabilityModel}.

    \item $R_i^{\mathrm{thr}}$: reliability threshold of $\tau_i$ (i.e., lowest required reliability).

    \item $\zeta_{i\hspace{-0.5pt}, \mu k\hspace{-0.5pt}.\hspace{-0.5pt}q}$: binary parameter indicating whether the reliability of $\hspace{-1pt} \tau_{i}$ on $\hspace{-1pt} p_{\mu k\hspace{-0.5pt}.\hspace{-0.5pt}q}$ is below its reliability threshold $R_i^{\mathrm{thr}}$:
    \begin{equation}
    \label{eq:reqReplica}
        \zeta_{i\hspace{-0.5pt}, \mu k\hspace{-0.5pt}.\hspace{-0.5pt}q} = 
        \begin{cases}
            1, & \text{if $R_{i\hspace{-0.5pt}, \mu k\hspace{-0.5pt}.\hspace{-0.5pt}q} < R_i^{\mathrm{thr}}$},\\
            0, & \text{if $R_{i\hspace{-0.5pt}, \mu k\hspace{-0.5pt}.\hspace{-0.5pt}q} \geq R_i^{\mathrm{thr}}$}.
        \end{cases}
    \end{equation}
\end{itemize}

\subsubsection{TAG Arc Parameters}
\label{subsubsec:TAGarcParams}
An arc $A_{i\hspace{-0.5pt}, \mu k\hspace{-0.5pt}.\hspace{-0.5pt}q \rightarrow\hspace{-0.5pt}j\hspace{-0.5pt}, \nu l\hspace{-0.5pt}.\hspace{-0.5pt}r} \in G^{\prime}$ has the following parameters:
\begin{itemize}
    \item  $\theta_{i\hspace{-0.5pt}, \mu k\hspace{-0.5pt}.\hspace{-0.5pt}q \rightarrow\hspace{-0.5pt}j\hspace{-0.5pt}, \nu l\hspace{-0.5pt}.\hspace{-0.5pt}r}^{\xi m}$: binary parameter denoting whether $A_{i\hspace{-0.5pt}, \mu k\hspace{-0.5pt}.\hspace{-0.5pt}q \rightarrow\hspace{-0.5pt}j\hspace{-0.5pt}, \nu l\hspace{-0.5pt}.\hspace{-0.5pt}r}$ involves indirect communication between devices $u_{\mu k}$ and $u_{\nu l}$ through device $u_{\xi m}$:
    \begin{equation}
    \label{eq:virtualNodeParam}
        \theta_{i\hspace{-0.5pt}, \mu k\hspace{-0.5pt}.\hspace{-0.5pt}q \rightarrow\hspace{-0.5pt}j\hspace{-0.5pt}, \nu l\hspace{-0.5pt}.\hspace{-0.5pt}r}^{\xi m} = 
        \begin{cases}
            1, & \text{if $\mathcal{I}_{\mu k\hspace{-0.5pt}, \nu l}=\{ u_{\xi m} \}$},\\
            0, & \text{if $\mathcal{I}_{\mu k\hspace{-0.5pt}, \nu l}= \varnothing$}.
        \end{cases}
    \end{equation}

    \item $CL_{i\hspace{-0.5pt}, \mu k\hspace{-0.5pt}.\hspace{-0.5pt}q \rightarrow\hspace{-0.5pt}j\hspace{-0.5pt}, \nu l\hspace{-0.5pt}.\hspace{-0.5pt}r}$: time required to transfer the output data $D_{i}$ of task $\tau_i$ that is allocated on $p_{\mu k\hspace{-0.5pt}.\hspace{-0.5pt}q}$, to task $\tau_j$ that is allocated on $p_{\nu l\hspace{-0.5pt}.\hspace{-0.5pt}r}$:
    \begin{equation}
    \label{eq:arcLatency}
        CL_{i\hspace{-0.5pt}, \mu k\hspace{-0.5pt}.\hspace{-0.5pt}q \rightarrow\hspace{-0.5pt}j\hspace{-0.5pt}, \nu l\hspace{-0.5pt}.\hspace{-0.5pt}r} =  
        \begin{cases}
            \frac {D_i}{\epsilon_{\mu k\hspace{-0.5pt}, \nu l}}, & \hspace{-58pt} \text{if $\theta_{i\hspace{-0.5pt}, \mu k\hspace{-0.5pt}.\hspace{-0.5pt}q \rightarrow\hspace{-0.5pt}j\hspace{-0.5pt}, \nu l\hspace{-0.5pt}.\hspace{-0.5pt}r}^{\xi m} = 0$},\\ 
             & \hspace{-58pt} \text{$(\mu, k) \neq (\nu, l)$},\\
             D_i \left( \frac{1}{\epsilon_{\mu k\hspace{-0.5pt}, \xi m}} + \frac{1}{\epsilon_{\xi m\hspace{-0.5pt}, \nu l}} \right), &\\
             & \hspace{-58pt} \text{if $\theta_{i\hspace{-0.5pt}, \mu k\hspace{-0.5pt}.\hspace{-0.5pt}q \rightarrow\hspace{-0.5pt}j\hspace{-0.5pt}, \nu l\hspace{-0.5pt}.\hspace{-0.5pt}r}^{\xi m} = 1$},\\
             0, & \hspace{-58pt} \text{if $(\mu, k) = (\nu, l)$}.
        \end{cases}
    \end{equation}

    \item $CE_{i\hspace{-0.5pt}, \mu k\hspace{-0.5pt}.\hspace{-0.5pt}q \rightarrow\hspace{-0.5pt}j\hspace{-0.5pt}, \nu l\hspace{-0.5pt}.\hspace{-0.5pt}r}$: energy required to transfer $D_{i}$ from task $\tau_i$ (allocated on $p_{\mu k\hspace{-0.5pt}.\hspace{-0.5pt}q}$) to task $\tau_j$ (allocated on $p_{\nu l\hspace{-0.5pt}.\hspace{-0.5pt}r}$). It is given by \eqref{eq:arcEnergy}, based on the energy model in \cref{subsec:energyModel}.
\end{itemize}

\begin{figure*}[!t]
    \centering
    \includegraphics[width=\textwidth]{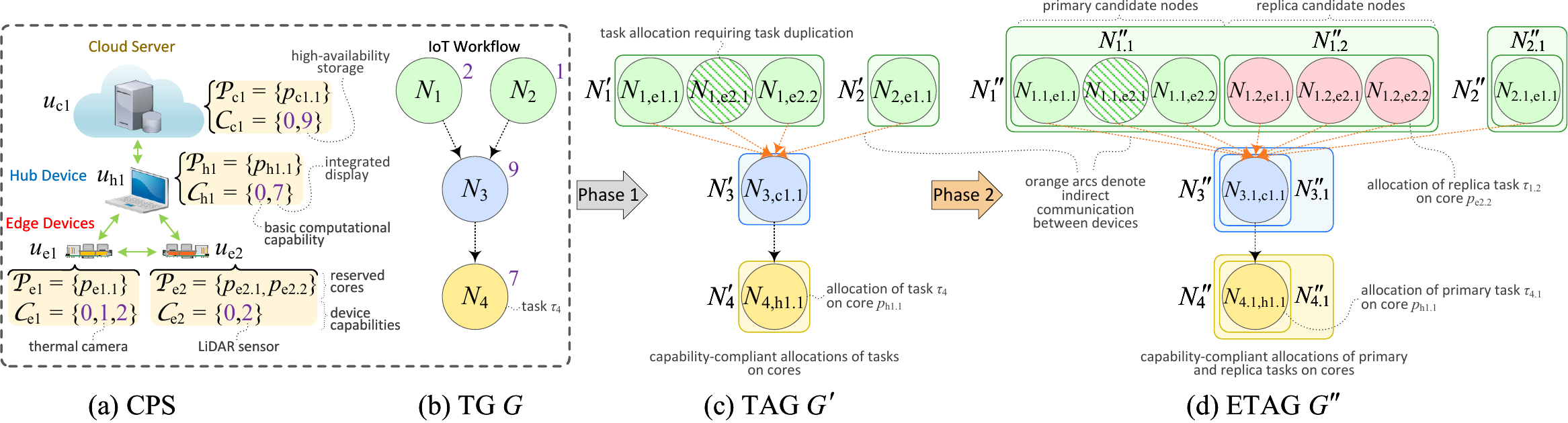}
    \caption{TG transformation example: (a) considered CPS, (b) TG $G$, (c) TAG $G^{\prime}$ (transformation phase 1), and (d) ETAG $G^{\prime \prime}$ (transformation phase 2).}
    \label{fig:example}
\end{figure*}

\subsubsection{Transformation Example (Phase 1)}
\label{subsubsec:examplePhase1}
\cref{fig:example} depicts an example of transforming a TG $G$ (\cref{fig:example}b), first into TAG $G^{\prime}$ (\cref{fig:example}c) and subsequently into ETAG $G^{\prime \prime}$ (\cref{fig:example}d), considering a CPS with two edge devices $u_{\mathrm{e}1}$ and $u_{\mathrm{e}2}$, a hub device $u_{\mathrm{h}1}$, and a cloud server $u_{\mathrm{c}1}$, communicating as shown in \cref{fig:example}a.
The sets of reserved cores $\mathcal{P}_{\mu k}$ and capabilities $\mathcal{C}_{\mu k}$ are also specified for each device $u_{\mu k}$.
These capabilities (denoted by purple integers) are a subset of those defined in the real-world use case, as shown in \cref{tab:capabilities}.
TG $G$ (\cref{fig:example}b) comprises four tasks, with entry and exit tasks shown in green and yellow, respectively, and intermediate tasks depicted in blue. 
The required capability of each task is indicated by its adjacent purple integer.
In phase 1 of the transformation (\cref{fig:example}c), TAG $G^{\prime}$ results from all applicable allocations of the tasks on the cores of system devices, taking into account the capability requirement of each task and the capabilities featured by each device.
Phase 2 is described in \cref{subsubsec:examplePhase2}.

\subsection{Energy Model}
\label{subsec:energyModel}
The energy required to execute task $\tau_i$ on core $p_{\mu k\hspace{-0.5pt}.\hspace{-0.5pt}q}$ is given by:
\begin{equation}
\label{eq:nodeEnergy}
E_{i\hspace{-0.5pt}, \mu k\hspace{-0.5pt}.\hspace{-0.5pt}q} = P_{i\hspace{-0.5pt}, \mu k\hspace{-0.5pt}.\hspace{-0.5pt}q} \, L_{i\hspace{-0.5pt}, \mu k\hspace{-0.5pt}.\hspace{-0.5pt}q}.
\end{equation}
The energy required to transfer the output data $D_{i}$ of parent task $\tau_i$ that is allocated on $p_{\mu k\hspace{-0.5pt}.\hspace{-0.5pt}q}$ to child task $\tau_j$ that is allocated on $p_{\nu l\hspace{-0.5pt}.\hspace{-0.5pt}r}$ is defined as:
\begin{equation}
\label{eq:arcEnergy}
    CE_{i\hspace{-0.5pt}, \mu k\hspace{-0.5pt}.\hspace{-0.5pt}q \rightarrow\hspace{-0.5pt}j\hspace{-0.5pt}, \nu l\hspace{-0.5pt}.\hspace{-0.5pt}r} \hspace{-1pt} = \hspace{-2pt} 
    \begin{cases}
         \hspace{-1pt} D_i \left( \pi_{\mu k\hspace{-0.5pt}, \nu l} \hspace{-1pt} + \hspace{-1pt} \sigma_{\mu k\hspace{-0.5pt}, \nu l} \right), & \hspace{-83pt} \text{if $\theta_{i\hspace{-0.5pt}, \mu k\hspace{-0.5pt}.\hspace{-0.5pt}q \rightarrow\hspace{-0.5pt}j\hspace{-0.5pt}, \nu l\hspace{-0.5pt}.\hspace{-0.5pt}r}^{\xi m} \hspace{-1pt} = \hspace{-1pt} 0$},\\
         & \hspace{-83pt} \text{$(\mu, k) \neq (\nu, l)$},\\

         \hspace{-1pt} D_i ( \pi_{\mu k\hspace{-0.5pt}, \xi m} \hspace{-1pt} + \hspace{-1pt} \sigma_{\mu k\hspace{-0.5pt}, \xi m} \hspace{-1pt} + \hspace{-1pt} \pi_{\xi m\hspace{-0.5pt}, \nu l} \hspace{-1pt}  + \hspace{-1pt}  \sigma_{\xi m\hspace{-0.5pt}, \nu l} ), &\\
         & \hspace{-83pt} \text{if $\theta_{i\hspace{-0.5pt}, \mu k\hspace{-0.5pt}.\hspace{-0.5pt}q \rightarrow\hspace{-0.5pt}j\hspace{-0.5pt}, \nu l\hspace{-0.5pt}.\hspace{-0.5pt}r}^{\xi m} \hspace{-1pt} = \hspace{-1pt} 1$},\\
         
         \hspace{-1pt} 0, & \hspace{-83pt} \text{if $(\mu, k) \hspace{-1pt} = \hspace{-1pt} (\nu, l)$}.
    \end{cases}
\end{equation}
This energy model captures both the computational and communication energy required for workflow execution in the considered edge-hub-cloud architecture, while remaining independent of low-level power management mechanisms to ensure broader applicability \cite{Tuli2022, Saeedi2020}.

\subsection{Reliability Model}
\label{subsec:reliabilityModel}
Cores are subject to transient failures caused by hardware faults (e.g., soft errors), which may disrupt task execution.
Such failures are typically modeled using the Poisson distribution \cite{Tang2021, Mo2022, Ye2025}.
This model is standard and appropriate for such faults, whose occurrence is dominated by stochastic phenomena (e.g., cosmic rays) rather than cumulative wear-out effects.
We assume that failures are detected by the underlying hardware or runtime environment \cite{Liu2019, Mo2022, Mo2023, Ye2025, Tang2021}.
This commonly used assumption allows us to focus on design-time reliability enhancement through selective task duplication, while abstracting from runtime-specific failure handling mechanisms.  
The reliability of task $\tau_i$ on core $p_{\mu k\hspace{-0.5pt}.\hspace{-0.5pt}q}$ over the duration of its execution $L_{i\hspace{-0.5pt}, \mu k\hspace{-0.5pt}.\hspace{-0.5pt}q}$ is given by \cite{Biswas2024}:
\begin{equation}
\label{eq:nodeReliability}
R_{i\hspace{-0.5pt}, \mu k\hspace{-0.5pt}.\hspace{-0.5pt}q} = e^{- \lambda_{\mu k\hspace{-0.5pt}.\hspace{-0.5pt}q} \, L_{i\hspace{-0.5pt}, \mu k\hspace{-0.5pt}.\hspace{-0.5pt}q}},
\end{equation}
where $\lambda_{\mu k\hspace{-0.5pt}.\hspace{-0.5pt}q}$ is the failure rate of $p_{\mu k\hspace{-0.5pt}.\hspace{-0.5pt}q}$.
$R_{i\hspace{-0.5pt}, \mu k\hspace{-0.5pt}.\hspace{-0.5pt}q}$ denotes the probability of executing $\tau_i$ on $p_{\mu k\hspace{-0.5pt}.\hspace{-0.5pt}q}$ without any failures. 
From \eqref{eq:nodeReliability}, it holds that $R_{i\hspace{-0.5pt}, \mu k\hspace{-0.5pt}.\hspace{-0.5pt}q} \in (0,1]$.
The probability that the execution of $\tau_i$ on $p_{\mu k\hspace{-0.5pt}.\hspace{-0.5pt}q}$ fails is equal to $1 - R_{i\hspace{-0.5pt}, \mu k\hspace{-0.5pt}.\hspace{-0.5pt}q} \in [0,1)$. 
If the reliability of $\tau_i$ on $p_{\mu k\hspace{-0.5pt}.\hspace{-0.5pt}q}$ is below its reliability threshold ($\zeta_{i\hspace{-0.5pt}, \mu k\hspace{-0.5pt}.\hspace{-0.5pt}q}=1$), we duplicate $\tau_i$, yielding a \emph{primary} task $\tau_{i\hspace{-0.5pt}.\hspace{-0.5pt}1}$ and a \emph{replica} task $\tau_{i\hspace{-0.5pt}.\hspace{-0.5pt}2}$. 
In this case, the total reliability of $\tau_i$ is defined as \cite{Xie2020}:
\begin{equation}
\label{eq:duplicationReliability}
R_{i\hspace{-0.5pt}, \mu k\hspace{-0.5pt}.\hspace{-0.5pt}q\hspace{-0.5pt}, \nu l\hspace{-0.5pt}.\hspace{-0.5pt}r} = 1 - \left( 1 - R_{i\hspace{-0.5pt}, \mu k\hspace{-0.5pt}.\hspace{-0.5pt}q} \right) \left( 1 - R_{i\hspace{-0.5pt}, \nu l\hspace{-0.5pt}.\hspace{-0.5pt}r} \right),
\end{equation}
where $R_{i\hspace{-0.5pt}, \mu k\hspace{-0.5pt}.\hspace{-0.5pt}q}$ and $R_{i\hspace{-0.5pt}, \nu l\hspace{-0.5pt}.\hspace{-0.5pt}r}$ denote the reliabilities of $\tau_{i\hspace{-0.5pt}.\hspace{-0.5pt}1}$ (when allocated on $p_{\mu k\hspace{-0.5pt}.\hspace{-0.5pt}q}$) and $\tau_{i\hspace{-0.5pt}.\hspace{-0.5pt}2}$ (when allocated on $p_{\nu l\hspace{-0.5pt}.\hspace{-0.5pt}r}$), respectively. 
Thus, $R_{i\hspace{-0.5pt}, \mu k\hspace{-0.5pt}.\hspace{-0.5pt}q\hspace{-0.5pt}, \nu l\hspace{-0.5pt}.\hspace{-0.5pt}r}$ represents the probability that at least one of $\tau_{i\hspace{-0.5pt}.\hspace{-0.5pt}1}$ and $\tau_{i\hspace{-0.5pt}.\hspace{-0.5pt}2}$ executes without failure.
By definition, if $R_{i\hspace{-0.5pt}, \mu k\hspace{-0.5pt}.\hspace{-0.5pt}q} < 1$ and $R_{i\hspace{-0.5pt}, \nu l\hspace{-0.5pt}.\hspace{-0.5pt}r} < 1$, then the total reliability of $\tau_i$ is $R_{i\hspace{-0.5pt}, \mu k\hspace{-0.5pt}.\hspace{-0.5pt}q\hspace{-0.5pt}, \nu l\hspace{-0.5pt}.\hspace{-0.5pt}r} > \max (R_{i\hspace{-0.5pt}, \mu k\hspace{-0.5pt}.\hspace{-0.5pt}q}, R_{i\hspace{-0.5pt}, \nu l\hspace{-0.5pt}.\hspace{-0.5pt}r})$. Otherwise, if $R_{i\hspace{-0.5pt}, \mu k\hspace{-0.5pt}.\hspace{-0.5pt}q} = 1$ or $R_{i\hspace{-0.5pt}, \nu l\hspace{-0.5pt}.\hspace{-0.5pt}r} = 1$, then $R_{i\hspace{-0.5pt}, \mu k\hspace{-0.5pt}.\hspace{-0.5pt}q\hspace{-0.5pt}, \nu l\hspace{-0.5pt}.\hspace{-0.5pt}r} = 1$. Consequently, by duplicating task $\tau_i$ we probabilistically improve its reliability.

Both the primary task $\tau_{i\hspace{-0.5pt}.\hspace{-0.5pt}1}$ and its replica $\tau_{i\hspace{-0.5pt}.\hspace{-0.5pt}2}$ have the same resource and capability requirements, and share identical communication and precedence relationships with the same parent/child tasks.
Consequently, the replica task $\tau_{i\hspace{-0.5pt}.\hspace{-0.5pt}2}$ is scheduled under the same constraints as its primary counterpart $\tau_{i\hspace{-0.5pt}.\hspace{-0.5pt}1}$, except for an additional restriction applied when $\tau_{i}$ is an exit task ($\phi_i = 1$). In that case, $\tau_{i\hspace{-0.5pt}.\hspace{-0.5pt}1}$ and $\tau_{i\hspace{-0.5pt}.\hspace{-0.5pt}2}$ must be allocated on the same device to ensure that a single final output is retained or a single actuator action is performed.
Otherwise, if $\tau_{i}$ is not an exit task ($\phi_i = 0$), $\tau_{i\hspace{-0.5pt}.\hspace{-0.5pt}1}$ and $\tau_{i\hspace{-0.5pt}.\hspace{-0.5pt}2}$ may be allocated on the same or different devices/cores. 
It is noted that if the reliability of $\tau_i$ on $p_{\mu k\hspace{-0.5pt}.\hspace{-0.5pt}q}$ is greater than or equal to its reliability threshold ($\zeta_{i\hspace{-0.5pt}, \mu k\hspace{-0.5pt}.\hspace{-0.5pt}q}=0$), no task duplication is required.
Hence, a selective task duplication strategy is employed, avoiding unnecessary task replicas and the additional overhead they entail.

\subsection{TG Transformation Phase 2 (ETAG)}
\label{subsec:ETAG}
We transform TAG $G^{\prime}$ into ETAG $G^{\prime \prime}=\left(\mathcal{N}^{\prime \prime}, \mathcal{A}^{\prime \prime}\right)$, by applying the selective task duplication technique described in \cref{subsec:reliabilityModel}.
Specifically, we transform each composite node $N_i^{\prime} \in \mathcal{N}^{\prime}$ in $G^{\prime}$ into another composite node $N_i^{\prime \prime} \in \mathcal{N}^{\prime \prime}$ in $G^{\prime \prime}$, such that:
\begin{equation}
\label{eq:ETAGnodes}
     N_i^{\prime \prime} = N_{i\hspace{-0.5pt}.\hspace{-0.5pt}1}^{\prime \prime} \cup N_{i\hspace{-0.5pt}.\hspace{-0.5pt}2}^{\prime \prime}.
\end{equation}
Similar to $N_i^{\prime}$ in \eqref{eq:TAGnodes}, $N_{i\hspace{-0.5pt}.\hspace{-0.5pt}1}^{\prime \prime}$ is the set of nodes that represent the applicable allocations of primary task $\tau_{i\hspace{-0.5pt}.\hspace{-0.5pt}1}$ on the reserved cores of the devices featuring its required capability. Thus, $N_{i\hspace{-0.5pt}.\hspace{-0.5pt}1}^{\prime \prime}$ is defined as:
\begin{equation}
\label{eq:ETAGprimaryNodes}
N_{i\hspace{-0.5pt}.\hspace{-0.5pt}1}^{\prime \prime} = \left\{ N_{i\hspace{-0.5pt}.\hspace{-0.5pt}1, \mu k\hspace{-0.5pt}.\hspace{-0.5pt}q} \, | \, N_{i\hspace{-0.5pt}, \mu k\hspace{-0.5pt}.\hspace{-0.5pt}q} \in N_i^{\prime} \right\}.
\end{equation}
$N_{i\hspace{-0.5pt}.\hspace{-0.5pt}2}^{\prime \prime}$ is the set of nodes denoting the allocations of replica $\tau_{i\hspace{-0.5pt}.\hspace{-0.5pt}2}$ on the same cores as primary task $\tau_{i\hspace{-0.5pt}.\hspace{-0.5pt}1}$, i.e.:
\begin{equation}
\label{eq:ETAGreplicaNodes}
\hspace{-6pt} N_{i\hspace{-0.5pt}.\hspace{-0.5pt}2}^{\prime \prime} \hspace{-2pt} = \hspace{-2.5pt} \{ \hspace{-0.5pt} N_{i\hspace{-0.5pt}.\hspace{-0.5pt}2\hspace{-0.5pt}, \mu k\hspace{-0.5pt}.\hspace{-0.5pt}q} \, | \, N_{i\hspace{-0.5pt}.\hspace{-0.5pt}1\hspace{-0.5pt}, \mu k\hspace{-0.5pt}.\hspace{-0.5pt}q} \hspace{-2.5pt} \in \hspace{-2.5pt} N_{i\hspace{-0.5pt}.\hspace{-0.5pt}1}^{\prime \prime}\hspace{-0.5pt}, \exists \, N_{i\hspace{-0.5pt}, \nu l\hspace{-0.5pt}.\hspace{-0.5pt}r} \hspace{-2.5pt} \in \hspace{-2.5pt} N_i^{\prime} \hspace{-2.5pt} : \hspace{-1.5pt} \zeta_{i\hspace{-0.5pt}, \nu l\hspace{-0.5pt}.\hspace{-0.5pt}r} \hspace{-2.5pt} = \hspace{-2pt} 1 \hspace{-0.5pt}\}.
\end{equation}
Hence, if there exists at least one node in $N_i^{\prime}$ (i.e., at least one allocation of $\tau_i$) that requires task duplication, then $N_{i\hspace{-0.5pt}.\hspace{-0.5pt}2}^{\prime \prime} \neq \varnothing$, otherwise $N_{i\hspace{-0.5pt}.\hspace{-0.5pt}2}^{\prime \prime} = \varnothing$.
An individual node $N_{i\hspace{-0.5pt}.\hspace{-0.5pt}n\hspace{-0.5pt}, \mu k\hspace{-0.5pt}.\hspace{-0.5pt}q} \in N_{i\hspace{-0.5pt}.\hspace{-0.5pt}n}^{\prime \prime}$, where  $n \in \{1,2\}$, represents the possible allocation of primary (if $n=1$) or replica (if $n=2$) task $\tau_{i\hspace{-0.5pt}.\hspace{-0.5pt}n}$ on a specific core $p_{\mu k\hspace{-0.5pt}.\hspace{-0.5pt}q}$. 
We refer to $N_{i\hspace{-0.5pt}.\hspace{-0.5pt}n\hspace{-0.5pt}, \mu k\hspace{-0.5pt}.\hspace{-0.5pt}q}$ as a \emph{candidate node}. If $n=1$, $N_{i\hspace{-0.5pt}.\hspace{-0.5pt}n\hspace{-0.5pt}, \mu k\hspace{-0.5pt}.\hspace{-0.5pt}q}$ is a primary candidate node, else if $n=2$, $N_{i\hspace{-0.5pt}.\hspace{-0.5pt}n\hspace{-0.5pt}, \mu k\hspace{-0.5pt}.\hspace{-0.5pt}q}$ is a replica candidate node.

We transform each composite arc $A_{i \rightarrow\hspace{-0.5pt}j}^{\prime} \in \mathcal{A}^{\prime}$ in $G^{\prime}$ into another composite arc $A^{\prime \prime}_{i \rightarrow\hspace{-0.5pt}j} \in \mathcal{A}^{\prime \prime}$ in $G^{\prime \prime}$, such that:
\begin{equation}
 \label{eq:ETAGarcs}
 \begin{split}
     A^{\prime \prime}_{i \rightarrow\hspace{-0.5pt}j}  \hspace{-2pt} = & \{ \hspace{-0.5pt} A_{i\hspace{-0.5pt}.\hspace{-0.5pt}n\hspace{-0.5pt}, \mu k\hspace{-0.5pt}.\hspace{-0.5pt}q \rightarrow\hspace{-0.5pt}j\hspace{-0.5pt}.\hspace{-0.5pt}o\hspace{-0.5pt}, \nu l\hspace{-0.5pt}.\hspace{-0.5pt}r} \hspace{-2pt} = \hspace{-2.5pt} N_{i\hspace{-0.5pt}.\hspace{-0.5pt}n\hspace{-0.5pt}, \mu k\hspace{-0.5pt}.\hspace{-0.5pt}q} \hspace{-3pt} \rightarrow \hspace{-3pt} N_{j\hspace{-0.5pt}.\hspace{-0.5pt}o\hspace{-0.5pt}, \nu l\hspace{-0.5pt}.\hspace{-0.5pt}r}\\
     &\,\,\, | \, A_{i\hspace{-0.5pt}, \mu k\hspace{-0.5pt}.\hspace{-0.5pt}q \rightarrow\hspace{-0.5pt}j\hspace{-0.5pt}, \nu l\hspace{-0.5pt}.\hspace{-0.5pt}r} \hspace{-2.5pt} \in \hspace{-2.5pt} A_{i \rightarrow\hspace{-0.5pt}j}^{\prime}\hspace{-0.5pt}, \, N_{i\hspace{-0.5pt}.\hspace{-0.5pt}n\hspace{-0.5pt}, \mu k\hspace{-0.5pt}.\hspace{-0.5pt}q} \hspace{-2.5pt} \in \hspace{-2.5pt} N_i^{\prime \prime}\hspace{-1pt}, \, N_{j\hspace{-0.5pt}.\hspace{-0.5pt}o\hspace{-0.5pt}, \nu l\hspace{-0.5pt}.\hspace{-0.5pt}r} \hspace{-2.5pt} \in \hspace{-2.5pt} N_j^{\prime \prime} \}.
\end{split}
\end{equation}
$N_i^{\prime \prime}$ and $N_j^{\prime \prime}$ include the primary and replica candidate nodes of parent task $\tau_i$ and child task $\tau_j$, respectively. 
Thus, each pair of candidate nodes of tasks $\tau_i$ and $\tau_j$ is connected by an individual arc in $A^{\prime \prime}_{i \rightarrow\hspace{-0.5pt}j}$.

\subsubsection{ETAG Candidate Node Parameters}
\label{subsubsec:ETAGnodeParams}
A primary or replica candidate node $N_{i\hspace{-0.5pt}.\hspace{-0.5pt}n\hspace{-0.5pt}, \mu k\hspace{-0.5pt}.\hspace{-0.5pt}q} \hspace{-2pt} \in \hspace{-2pt} G^{\prime \prime}$ has the same parameters as $N_{i\hspace{-0.5pt}, \mu k\hspace{-0.5pt}.\hspace{-0.5pt}q} \hspace{-2pt} \in \hspace{-2pt} G^{\prime}$ (\cref{subsubsec:TAGnodeParams}). 
Additionally, if $N_{i\hspace{-0.5pt}.\hspace{-0.5pt}n\hspace{-0.5pt}, \mu k\hspace{-0.5pt}.\hspace{-0.5pt}q}$ is a primary candidate node ($n \hspace{-1pt} = \hspace{-1pt} 1$), it has the following parameter, which denotes the total reliability of task $\tau_{i}$:
\begin{equation}
\label{eq:totalReliability}
    \hat{R}_{i\hspace{-0.5pt}, \mu k\hspace{-0.5pt}.\hspace{-0.5pt}q} =
    \begin{cases}
    R_{i\hspace{-0.5pt}, \mu k\hspace{-0.5pt}.\hspace{-0.5pt}q\hspace{-0.5pt}, \nu l\hspace{-0.5pt}.\hspace{-0.5pt}r}, & \text{if $\zeta_{i\hspace{-0.5pt}, \mu k\hspace{-0.5pt}.\hspace{-0.5pt}q} = 1, \, N_{i\hspace{-0.5pt}.\hspace{-0.5pt}2\hspace{-0.5pt}, \nu l\hspace{-0.5pt}.\hspace{-0.5pt}r} \in N_{i\hspace{-0.5pt}.\hspace{-0.5pt}2}^{\prime \prime}$},\\
    R_{i\hspace{-0.5pt}, \mu k\hspace{-0.5pt}.\hspace{-0.5pt}q}, & \text{if $\zeta_{i\hspace{-0.5pt}, \mu k\hspace{-0.5pt}.\hspace{-0.5pt}q} = 0$},
    \end{cases}
\end{equation}
where $R_{i\hspace{-0.5pt}, \mu k\hspace{-0.5pt}.\hspace{-0.5pt}q\hspace{-0.5pt}, \nu l\hspace{-0.5pt}.\hspace{-0.5pt}r}$ is given by \eqref{eq:duplicationReliability}. 
Hence, if task duplication is required ($\zeta_{i\hspace{-0.5pt}, \mu k\hspace{-0.5pt}.\hspace{-0.5pt}q} = 1$), $\hat{R}_{i\hspace{-0.5pt}, \mu k\hspace{-0.5pt}.\hspace{-0.5pt}q}$ is the probability that $\tau_{i\hspace{-0.5pt}.\hspace{-0.5pt}1}$, $\tau_{i\hspace{-0.5pt}.\hspace{-0.5pt}2}$, or both execute without failure. Otherwise ($\zeta_{i\hspace{-0.5pt}, \mu k\hspace{-0.5pt}.\hspace{-0.5pt}q} = 0$), $\hat{R}_{i\hspace{-0.5pt}, \mu k\hspace{-0.5pt}.\hspace{-0.5pt}q}$ is the probability that solely $\tau_{i\hspace{-0.5pt}.\hspace{-0.5pt}1}$ executes without failure.

\subsubsection{ETAG Arc Parameters}
\label{subsubsec:ETAGarcParams}
An arc $A_{i\hspace{-0.5pt}.\hspace{-0.5pt}n\hspace{-0.5pt}, \mu k\hspace{-0.5pt}.\hspace{-0.5pt}q \rightarrow\hspace{-0.5pt}j\hspace{-0.5pt}.\hspace{-0.5pt}o\hspace{-0.5pt}, \nu l\hspace{-0.5pt}.\hspace{-0.5pt}r} \in G^{\prime \prime}$ has the same parameters as $A_{i\hspace{-0.5pt}, \mu k\hspace{-0.5pt}.\hspace{-0.5pt}q \rightarrow\hspace{-0.5pt}j\hspace{-0.5pt}, \nu l\hspace{-0.5pt}.\hspace{-0.5pt}r} \in G^{\prime}$.

\subsubsection{Transformation Example (Phase 2)}
\label{subsubsec:examplePhase2}
Continuing from \cref{subsubsec:examplePhase1}, TAG $G^{\prime}$ (\cref{fig:example}c) is transformed into ETAG $G^{\prime \prime}$ (\cref{fig:example}d), driven by the reliability requirements of the tasks. 
Specifically, as task allocation represented by node $N_{1\hspace{-0.5pt},\mathrm{e}2\hspace{-0.5pt}.\hspace{-0.5pt}1} \in G^{\prime}$ requires task duplication, composite node $N_1^{\prime} \hspace{-2pt} \in \hspace{-2pt} G^{\prime}$ is transformed into composite node $N_1^{\prime \prime} \hspace{-2pt} \in \hspace{-2pt} G^{\prime \prime}$, comprising the set of primary candidate nodes $N_{1\hspace{-0.5pt}.\hspace{-0.5pt}1}^{\prime \prime} \hspace{-2pt} = \hspace{-2pt} \{ N_{1\hspace{-0.5pt}.\hspace{-0.5pt}1\hspace{-0.5pt}, \mathrm{e}1\hspace{-0.5pt}.\hspace{-0.5pt}1}, \allowbreak N_{1\hspace{-0.5pt}.\hspace{-0.5pt}1\hspace{-0.5pt}, \mathrm{e}2\hspace{-0.5pt}.\hspace{-0.5pt}1}, \allowbreak 
N_{1\hspace{-0.5pt}.\hspace{-0.5pt}1\hspace{-0.5pt}, \mathrm{e}2\hspace{-0.5pt}.\hspace{-0.5pt}2} \}$
and the set of replica candidate nodes 
$N_{1\hspace{-0.5pt}.\hspace{-0.5pt}2}^{\prime \prime} \hspace{-2pt} = \hspace{-2pt} \{ N_{1\hspace{-0.5pt}.\hspace{-0.5pt}2\hspace{-0.5pt}, \mathrm{e}1\hspace{-0.5pt}.\hspace{-0.5pt}1}, \allowbreak N_{1\hspace{-0.5pt}.\hspace{-0.5pt}2\hspace{-0.5pt}, \mathrm{e}2\hspace{-0.5pt}.\hspace{-0.5pt}1}, \allowbreak 
N_{1\hspace{-0.5pt}.\hspace{-0.5pt}2\hspace{-0.5pt}, \mathrm{e}2\hspace{-0.5pt}.\hspace{-0.5pt}2} \}$ 
(depicted in pink). 
As there is no capability-compliant allocation of the remaining tasks that requires task duplication, only a set of primary candidate nodes is generated in $G^{\prime \prime}$ for these tasks.

\subsubsection{ETAG Size Analysis}
\label{subsubsec:etagSize}
In the worst case of our transformation technique, where all devices feature all capabilities in $\mathcal{C}$ and all tasks require duplication, for $|\mathcal{P}|$ cores, the number of candidate nodes and arcs in ETAG $G^{\prime \prime}$ increases by $2|\mathcal{P}|$ and $4|\mathcal{P}|^2$ times, respectively, compared to those in TG $G$.

\subsection{MILP Problem Formulation}
\label{subsec:milp}

\subsubsection{Conceptual Overview}
\label{subsubsec:formulationOverview}

We leverage the derived ETAG $G^{\prime \prime}$ to formulate the problem as a continuous-time multi-objective MILP model. The proposed formulation simultaneously decides (a) where (on which core) a task (and its potential replica) is allocated, (b) whether each task allocated on a specific core requires duplication to satisfy its reliability requirements, and (c) when each task (and its potential replica) starts execution, while minimizing latency and energy and maximizing reliability, subject to timing and resource constraints.
We use binary decision variables to select exactly one feasible allocation for each primary task and, when required, at most one replica allocation. Continuous variables determine task start times, while auxiliary binary variables are introduced to preserve linearity when modeling conditional relationships and ordering decisions between tasks, and time-dependent cumulative resource constraints.
We employ a combined objective function that captures the inherent trade-offs between latency, energy, and reliability. Selection constraints ensure valid task allocations, reliability constraints activate task duplication only when necessary, temporal constraints preserve precedence and prevent task overlap on shared cores, and cumulative resource constraints enforce device capability, memory, storage, and energy limits over time.

\subsubsection{Key Design Choices \& Modeling Rationale}
\label{subsubsec:designChoices}
To achieve a realistic model, the proposed approach adopts a continuous-time formulation rather than a discrete-time one. Discrete-time formulations restrict events to predefined time slots, which may not adequately capture the continuous nature of task execution and resource usage.
In the proposed formulation, time-dependent cumulative constraints for the concurrent use of limited resources such as device capabilities, memory, and storage are enforced only at task start times. This event-driven enforcement is sufficient, as violations of such constraints may occur only when a task begins execution. Consequently, we avoid partitioning the entire scheduling horizon into time intervals and thus introducing unnecessary variables.
The size of the resulting ETAG increases in the worst case as analyzed in \cref{subsubsec:etagSize}. However, in practice, this growth is significantly constrained, as capability requirements prune infeasible task allocations during the initial transformation phase (\cref{subsec:TAG}), and selective task duplication generates replicas only for task allocations that violate reliability thresholds during the second transformation phase (\cref{subsec:ETAG}). As shown in our experiments in \cref{subsubsec:syntheticResults}, the resulting ETAGs have substantially fewer nodes and arcs than the theoretical worst case, ensuring that the MILP model is practically tractable for problem sizes relevant to the targeted applications.

\subsubsection{Decision Variables}
\label{subsubsec:variables}
We employ the following variables:
\begin{itemize}
    \item $x_{i\hspace{-0.5pt}.\hspace{-0.5pt}n\hspace{-0.5pt}, \mu k\hspace{-0.5pt}.\hspace{-0.5pt}q}$: binary variable corresponding to a (primary or replica) candidate node\,$N_{i\hspace{-0.5pt}.\hspace{-0.5pt}n\hspace{-0.5pt}, \mu k\hspace{-0.5pt}.\hspace{-0.5pt}q} \hspace{-1.5pt} \in \hspace{-1.5pt} G^{\prime \prime}$\!,\hspace{1.5pt}such\,that $x_{i\hspace{-0.5pt}.\hspace{-0.5pt}n\hspace{-0.5pt}, \mu k\hspace{-0.5pt}.\hspace{-0.5pt}q} \hspace{-1.5pt} = \hspace{-2pt} 1$ if $N_{i\hspace{-0.5pt}.\hspace{-0.5pt}n\hspace{-0.5pt}, \mu k\hspace{-0.5pt}.\hspace{-0.5pt}q}$ is selected, and 0 otherwise.
   
    \item $x_{i\hspace{-0.5pt}.\hspace{-0.5pt}n\hspace{-0.5pt}, \mu k\hspace{-0.5pt}.\hspace{-0.5pt}q \rightarrow\hspace{-0.5pt}j\hspace{-0.5pt}.\hspace{-0.5pt}o\hspace{-0.5pt}, \nu l\hspace{-0.5pt}.\hspace{-0.5pt}r}$: binary variable corresponding to an arc $A_{i\hspace{-0.5pt}.\hspace{-0.5pt}n\hspace{-0.5pt}, \mu k\hspace{-0.5pt}.\hspace{-0.5pt}q \rightarrow\hspace{-0.5pt}j\hspace{-0.5pt}.\hspace{-0.5pt}o\hspace{-0.5pt}, \nu l\hspace{-0.5pt}.\hspace{-0.5pt}r} \hspace{-2.5pt} \in \hspace{-2.5pt} G^{\prime \prime}$, such that $x_{i\hspace{-0.5pt}.\hspace{-0.5pt}n\hspace{-0.5pt}, \mu k\hspace{-0.5pt}.\hspace{-0.5pt}q \rightarrow\hspace{-0.5pt}j\hspace{-0.5pt}.\hspace{-0.5pt}o\hspace{-0.5pt}, \nu l\hspace{-0.5pt}.\hspace{-0.5pt}r} \hspace{-2pt} = \hspace{-2pt} 1$ if $A_{i\hspace{-0.5pt}.\hspace{-0.5pt}n\hspace{-0.5pt}, \mu k\hspace{-0.5pt}.\hspace{-0.5pt}q \rightarrow\hspace{-0.5pt}j\hspace{-0.5pt}.\hspace{-0.5pt}o\hspace{-0.5pt}, \nu l\hspace{-0.5pt}.\hspace{-0.5pt}r}$ is selected, and 0 otherwise.

    \item $t_{i\hspace{-0.5pt}.\hspace{-0.5pt}n}$: continuous variable denoting the start time of (primary or replica) task $\tau_{i\hspace{-0.5pt}.\hspace{-0.5pt}n}$, corresponding to set $N_{i\hspace{-0.5pt}.\hspace{-0.5pt}n}^{\prime \prime} \hspace{-2pt} \in \hspace{-1pt} G^{\prime \prime}$.

    \item $T$: continuous variable indicating the completion time of the application.

    \item $x_{i\hspace{-0.5pt}, \mu k\hspace{-0.5pt}.\hspace{-0.5pt}q\hspace{-0.5pt}, \nu l\hspace{-0.5pt}.\hspace{-0.5pt}r}$: auxiliary binary variable corresponding to a primary candidate node $N_{i\hspace{-0.5pt}.\hspace{-0.5pt}1\hspace{-0.5pt}, \mu k\hspace{-0.5pt}.\hspace{-0.5pt}q}$ and a replica candidate node $N_{i\hspace{-0.5pt}.\hspace{-0.5pt}2\hspace{-0.5pt}, \nu l\hspace{-0.5pt}.\hspace{-0.5pt}r}$ of task $\tau_i$, such that $x_{i\hspace{-0.5pt}, \mu k\hspace{-0.5pt}.\hspace{-0.5pt}q\hspace{-0.5pt}, \nu l\hspace{-0.5pt}.\hspace{-0.5pt}r} = 1$ if both nodes are selected, and 0 otherwise. 
    It is used when task duplication is required and a primary and a replica candidate node of $\tau_i$ should be jointly considered, while preserving the linearity of the model. 

    \item $x_{i\hspace{-0.5pt}.\hspace{-0.5pt}n\hspace{-0.5pt}, j\hspace{-0.5pt}.\hspace{-0.5pt}o}$: auxiliary binary variable denoting whether task $\tau_{i\hspace{-0.5pt}.\hspace{-0.5pt}n}$ will be executed before task $\tau_{j\hspace{-0.5pt}.\hspace{-0.5pt}o}$ ($x_{i\hspace{-0.5pt}.\hspace{-0.5pt}n\hspace{-0.5pt}, j\hspace{-0.5pt}.\hspace{-0.5pt}o} = 1$) or not ($x_{i\hspace{-0.5pt}.\hspace{-0.5pt}n\hspace{-0.5pt}, j\hspace{-0.5pt}.\hspace{-0.5pt}o} = 0$). It is used when both tasks are allocated on the same core and there is no precedence relationship between them, and thus their order of execution should be determined to avoid overlap.

    \item $x_h^{i\hspace{-0.5pt}.\hspace{-0.5pt}n\hspace{-0.5pt}, \mu k\hspace{-0.5pt}.\hspace{-0.5pt}q}$: auxiliary binary variable indicating whether task $\tau_{i\hspace{-0.5pt}.\hspace{-0.5pt}n}$ will be in execution on its allocated core $p_{\mu k\hspace{-0.5pt}.\hspace{-0.5pt}q}$ at time $s_h$ ($x_h^{i\hspace{-0.5pt}.\hspace{-0.5pt}n\hspace{-0.5pt}, \mu k\hspace{-0.5pt}.\hspace{-0.5pt}q} \hspace{-0.5pt} = \hspace{-0.5pt} 1$) or not ($x_h^{i\hspace{-0.5pt}.\hspace{-0.5pt}n\hspace{-0.5pt}, \mu k\hspace{-0.5pt}.\hspace{-0.5pt}q} \hspace{-0.5pt} = \hspace{-0.5pt} 0$). It is used to prevent exceeding the capability, memory, and storage capacity of a device during the execution of its assigned tasks. 
    For this purpose, given that the time interval in which a task $\tau_{i\hspace{-0.5pt}.\hspace{-0.5pt}n}$ will be in execution on $p_{\mu k\hspace{-0.5pt}.\hspace{-0.5pt}q}$ is $[t_{i\hspace{-0.5pt}.\hspace{-0.5pt}n}, t_{i\hspace{-0.5pt}.\hspace{-0.5pt}n} + L_{i\hspace{-0.5pt}, \mu k\hspace{-0.5pt}.\hspace{-0.5pt}q})$, we consider a set encompassing the start times of all primary and replica tasks, i.e., $\mathcal{S} \hspace{-1pt} = \hspace{-1pt} \{s_h \,| \, s_h \hspace{-1pt} = \hspace{-1pt} t_{i\hspace{-0.5pt}.\hspace{-0.5pt}n}, \allowbreak \, N_{i\hspace{-0.5pt}.\hspace{-0.5pt}n}^{\prime \prime} \hspace{-1pt} \in \hspace{-1pt} \mathcal{N}^{\prime \prime} \}$.
    To preserve the linearity of the model, we use an additional related auxiliary binary variable $\hat{x}_h^{i\hspace{-0.5pt}.\hspace{-0.5pt}n\hspace{-0.5pt}, \mu k\hspace{-0.5pt}.\hspace{-0.5pt}q}$.
\end{itemize}

\subsubsection{Objectives}
\label{subsubsec:objectives}
We aim to minimize the overall latency and energy consumption, and maximize the overall reliability of the IoT workflow in the considered CPS.

\paragraph{Overall latency}
\label{par:latencyObjective}
It is equal to the completion time of the application:
\begin{equation}
\label{eq:latencyObjective}
    f_{\mathrm{lat}} = T.
\end{equation}

\paragraph{Overall energy}
\label{par:energyObjective}
It encompasses both the computational and communication energy required for the execution of the application (first and second terms in \eqref{eq:energyObjective}, respectively):
\begin{equation}
\label{eq:energyObjective}
\begin{split}
    f_{\mathrm{en}} = 
    & \hspace{-2.5pt} \sum_{N_{i\hspace{-0.5pt}.\hspace{-0.5pt}n\hspace{-0.5pt}, \mu k\hspace{-0.5pt}.\hspace{-0.5pt}q} \in \mathcal{N}^{\prime \prime}} \hspace{-13pt} E_{i\hspace{-0.5pt}, \mu k\hspace{-0.5pt}.\hspace{-0.5pt}q} \, x_{i\hspace{-0.5pt}.\hspace{-0.5pt}n\hspace{-0.5pt}, \mu k\hspace{-0.5pt}.\hspace{-0.5pt}q}\\ 
    & + \hspace{-26pt} \sum_{A_{i\hspace{-0.5pt}.\hspace{-0.5pt}n\hspace{-0.5pt}, \mu k\hspace{-0.5pt}.\hspace{-0.5pt}q \rightarrow\hspace{-0.5pt}j\hspace{-0.5pt}.\hspace{-0.5pt}o\hspace{-0.5pt}, \nu l\hspace{-0.5pt}.\hspace{-0.5pt}r} \in \mathcal{A}^{\prime \prime}} \hspace{-26pt} CE_{i\hspace{-0.5pt}, \mu k\hspace{-0.5pt}.\hspace{-0.5pt}q \rightarrow\hspace{-0.5pt}j\hspace{-0.5pt}, \nu l\hspace{-0.5pt}.\hspace{-0.5pt}r} \, x_{i\hspace{-0.5pt}.\hspace{-0.5pt}n\hspace{-0.5pt}, \mu k\hspace{-0.5pt}.\hspace{-0.5pt}q \rightarrow\hspace{-0.5pt}j\hspace{-0.5pt}.\hspace{-0.5pt}o\hspace{-0.5pt}, \nu l\hspace{-0.5pt}.\hspace{-0.5pt}r}.
\end{split}
\end{equation}

\paragraph{Overall reliability}
\label{par:reliabilityObjective}
It is defined as the probability that all tasks of the application are executed such that: (a) if a task is duplicated, the primary task, its replica, or both are executed without any failures (first term in \eqref{eq:overallReliability}), and (b) if a task is not duplicated, the primary task is executed without any failures (second term in \eqref{eq:overallReliability}) \cite{Xie2020}:
\begin{equation}
\label{eq:overallReliability}
\hat{R} \left( G^{\prime \prime} \right) = \hspace{-8pt} \prod_{\substack{N_{i\hspace{-0.5pt}.\hspace{-0.5pt}1\hspace{-0.5pt}, \mu k\hspace{-0.5pt}.\hspace{-0.5pt}q},\\ N_{i\hspace{-0.5pt}.\hspace{-0.5pt}2\hspace{-0.5pt}, \nu l\hspace{-0.5pt}.\hspace{-0.5pt}r} \in \mathcal{N}^{\prime \prime},\\ \zeta_{i\hspace{-0.5pt}, \mu k\hspace{-0.5pt}.\hspace{-0.5pt}q} = 1}} \hspace{-16pt} \left( R_{i\hspace{-0.5pt}, \mu k\hspace{-0.5pt}.\hspace{-0.5pt}q\hspace{-0.5pt}, \nu l\hspace{-0.5pt}.\hspace{-0.5pt}r} \right)^{x_{i\hspace{-0.5pt}, \mu k\hspace{-0.5pt}.\hspace{-0.5pt}q\hspace{-0.5pt}, \nu l\hspace{-0.5pt}.\hspace{-0.5pt}r}} \hspace{-16pt} \prod_{\substack{N_{i\hspace{-0.5pt}.\hspace{-0.5pt}1\hspace{-0.5pt}, \mu k\hspace{-0.5pt}.\hspace{-0.5pt}q} \in \mathcal{N}^{\prime \prime},\\ \zeta_{i\hspace{-0.5pt}, \mu k\hspace{-0.5pt}.\hspace{-0.5pt}q} = 0}} \hspace{-17pt} \left( R_{i\hspace{-0.5pt}, \mu k\hspace{-0.5pt}.\hspace{-0.5pt}q} \right)^{x_{i\hspace{-0.5pt}.\hspace{-0.5pt}1\hspace{-0.5pt}, \mu k\hspace{-0.5pt}.\hspace{-0.5pt}q}}.
\end{equation}

To define the objective function, we linearize \eqref{eq:overallReliability} by taking its natural logarithm \cite{Kherraf2019}:
\begin{equation}
\label{eq:reliabilityObjective}
\begin{split}
     f_{\mathrm{rel}} & = \ln \Big( \hat{R} \left( G^{\prime \prime} \right) \Big)\\
     & = \hspace{-4.4pt} \sum_{\substack{N_{i\hspace{-0.5pt}.\hspace{-0.5pt}1\hspace{-0.5pt}, \mu k\hspace{-0.5pt}.\hspace{-0.5pt}q},\\ N_{i\hspace{-0.5pt}.\hspace{-0.5pt}2\hspace{-0.5pt}, \nu l\hspace{-0.5pt}.\hspace{-0.5pt}r} \in \mathcal{N}^{\prime \prime},\\ \zeta_{i\hspace{-0.5pt}, \mu k\hspace{-0.5pt}.\hspace{-0.5pt}q} = 1}} \hspace{-13.25pt} \ln \left( R_{i\hspace{-0.5pt}, \mu k\hspace{-0.5pt}.\hspace{-0.5pt}q\hspace{-0.5pt}, \nu l\hspace{-0.5pt}.\hspace{-0.5pt}r} \right) x_{i\hspace{-0.5pt}, \mu k\hspace{-0.5pt}.\hspace{-0.5pt}q\hspace{-0.5pt}, \nu l\hspace{-0.5pt}.\hspace{-0.5pt}r}\\
     & \hspace{10pt} + \hspace{-14pt} \sum_{\substack{N_{i\hspace{-0.5pt}.\hspace{-0.5pt}1\hspace{-0.5pt}, \mu k\hspace{-0.5pt}.\hspace{-0.5pt}q} \in \mathcal{N}^{\prime \prime},\\ \zeta_{i\hspace{-0.5pt}, \mu k\hspace{-0.5pt}.\hspace{-0.5pt}q} = 0}} \hspace{-14pt} \ln \left( R_{i\hspace{-0.5pt}, \mu k\hspace{-0.5pt}.\hspace{-0.5pt}q} \right) x_{i\hspace{-0.5pt}.\hspace{-0.5pt}1\hspace{-0.5pt}, \mu k\hspace{-0.5pt}.\hspace{-0.5pt}q}.
\end{split}
\end{equation}

\paragraph{Multi-objective} 
\label{par:multiObjective}
Since latency, energy, and reliability vary in magnitude, we normalize their corresponding functions \eqref{eq:latencyObjective}, \eqref{eq:energyObjective}, and \eqref{eq:reliabilityObjective} in the interval $[0,1]$ \cite{Grodzevich2006}:
\begin{equation}
\label{eq:normalization}
\dot{f} = \frac{f - \min \left( f \right)} {\max \left( f \right)  - \min \left( f \right)},
\end{equation}
where $f$ represents $f_{\mathrm{lat}}$, $f_{\mathrm{en}}$, or $f_{\mathrm{rel}}$. Similarly, $\dot{f}$ denotes the normalized form of the respective function, i.e., $\dot{f}_{\mathrm{lat}}$, $\dot{f}_{\mathrm{en}}$, or $\dot{f}_{\mathrm{rel}}$.
Given that the first two objectives aim to minimize latency and energy, while the third aims to maximize reliability, we convert the third objective into a minimization problem by considering $\ddot{f}_{\mathrm{rel}} = - \dot{f}_{\mathrm{rel}}$.
We define the multi-objective function as the weighted sum of $\dot{f}_{\mathrm{lat}}$, $\dot{f}_{\mathrm{en}}$, and $\ddot{f}_{\mathrm{rel}}$ \cite{Grodzevich2006, Anka2025}:
\begin{equation}
\label{eq:multiObjective}
    g = w_{\mathrm{lat}} \, \dot{f}_{\mathrm{lat}} + w_{\mathrm{en}} \, \dot{f}_{\mathrm{en}} + w_{\mathrm{rel}} \, \ddot{f}_{\mathrm{rel}},
\end{equation}
where $w_{\mathrm{lat}}$, $w_{\mathrm{en}}$, and $w_{\mathrm{rel}}$ are predefined weights reflecting the relative importance of each objective ($0 \leq w_{\mathrm{lat}}, w_{\mathrm{en}}, w_{\mathrm{rel}} \leq 1$,  $w_{\mathrm{lat}} + w_{\mathrm{en}} + w_{\mathrm{rel}} = 1$). 
Hence, the problem is formulated as:
\begin{equation}
\label{eq:problem}
\min g
\end{equation}
subject to the constraints defined in \cref{subsubsec:constraints}.

\subsubsection{Constraints}
\label{subsubsec:constraints}
We consider the following constraints:

\paragraph{Candidate node selection constraints} 
Only one primary candidate node per task should be selected in $G^{\prime \prime}$:
\begin{equation}
\label{eq:one1}
    \sum_{N_{i\hspace{-0.5pt}.\hspace{-0.5pt}1\hspace{-0.5pt}, \mu k\hspace{-0.5pt}.\hspace{-0.5pt}q} \in N_i^{\prime \prime}} \hspace{-12pt} x_{i\hspace{-0.5pt}.\hspace{-0.5pt}1\hspace{-0.5pt}, \mu k\hspace{-0.5pt}.\hspace{-0.5pt}q} =  1, \, \forall \, N_i^{\prime \prime} \in \mathcal{N}^{\prime \prime}.
\end{equation}

\paragraph{Arc selection constraints}
If a parent and child candidate nodes are selected, their corresponding arc should be selected as well:
\begin{equation}
\label{eq:three1}
    x_{i\hspace{-0.5pt}.\hspace{-0.5pt}n\hspace{-0.5pt}, \mu k\hspace{-0.5pt}.\hspace{-0.5pt}q \rightarrow\hspace{-0.5pt}j\hspace{-0.5pt}.\hspace{-0.5pt}o\hspace{-0.5pt}, \nu l\hspace{-0.5pt}.\hspace{-0.5pt}r} \leq x_{i\hspace{-0.5pt}.\hspace{-0.5pt}n\hspace{-0.5pt}, \mu k\hspace{-0.5pt}.\hspace{-0.5pt}q}, \, \forall \, A_{i\hspace{-0.5pt}.\hspace{-0.5pt}n\hspace{-0.5pt}, \mu k\hspace{-0.5pt}.\hspace{-0.5pt}q \rightarrow\hspace{-0.5pt}j\hspace{-0.5pt}.\hspace{-0.5pt}o\hspace{-0.5pt}, \nu l\hspace{-0.5pt}.\hspace{-0.5pt}r} \in \mathcal{A}^{\prime \prime},
\end{equation}
\begin{equation}
\label{eq:three2}
    x_{i\hspace{-0.5pt}.\hspace{-0.5pt}n\hspace{-0.5pt}, \mu k\hspace{-0.5pt}.\hspace{-0.5pt}q \rightarrow\hspace{-0.5pt}j\hspace{-0.5pt}.\hspace{-0.5pt}o\hspace{-0.5pt}, \nu l\hspace{-0.5pt}.\hspace{-0.5pt}r} \leq x_{j\hspace{-0.5pt}.\hspace{-0.5pt}o\hspace{-0.5pt}, \nu l\hspace{-0.5pt}.\hspace{-0.5pt}r}, \, \forall \, A_{i\hspace{-0.5pt}.\hspace{-0.5pt}n\hspace{-0.5pt}, \mu k\hspace{-0.5pt}.\hspace{-0.5pt}q \rightarrow\hspace{-0.5pt}j\hspace{-0.5pt}.\hspace{-0.5pt}o\hspace{-0.5pt}, \nu l\hspace{-0.5pt}.\hspace{-0.5pt}r} \in \mathcal{A}^{\prime \prime},
\end{equation}
\begin{equation}
\label{eq:three3}
\begin{split}
    x_{i\hspace{-0.5pt}.\hspace{-0.5pt}n\hspace{-0.5pt}, \mu k\hspace{-0.5pt}.\hspace{-0.5pt}q \rightarrow\hspace{-0.5pt}j\hspace{-0.5pt}.\hspace{-0.5pt}o\hspace{-0.5pt}, \nu l\hspace{-0.5pt}.\hspace{-0.5pt}r} \geq x_{i\hspace{-0.5pt}.\hspace{-0.5pt}n\hspace{-0.5pt}, \mu k\hspace{-0.5pt}.\hspace{-0.5pt}q} + x_{j\hspace{-0.5pt}.\hspace{-0.5pt}o\hspace{-0.5pt}, \nu l\hspace{-0.5pt}.\hspace{-0.5pt}r} - 1, \\
    \forall \, A_{i\hspace{-0.5pt}.\hspace{-0.5pt}n\hspace{-0.5pt}, \mu k\hspace{-0.5pt}.\hspace{-0.5pt}q \rightarrow\hspace{-0.5pt}j\hspace{-0.5pt}.\hspace{-0.5pt}o\hspace{-0.5pt}, \nu l\hspace{-0.5pt}.\hspace{-0.5pt}r} \in \mathcal{A}^{\prime \prime}.
\end{split}
\end{equation}

\paragraph{Task reliability constraints}
If the selected primary candidate node requires task duplication, only one replica candidate node should be selected for the specific task. 
Otherwise, no replica candidate node should be selected:
\begin{equation}
\label{eq:one2}
    \sum_{N_{i\hspace{-0.5pt}.\hspace{-0.5pt}1\hspace{-0.5pt}, \mu k\hspace{-0.5pt}.\hspace{-0.5pt}q} \in N_i^{\prime \prime}} \hspace{-12pt} \zeta_{i\hspace{-0.5pt}, \mu k\hspace{-0.5pt}.\hspace{-0.5pt}q} \, x_{i\hspace{-0.5pt}.\hspace{-0.5pt}1\hspace{-0.5pt}, \mu k\hspace{-0.5pt}.\hspace{-0.5pt}q} = \hspace{-10pt} \sum_{N_{i\hspace{-0.5pt}.\hspace{-0.5pt}2\hspace{-0.5pt}, \nu l\hspace{-0.5pt}.\hspace{-0.5pt}r} \in N_i^{\prime \prime}} \hspace{-12pt} x_{i\hspace{-0.5pt}.\hspace{-0.5pt}2\hspace{-0.5pt}, \nu l\hspace{-0.5pt}.\hspace{-0.5pt}r}, \, \forall \, N_i^{\prime \prime} \in \mathcal{N}^{\prime \prime}. 
\end{equation}
If a primary candidate node corresponds to an exit task and requires task duplication, then the replica should be allocated on the same device as the primary task to ensure that a single final output is retained or a single actuator action is performed:
\begin{equation}
\label{eq:one3}
   x_{i\hspace{-0.5pt}.\hspace{-0.5pt}1\hspace{-0.5pt}, \mu k\hspace{-0.5pt}.\hspace{-0.5pt}q} \leq \hspace{-10pt} \sum_{N_{i\hspace{-0.5pt}.\hspace{-0.5pt}2\hspace{-0.5pt}, \mu k\hspace{-0.5pt}.\hspace{-0.5pt}r} \in \mathcal{N}^{\prime \prime}} \hspace{-12pt} x_{i\hspace{-0.5pt}.\hspace{-0.5pt}2\hspace{-0.5pt}, \mu k\hspace{-0.5pt}.\hspace{-0.5pt}r}, \, \forall \, N_{i\hspace{-0.5pt}.\hspace{-0.5pt}1\hspace{-0.5pt}, \mu k\hspace{-0.5pt}.\hspace{-0.5pt}q} \in \mathcal{N}^{\prime \prime}, \, \zeta_{i\hspace{-0.5pt}, \mu k\hspace{-0.5pt}.\hspace{-0.5pt}q} \hspace{-0.5pt} = \hspace{-0.5pt} \phi_i \hspace{-0.5pt} = \hspace{-0.5pt} 1.
\end{equation}

Moreover, if a primary candidate node requires task duplication, the selected replica candidate node should ensure that the total reliability of the task is greater than or equal to its reliability threshold: 
\begin{equation}
\label{eq:three4}
\begin{split}
    R_{i\hspace{-0.5pt}, \mu k\hspace{-0.5pt}.\hspace{-0.5pt}q\hspace{-0.5pt}, \nu l\hspace{-0.5pt}.\hspace{-0.5pt}r} \, x_{i\hspace{-0.5pt}, \mu k\hspace{-0.5pt}.\hspace{-0.5pt}q\hspace{-0.5pt}, \nu l\hspace{-0.5pt}.\hspace{-0.5pt}r} + (1 - x_{i\hspace{-0.5pt}, \mu k\hspace{-0.5pt}.\hspace{-0.5pt}q\hspace{-0.5pt}, \nu l\hspace{-0.5pt}.\hspace{-0.5pt}r}) \mathit{\Omega} \geq R_i^{\mathrm{thr}},\\
    \forall \, N_{i\hspace{-0.5pt}.\hspace{-0.5pt}1\hspace{-0.5pt}, \mu k\hspace{-0.5pt}.\hspace{-0.5pt}q}, N_{i\hspace{-0.5pt}.\hspace{-0.5pt}2\hspace{-0.5pt}, \nu l\hspace{-0.5pt}.\hspace{-0.5pt}r} \in \mathcal{N}^{\prime \prime}, \, \zeta_{i\hspace{-0.5pt}, \mu k\hspace{-0.5pt}.\hspace{-0.5pt}q} = 1.
\end{split}
\end{equation}
In \eqref{eq:three4}, we use a sufficiently large constant $\mathit{\Omega}$ ($\mathit{\Omega} > R_i^{\mathrm{thr}}$) to formulate the conditional aspect of the constraint in linear form. 
Specifically, the constraint becomes meaningful ($R_{i\hspace{-0.5pt}, \mu k\hspace{-0.5pt}.\hspace{-0.5pt}q\hspace{-0.5pt}, \nu l\hspace{-0.5pt}.\hspace{-0.5pt}r} \geq R_i^{\mathrm{thr}}$) only if both the primary and replica candidate nodes $N_{i\hspace{-0.5pt}.\hspace{-0.5pt}1\hspace{-0.5pt}, \mu k\hspace{-0.5pt}.\hspace{-0.5pt}q}$ and $N_{i\hspace{-0.5pt}.\hspace{-0.5pt}2\hspace{-0.5pt}, \nu l\hspace{-0.5pt}.\hspace{-0.5pt}r}$, respectively, are selected for task $\tau_i$, i.e., only if $x_{i\hspace{-0.5pt}, \mu k\hspace{-0.5pt}.\hspace{-0.5pt}q\hspace{-0.5pt}, \nu l\hspace{-0.5pt}.\hspace{-0.5pt}r}=1$. Otherwise, if $x_{i\hspace{-0.5pt}, \mu k\hspace{-0.5pt}.\hspace{-0.5pt}q\hspace{-0.5pt}, \nu l\hspace{-0.5pt}.\hspace{-0.5pt}r}=0$, the constraint becomes irrelevant ($\mathit{\Omega} \geq R_i^{\mathrm{thr}}$), as it is always true. 
By definition, $x_{i\hspace{-0.5pt}, \mu k\hspace{-0.5pt}.\hspace{-0.5pt}q\hspace{-0.5pt}, \nu l\hspace{-0.5pt}.\hspace{-0.5pt}r} = 1$ if both $N_{i\hspace{-0.5pt}.\hspace{-0.5pt}1\hspace{-0.5pt}, \mu k\hspace{-0.5pt}.\hspace{-0.5pt}q}$ and $N_{i\hspace{-0.5pt}.\hspace{-0.5pt}2\hspace{-0.5pt}, \nu l\hspace{-0.5pt}.\hspace{-0.5pt}r}$ are selected. 
We express this in a similar manner to \eqref{eq:three1}--\eqref{eq:three3}, as follows:
\begin{equation}
\label{eq:three5}
    \hspace{-3.6pt} x_{i\hspace{-0.5pt}, \mu k\hspace{-0.5pt}.\hspace{-0.5pt}q\hspace{-0.5pt}, \nu l\hspace{-0.5pt}.\hspace{-0.5pt}r} \hspace{-1.5pt} \leq \hspace{-1pt} x_{i\hspace{-0.5pt}.\hspace{-0.5pt}1\hspace{-0.5pt}, \mu k\hspace{-0.5pt}.\hspace{-0.5pt}q}, \forall \hspace{0.5pt} N_{i\hspace{-0.5pt}.\hspace{-0.5pt}1\hspace{-0.5pt}, \mu k\hspace{-0.5pt}.\hspace{-0.5pt}q}, \hspace{-0.5pt} N_{i\hspace{-0.5pt}.\hspace{-0.5pt}2\hspace{-0.5pt}, \nu l\hspace{-0.5pt}.\hspace{-0.5pt}r} \hspace{-1.5pt} \in \hspace{-1.5pt} \mathcal{N}^{\prime \prime}\hspace{-1.5pt}, \zeta_{i\hspace{-0.5pt}, \mu k\hspace{-0.5pt}.\hspace{-0.5pt}q} \hspace{-2pt} = \hspace{-2pt} 1,
\end{equation}
\begin{equation}
\label{eq:three7}
    \hspace{-3.6pt} x_{i\hspace{-0.5pt}, \mu k\hspace{-0.5pt}.\hspace{-0.5pt}q\hspace{-0.5pt}, \nu l\hspace{-0.5pt}.\hspace{-0.5pt}r} \hspace{-1.5pt} \leq \hspace{-1pt} x_{i\hspace{-0.5pt}.\hspace{-0.5pt}2\hspace{-0.5pt}, \nu l\hspace{-0.5pt}.\hspace{-0.5pt}r}, \forall \hspace{0.5pt} N_{i\hspace{-0.5pt}.\hspace{-0.5pt}1\hspace{-0.5pt}, \mu k\hspace{-0.5pt}.\hspace{-0.5pt}q}, N_{i\hspace{-0.5pt}.\hspace{-0.5pt}2\hspace{-0.5pt}, \nu l\hspace{-0.5pt}.\hspace{-0.5pt}r} \hspace{-1.5pt} \in \hspace{-1.5pt} \mathcal{N}^{\prime \prime}\hspace{-1.5pt}, \zeta_{i\hspace{-0.5pt}, \mu k\hspace{-0.5pt}.\hspace{-0.5pt}q} \hspace{-2pt} = \hspace{-2pt} 1,
\end{equation}
\begin{equation}
\label{eq:three8}
\begin{split}
    x_{i\hspace{-0.5pt}, \mu k\hspace{-0.5pt}.\hspace{-0.5pt}q\hspace{-0.5pt}, \nu l\hspace{-0.5pt}.\hspace{-0.5pt}r} \geq x_{i\hspace{-0.5pt}.\hspace{-0.5pt}1\hspace{-0.5pt}, \mu k\hspace{-0.5pt}.\hspace{-0.5pt}q} + x_{i\hspace{-0.5pt}.\hspace{-0.5pt}2\hspace{-0.5pt}, \nu l\hspace{-0.5pt}.\hspace{-0.5pt}r} - 1,\\
    \forall \, N_{i\hspace{-0.5pt}.\hspace{-0.5pt}1\hspace{-0.5pt}, \mu k\hspace{-0.5pt}.\hspace{-0.5pt}q}, N_{i\hspace{-0.5pt}.\hspace{-0.5pt}2\hspace{-0.5pt}, \nu l\hspace{-0.5pt}.\hspace{-0.5pt}r} \in \mathcal{N}^{\prime \prime}, \, \zeta_{i\hspace{-0.5pt}, \mu k\hspace{-0.5pt}.\hspace{-0.5pt}q} = 1.
\end{split}
\end{equation}

\paragraph{Task precedence constraints} The precedence relationships among the tasks should be preserved:
\begin{equation}
\label{eq:four}
    \begin{split}
        t_{i\hspace{-0.5pt}.\hspace{-0.5pt}n} + L_{i\hspace{-0.5pt},\mu k\hspace{-0.5pt}.\hspace{-0.5pt}q} \, x_{i\hspace{-0.5pt}.\hspace{-0.5pt}n\hspace{-0.5pt},\mu k\hspace{-0.5pt}.\hspace{-0.5pt}q} + CL_{i\hspace{-0.5pt},\mu k\hspace{-0.5pt}.\hspace{-0.5pt}q \rightarrow\hspace{-0.5pt}j\hspace{-0.5pt},\nu l\hspace{-0.5pt}.\hspace{-0.5pt}r} \, x_{i\hspace{-0.5pt}.\hspace{-0.5pt}n\hspace{-0.5pt},\mu k\hspace{-0.5pt}.\hspace{-0.5pt}q \rightarrow\hspace{-0.5pt}j\hspace{-0.5pt}.\hspace{-0.5pt}o\hspace{-0.5pt},\nu l\hspace{-0.5pt}.\hspace{-0.5pt}r} \leq t_{j\hspace{-0.5pt}.\hspace{-0.5pt}o},\\ 
        \forall \, A_{i\hspace{-0.5pt}.\hspace{-0.5pt}n\hspace{-0.5pt},\mu k\hspace{-0.5pt}.\hspace{-0.5pt}q \rightarrow\hspace{-0.5pt}j\hspace{-0.5pt}.\hspace{-0.5pt}o\hspace{-0.5pt},\nu l\hspace{-0.5pt}.\hspace{-0.5pt}r} \in \mathcal{A}^{\prime \prime}.
    \end{split}
\end{equation}

\paragraph{Application completion time \& deadline constraints} 
The completion time of the application should be equal to the completion time of its last task: 
\begin{equation}
\label{eq:five}
    t_{i\hspace{-0.5pt}.\hspace{-0.5pt}n} + L_{i\hspace{-0.5pt}, \mu k\hspace{-0.5pt}.\hspace{-0.5pt}q} \, x_{i\hspace{-0.5pt}.\hspace{-0.5pt}n\hspace{-0.5pt}, \mu k\hspace{-0.5pt}.\hspace{-0.5pt}q} \leq T, \, \forall \, N_{i\hspace{-0.5pt}.\hspace{-0.5pt}n\hspace{-0.5pt}, \mu k\hspace{-0.5pt}.\hspace{-0.5pt}q} \in \mathcal{N}^{\prime \prime}.
\end{equation}
Moreover, it should be within its predefined deadline:
\begin{equation}
\label{eq:six}
    T \leq L_{\mathrm{thr}}.
\end{equation}

\paragraph{Task non-overlapping constraints} 
Any two tasks without a precedence relationship between them, allocated on the same core, should not be executed at the same time, as each core can process only one task at a time:
\begin{equation}
\label{eq:eight1}
\begin{split}
    t_{i\hspace{-0.5pt}.\hspace{-0.5pt}n} \hspace{-1.75pt} + \hspace{-1.25pt} L_{i\hspace{-0.5pt}, \mu k\hspace{-0.5pt}.\hspace{-0.5pt}q} \hspace{0.7pt} x_{i\hspace{-0.5pt}.\hspace{-0.5pt}n\hspace{-0.5pt}, \mu k\hspace{-0.5pt}.\hspace{-0.5pt}q} \hspace{-2pt} \leq \hspace{-1.5pt} t_{j\hspace{-0.5pt}.\hspace{-0.5pt}o} \hspace{-1.75pt} + \hspace{-1.75pt} ( 3 \hspace{-2pt} - \hspace{-1.75pt} x_{i\hspace{-0.5pt}.\hspace{-0.5pt}n\hspace{-0.5pt}, \mu k\hspace{-0.5pt}.\hspace{-0.5pt}q} \hspace{-2pt} - \hspace{-1.75pt} x_{j\hspace{-0.5pt}.\hspace{-0.5pt}o\hspace{-0.5pt}, \mu k\hspace{-0.5pt}.\hspace{-0.5pt}q} \hspace{-2pt} - \hspace{-1.5pt} x_{i\hspace{-0.5pt}.\hspace{-0.5pt}n\hspace{-0.5pt}, j\hspace{-0.5pt}.\hspace{-0.5pt}o} \hspace{-0.75pt}) \mathit{\Omega},\\
    \forall \hspace{0.75pt} N_{i\hspace{-0.5pt}.\hspace{-0.5pt}n\hspace{-0.5pt}, \mu k\hspace{-0.5pt}.\hspace{-0.5pt}q}, N_{j\hspace{-0.5pt}.\hspace{-0.5pt}o\hspace{-0.5pt}, \mu k\hspace{-0.5pt}.\hspace{-0.5pt}q} \hspace{-2pt} \in \hspace{-1.5pt} \mathcal{N}^{\prime \prime}\hspace{-1.5pt}, \hspace{0.5pt} i \hspace{-1.75pt} < \hspace{-1.75pt} j \hspace{-1.5pt} \lor \hspace{-1.5pt} (i \hspace{-2pt} = \hspace{-2pt} j \hspace{-1.5pt} \land \hspace{-1.5pt} n \hspace{-1.75pt} < \hspace{-1.75pt} o)\hspace{-0.5pt}, \tau_{\hspace{-0.5pt}i} \hspace{-2pt} \notin \hspace{-2.5pt} \mathcal{Q}_j\hspace{-0.5pt}, \tau_{\hspace{-0.5pt}j} \hspace{-2pt} \notin \hspace{-2.5pt} \mathcal{Q}_i,
\end{split}    
\end{equation}
\begin{equation}
\label{eq:eight2}
\begin{split}
    t_{j\hspace{-0.5pt}.\hspace{-0.5pt}o} \hspace{-1.75pt} + \hspace{-1.25pt} L_{j\hspace{-0.5pt}, \mu k\hspace{-0.5pt}.\hspace{-0.5pt}q} \hspace{0.7pt} x_{j\hspace{-0.5pt}.\hspace{-0.5pt}o\hspace{-0.5pt}, \mu k\hspace{-0.5pt}.\hspace{-0.5pt}q} \hspace{-2pt} \leq \hspace{-1.5pt} t_{i\hspace{-0.5pt}.\hspace{-0.5pt}n} \hspace{-1.75pt} + \hspace{-1.75pt} ( 2 \hspace{-2pt} - \hspace{-1.75pt} x_{i\hspace{-0.5pt}.\hspace{-0.5pt}n\hspace{-0.5pt}, \mu k\hspace{-0.5pt}.\hspace{-0.5pt}q} \hspace{-2pt} - \hspace{-1.75pt} x_{j\hspace{-0.5pt}.\hspace{-0.5pt}o\hspace{-0.5pt}, \mu k\hspace{-0.5pt}.\hspace{-0.5pt}q} \hspace{-2pt} + \hspace{-1.5pt} x_{i\hspace{-0.5pt}.\hspace{-0.5pt}n\hspace{-0.5pt}, j\hspace{-0.5pt}.\hspace{-0.5pt}o} \hspace{-0.75pt}) \mathit{\Omega},\\
    \forall \hspace{0.75pt} N_{i\hspace{-0.5pt}.\hspace{-0.5pt}n\hspace{-0.5pt}, \mu k\hspace{-0.5pt}.\hspace{-0.5pt}q}, N_{j\hspace{-0.5pt}.\hspace{-0.5pt}o\hspace{-0.5pt}, \mu k\hspace{-0.5pt}.\hspace{-0.5pt}q} \hspace{-2pt} \in \hspace{-1.5pt} \mathcal{N}^{\prime \prime}\hspace{-1.5pt}, \hspace{0.5pt} i \hspace{-1.75pt} < \hspace{-1.75pt} j \hspace{-1.5pt} \lor \hspace{-1.5pt} (i \hspace{-2pt} = \hspace{-2pt} j \hspace{-1.5pt} \land \hspace{-1.5pt} n \hspace{-1.75pt} < \hspace{-1.75pt} o)\hspace{-0.5pt}, \tau_{\hspace{-0.5pt}i} \hspace{-2pt} \notin \hspace{-2.5pt} \mathcal{Q}_j\hspace{-0.5pt}, \tau_{\hspace{-0.5pt}j} \hspace{-2pt} \notin \hspace{-2.5pt} \mathcal{Q}_i.  
\end{split}    
\end{equation}
Similar to \eqref{eq:three4}, in \eqref{eq:eight1} and \eqref{eq:eight2} we utilize constant $\mathit{\Omega}$ ($\mathit{\Omega} > L_{\mathrm{thr}}$) to ensure that task $\tau_{i\hspace{-0.5pt}.\hspace{-0.5pt}n}$ is executed either before ($x_{i\hspace{-0.5pt}.\hspace{-0.5pt}n\hspace{-0.5pt}, j\hspace{-0.5pt}.\hspace{-0.5pt}o} \hspace{-1.5pt} = \hspace{-1.5pt} 1$) or after ($x_{i\hspace{-0.5pt}.\hspace{-0.5pt}n\hspace{-0.5pt}, j\hspace{-0.5pt}.\hspace{-0.5pt}o} \hspace{-1.5pt} = \hspace{-1.5pt} 0$) task $\tau_{j\hspace{-0.5pt}.\hspace{-0.5pt}o}$, respectively, when both tasks are allocated on the same core $p_{\mu k\hspace{-0.5pt}.\hspace{-0.5pt}q}$ (i.e., if $x_{i\hspace{-0.5pt}.\hspace{-0.5pt}n\hspace{-0.5pt}, \mu k\hspace{-0.5pt}.\hspace{-0.5pt}q} \hspace{-1.5pt} = \hspace{-1.5pt} x_{j\hspace{-0.5pt}.\hspace{-0.5pt}o\hspace{-0.5pt}, \mu k\hspace{-0.5pt}.\hspace{-0.5pt}q} \hspace{-1.5pt} = \hspace{-1.5pt} 1$).
Additionally, among the conditions of \eqref{eq:eight1} and \eqref{eq:eight2}, we use $i \hspace{-1.5pt} < \hspace{-1.5pt} j \hspace{-1.25pt} \lor \hspace{-1.25pt} (i \hspace{-1.5pt} = \hspace{-1.5pt} j \hspace{-1.25pt} \land \hspace{-1.25pt} n \hspace{-1.5pt} < \hspace{-1.5pt} o)$ to prevent the generation of redundant constraints.

\paragraph{Task execution constraints} 
The time instants $s_h \in \mathcal{S}$ at which each task will be in execution are determined by:
\begin{equation}
\label{eq:nine}
    x_h^{i\hspace{-0.5pt}.\hspace{-0.5pt}n\hspace{-0.5pt}, \mu k\hspace{-0.5pt}.\hspace{-0.5pt}q} \leq x_{i\hspace{-0.5pt}.\hspace{-0.5pt}n\hspace{-0.5pt}, \mu k\hspace{-0.5pt}.\hspace{-0.5pt}q}, \, \forall \, s_h \in \mathcal{S}, \, \forall \, N_{i\hspace{-0.5pt}.\hspace{-0.5pt}n\hspace{-0.5pt}, \mu k\hspace{-0.5pt}.\hspace{-0.5pt}q} \in \mathcal{N}^{\prime \prime}\hspace{-0.5pt},
\end{equation}
\begin{equation}
\label{eq:ten1}
\begin{split}
    t_{\hspace{-0.25pt}i\hspace{-0.5pt}.\hspace{-0.5pt}n} \hspace{-1.5pt} \leq \hspace{-1.5pt} s_{\hspace{-0.25pt}h} \hspace{-2pt} + \hspace{-1.5pt} \left( \hspace{-1pt} 2 \hspace{-2pt} - \hspace{-1.5pt} x_{\hspace{-0.25pt}i\hspace{-0.5pt}.\hspace{-0.5pt}n\hspace{-0.5pt}, \mu k\hspace{-0.5pt}.\hspace{-0.5pt}q} \hspace{-2pt} - \hspace{-1.5pt} x_{\hspace{-0.25pt}h}^{\hspace{-0.25pt}i\hspace{-0.5pt}.\hspace{-0.5pt}n\hspace{-0.5pt}, \mu k\hspace{-0.5pt}.\hspace{-0.5pt}q} \hspace{-1pt} \right) \hspace{-2pt} \mathit{\Omega}\hspace{-0.5pt},
    \hspace{0.75pt} \forall \hspace{0.75pt} s_{\hspace{-0.25pt}h} \hspace{-2pt} \in \hspace{-1.5pt} \mathcal{S}, \hspace{0.75pt} \forall \hspace{0.75pt} N_{\hspace{-0.25pt}i\hspace{-0.5pt}.\hspace{-0.5pt}n\hspace{-0.5pt}, \mu k\hspace{-0.5pt}.\hspace{-0.5pt}q} \hspace{-2pt} \in \hspace{-1.5pt} \mathcal{N}^{\prime \prime}\hspace{-0.5pt},
\end{split}
\end{equation}
\begin{equation}
\label{eq:ten2}
\begin{split}
    s_h \hspace{-1pt} + \hspace{-1pt} \omega \hspace{-1pt} \leq \hspace{-1pt} t_{i\hspace{-0.5pt}.\hspace{-0.5pt}n} \hspace{-1pt} + \hspace{-1pt} L_{i\hspace{-0.5pt}, \mu k\hspace{-0.5pt}.\hspace{-0.5pt}q} \, x_{i\hspace{-0.5pt}.\hspace{-0.5pt}n\hspace{-0.5pt}, \mu k\hspace{-0.5pt}.\hspace{-0.5pt}q} \hspace{-1pt} + \hspace{-1pt} \left( 2 \hspace{-1pt} - \hspace{-1pt} x_{i\hspace{-0.5pt}.\hspace{-0.5pt}n\hspace{-0.5pt}, \mu k\hspace{-0.5pt}.\hspace{-0.5pt}q} \hspace{-1pt} - \hspace{-1pt} x_h^{i\hspace{-0.5pt}.\hspace{-0.5pt}n\hspace{-0.5pt}, \mu k\hspace{-0.5pt}.\hspace{-0.5pt}q} \right) \hspace{-1.5pt} \mathit{\Omega},\\ 
    \forall \, s_h \hspace{-1pt} \in \hspace{-1pt} \mathcal{S}, \, \forall \, N_{i\hspace{-0.5pt}.\hspace{-0.5pt}n\hspace{-0.5pt}, \mu k\hspace{-0.5pt}.\hspace{-0.5pt}q} \hspace{-1pt} \in \hspace{-1pt} \mathcal{N}^{\prime \prime}\hspace{-0.5pt},
\end{split}
\end{equation}
\begin{equation}
\label{eq:ten3}
\begin{split}
    s_h \hspace{-1pt} + \hspace{-1pt} \omega \hspace{-1pt} \leq \hspace{-1pt} t_{i\hspace{-0.5pt}.\hspace{-0.5pt}n} \hspace{-1pt} + \hspace{-1pt} \left( 2 \hspace{-1pt} - \hspace{-1pt} x_{i\hspace{-0.5pt}.\hspace{-0.5pt}n\hspace{-0.5pt}, \mu k\hspace{-0.5pt}.\hspace{-0.5pt}q} \hspace{-1pt} + \hspace{-1pt} x_h^{i\hspace{-0.5pt}.\hspace{-0.5pt}n\hspace{-0.5pt}, \mu k\hspace{-0.5pt}.\hspace{-0.5pt}q} \hspace{-1pt} - \hspace{-1pt} \hat{x}_h^{i\hspace{-0.5pt}.\hspace{-0.5pt}n\hspace{-0.5pt}, \mu k\hspace{-0.5pt}.\hspace{-0.5pt}q} \right) \hspace{-1.5pt} \mathit{\Omega},\\ 
    \forall \, s_h \hspace{-1pt} \in \hspace{-1pt} \mathcal{S}, \, \forall \, N_{i\hspace{-0.5pt}.\hspace{-0.5pt}n\hspace{-0.5pt}, \mu k\hspace{-0.5pt}.\hspace{-0.5pt}q} \hspace{-1pt} \in \hspace{-1pt} \mathcal{N}^{\prime \prime}\hspace{-0.5pt},
\end{split}
\end{equation}
\begin{equation}
\label{eq:ten4}
\begin{split}
    t_{\hspace{-0.25pt}i\hspace{-0.5pt}.\hspace{-0.5pt}n} \hspace{-2pt} + \hspace{-1.75pt} L_{\hspace{-0.25pt}i\hspace{-0.5pt}, \mu k\hspace{-0.5pt}.\hspace{-0.5pt}q} \hspace{0.75pt} x_{\hspace{-0.25pt}i\hspace{-0.5pt}.\hspace{-0.5pt}n\hspace{-0.5pt}, \mu k\hspace{-0.5pt}.\hspace{-0.5pt}q} \hspace{-2pt} \leq \hspace{-2pt} s_{\hspace{-0.25pt}h} \hspace{-2pt} + \hspace{-2pt} \left( \hspace{-1pt} 1 \hspace{-2pt} - \hspace{-1.5pt} x_{\hspace{-0.25pt}i\hspace{-0.5pt}.\hspace{-0.5pt}n\hspace{-0.5pt}, \mu k\hspace{-0.5pt}.\hspace{-0.5pt}q} \hspace{-2pt} + \hspace{-1.5pt} x_{\hspace{-0.25pt}h}^{\hspace{-0.25pt}i\hspace{-0.5pt}.\hspace{-0.5pt}n\hspace{-0.5pt}, \mu k\hspace{-0.5pt}.\hspace{-0.5pt}q} \hspace{-2pt} + \hspace{-1.5pt} \hat{x}_{\hspace{-0.25pt}h}^{\hspace{-0.25pt}i\hspace{-0.5pt}.\hspace{-0.5pt}n\hspace{-0.5pt}, \mu k\hspace{-0.5pt}.\hspace{-0.5pt}q} \hspace{-1pt} \right) \hspace{-2.5pt} \mathit{\Omega}\hspace{-0.75pt},\\
    \forall \, s_h \hspace{-1pt} \in \hspace{-1pt} \mathcal{S}, \, \forall \, N_{i\hspace{-0.5pt}.\hspace{-0.5pt}n\hspace{-0.5pt}, \mu k\hspace{-0.5pt}.\hspace{-0.5pt}q} \hspace{-1pt} \in \hspace{-1pt} \mathcal{N}^{\prime \prime}\hspace{-0.5pt}.
\end{split}
\end{equation}
In \eqref{eq:nine}, we enforce $x_h^{i\hspace{-0.5pt}.\hspace{-0.5pt}n\hspace{-0.5pt}, \mu k\hspace{-0.5pt}.\hspace{-0.5pt}q} \hspace{-1.5pt} \leq \hspace{-1.5pt} 1$ only if task $\tau_{i\hspace{-0.5pt}.\hspace{-0.5pt}n}$ is allocated on $p_{\mu k\hspace{-0.5pt}.\hspace{-0.5pt}q}$, i.e., only if $x_{i\hspace{-0.5pt}.\hspace{-0.5pt}n\hspace{-0.5pt}, \mu k\hspace{-0.5pt}.\hspace{-0.5pt}q} \hspace{-1.5pt} = \hspace{-1.5pt} 1$. In \eqref{eq:ten1} and \eqref{eq:ten2}, we ensure $x_h^{i\hspace{-0.5pt}.\hspace{-0.5pt}n\hspace{-0.5pt}, \mu k\hspace{-0.5pt}.\hspace{-0.5pt}q} \hspace{-1.5pt} = \hspace{-1.5pt} 1$ if $t_{i\hspace{-0.5pt}.\hspace{-0.5pt}n} \hspace{-1.5pt} \leq \hspace{-1.5pt} s_h \hspace{-1.5pt} < \hspace{-1.5pt} t_{i\hspace{-0.5pt}.\hspace{-0.5pt}n} \hspace{-1.5pt} + \hspace{-1.5pt} L_{i\hspace{-0.5pt}, \mu k\hspace{-0.5pt}.\hspace{-0.5pt}q}$. 
Similarly, in \eqref{eq:ten3} and \eqref{eq:ten4} we enforce $x_h^{i\hspace{-0.5pt}.\hspace{-0.5pt}n\hspace{-0.5pt}, \mu k\hspace{-0.5pt}.\hspace{-0.5pt}q} \hspace{-1.5pt} = \hspace{-1.5pt} 0$ if either $s_h \hspace{-1.5pt} < \hspace{-1.5pt} t_{i\hspace{-0.5pt}.\hspace{-0.5pt}n}$ or $s_h \hspace{-1.5pt} \geq \hspace{-1.5pt} t_{i\hspace{-0.5pt}.\hspace{-0.5pt}n} \hspace{-1.5pt} + \hspace{-1.5pt} L_{i\hspace{-0.5pt}, \mu k\hspace{-0.5pt}.\hspace{-0.5pt}q}$, using the additional binary variable $\hat{x}_h^{i\hspace{-0.5pt}.\hspace{-0.5pt}n\hspace{-0.5pt}, \mu k\hspace{-0.5pt}.\hspace{-0.5pt}q}$. 
Conditional constraints \eqref{eq:ten1}--\eqref{eq:ten4} are defined in linear form using constant $\mathit{\Omega}$.
Strict inequalities (which are not supported in MILP) in \eqref{eq:ten2} and \eqref{eq:ten3} are converted to non-strict inequalities using a positive tolerance constant $\omega$, sufficiently smaller than all the variables and parameters of the model.

\paragraph{Device capability, memory \& storage constraints} 
At each  instant $s_h \hspace{-2.25pt} \in \hspace{-1.5pt} \mathcal{S}$, at most one task from those executed on a particular device should use a specific specialized capability:
\begin{equation}
\label{eq:eleven}
\begin{split}
    \sum_{N_{i\hspace{-0.5pt}.\hspace{-0.5pt}n\hspace{-0.5pt}, \mu k\hspace{-0.5pt}.\hspace{-0.5pt}q} \in \mathcal{N}^{\prime \prime}} \hspace{-12.25pt} x_{\hspace{-0.25pt}h}^{\hspace{-0.25pt}i\hspace{-0.5pt}.\hspace{-0.5pt}n\hspace{-0.5pt}, \mu k\hspace{-0.5pt}.\hspace{-0.5pt}q} \hspace{0.7pt} y_{\hspace{-0.5pt}\mu k}^{c_a} \hspace{0.7pt} z_{\hspace{-0.25pt}i}^{c_a} \hspace{-1.7pt} \leq \hspace{-2pt} 1\hspace{-0.25pt}, \hspace{0.7pt} \forall \hspace{0.7pt} s_{\hspace{-0.25pt}h} \hspace{-2.25pt} \in \hspace{-1.5pt} \mathcal{S}\hspace{-0.25pt}, \hspace{0.7pt} \forall \hspace{0.7pt} u_{\hspace{-0.25pt}\mu k} \hspace{-2.25pt} \in \hspace{-1.5pt} \mathcal{U}\hspace{-0.25pt}, \hspace{0.7pt} \forall \hspace{0.7pt} c_{\hspace{-0.25pt}a} \hspace{-2.25pt} \in \hspace{-1.5pt} \mathcal{C}\hspace{-0.25pt}, \hspace{0.7pt} a \hspace{-1.7pt} > \hspace{-2pt} 0\hspace{-0.25pt}.
\end{split}
\end{equation}
Similarly, at each instant $s_h \hspace{-2.25pt} \in \hspace{-1.5pt} \mathcal{S}$ the memory and storage budgets of each device should not be exceeded:
\begin{equation}
\label{eq:twelve}
    \sum_{N_{i\hspace{-0.5pt}.\hspace{-0.5pt}n\hspace{-0.5pt}, \mu k\hspace{-0.5pt}.\hspace{-0.5pt}q} \in \mathcal{N}^{\prime \prime}} \hspace{-12pt} M_i \, x_h^{i\hspace{-0.5pt}.\hspace{-0.5pt}n\hspace{-0.5pt}, \mu k\hspace{-0.5pt}.\hspace{-0.5pt}q} \hspace{-1pt} \leq \hspace{-1pt} M_{\mu k}^{\mathrm{bgt}}, \, \forall \, s_h \hspace{-1pt} \in \hspace{-1pt} \mathcal{S}, \, \forall \, u_{\mu k} \hspace{-1pt} \in \hspace{-1pt} \mathcal{U},
\end{equation}
\begin{equation}
\label{eq:thirteen}
    \sum_{N_{i\hspace{-0.5pt}.\hspace{-0.5pt}n\hspace{-0.5pt}, \mu k\hspace{-0.5pt}.\hspace{-0.5pt}q} \in \mathcal{N}^{\prime \prime}} \hspace{-12pt} S_i \, x_h^{i\hspace{-0.5pt}.\hspace{-0.5pt}n\hspace{-0.5pt}, \mu k\hspace{-0.5pt}.\hspace{-0.5pt}q} \hspace{-1pt} \leq \hspace{-1pt} S_{\mu k}^{\mathrm{bgt}}, \, \forall \, s_h \hspace{-1pt} \in \hspace{-1pt} \mathcal{S}, \, \forall \, u_{\mu k} \hspace{-1pt} \in \hspace{-1pt} \mathcal{U}.
\end{equation}

\paragraph{Device energy constraints} 
The energy budget of each device should not be exceeded for the execution of the application:
\begin{equation}
\label{eq:fourteen}
\begin{split}
    \sum_{N_{i\hspace{-0.5pt}.\hspace{-0.5pt}n\hspace{-0.5pt}, \mu k\hspace{-0.5pt}.\hspace{-0.5pt}q} \in \mathcal{N}^{\prime \prime}} \hspace{-13pt} E_{i\hspace{-0.5pt}, \mu k\hspace{-0.5pt}.\hspace{-0.5pt}q} \, x_{i\hspace{-0.5pt}.\hspace{-0.5pt}n\hspace{-0.5pt}, \mu k\hspace{-0.5pt}.\hspace{-0.5pt}q}\\     
    & \hspace{-82.5pt} + \hspace{-26pt} \sum_{A_{i\hspace{-0.5pt}.\hspace{-0.5pt}n\hspace{-0.5pt}, \mu k\hspace{-0.5pt}.\hspace{-0.5pt}q \rightarrow\hspace{-0.5pt}j\hspace{-0.5pt}.\hspace{-0.5pt}o\hspace{-0.5pt}, \nu l\hspace{-0.5pt}.\hspace{-0.5pt}r} \in \mathcal{A}^{\prime \prime}} \hspace{-27.5pt} D_i \, x_{i\hspace{-0.5pt}.\hspace{-0.5pt}n\hspace{-0.5pt}, \mu k\hspace{-0.5pt}.\hspace{-0.5pt}q \rightarrow\hspace{-0.5pt}j\hspace{-0.5pt}.\hspace{-0.5pt}o\hspace{-0.5pt}, \nu l\hspace{-0.5pt}.\hspace{-0.5pt}r} \bigg( \pi_{\mu k\hspace{-0.5pt}, \nu l} \left( 1 \hspace{-0.75pt} - \hspace{-0.75pt} \theta_{i\hspace{-0.5pt}, \mu k\hspace{-0.5pt}.\hspace{-0.5pt}q \rightarrow\hspace{-0.5pt}j\hspace{-0.5pt}, \nu l\hspace{-0.5pt}.\hspace{-0.5pt}r}^{\xi m} \right)\\
    & \hspace{40pt} + \hspace{-0.75pt} \pi_{\mu k\hspace{-0.5pt}, \xi m} \, \theta_{i\hspace{-0.5pt}, \mu k\hspace{-0.5pt}.\hspace{-0.5pt}q \rightarrow\hspace{-0.5pt}j\hspace{-0.5pt}, \nu l\hspace{-0.5pt}.\hspace{-0.5pt}r}^{\xi m} \bigg)\\ 
    & \hspace{-82.5pt} + \hspace{-26pt} \sum_{A_{j\hspace{-0.5pt}.\hspace{-0.5pt}o\hspace{-0.5pt}, \nu l\hspace{-0.5pt}.\hspace{-0.5pt}r \rightarrow\hspace{-0.5pt}i\hspace{-0.5pt}.\hspace{-0.5pt}n\hspace{-0.5pt}, \mu k\hspace{-0.5pt}.\hspace{-0.5pt}q} \in \mathcal{A}^{\prime \prime}} \hspace{-27.5pt} D_j \, x_{j\hspace{-0.5pt}.\hspace{-0.5pt}o\hspace{-0.5pt}, \nu l\hspace{-0.5pt}.\hspace{-0.5pt}r \rightarrow\hspace{-0.5pt}i\hspace{-0.5pt}.\hspace{-0.5pt}n\hspace{-0.5pt}, \mu k\hspace{-0.5pt}.\hspace{-0.5pt}q} \bigg( \sigma_{\nu l\hspace{-0.5pt}, \mu k} \left( 1 \hspace{-0.75pt} - \hspace{-0.75pt} \theta_{j\hspace{-0.5pt}, \nu l\hspace{-0.5pt}.\hspace{-0.5pt}r \rightarrow\hspace{-0.5pt}i\hspace{-0.5pt}, \mu k\hspace{-0.5pt}.\hspace{-0.5pt}q}^{\xi m} \right)\\
    & \hspace{40pt} + \hspace{-0.75pt} \sigma_{\xi m\hspace{-0.5pt}, \mu k} \, \theta_{j\hspace{-0.5pt}, \nu l\hspace{-0.5pt}.\hspace{-0.5pt}r \rightarrow\hspace{-0.5pt}i\hspace{-0.5pt}, \mu k\hspace{-0.5pt}.\hspace{-0.5pt}q}^{\xi m} \bigg)\\
    & \hspace{-82.5pt} + \hspace{-26.75pt} \sum_{A_{i\hspace{-0.5pt}.\hspace{-0.5pt}n\hspace{-0.5pt}, \nu l\hspace{-0.5pt}.\hspace{-0.5pt}q \rightarrow\hspace{-0.5pt}j\hspace{-0.5pt}.\hspace{-0.5pt}o\hspace{-0.5pt}, \xi m\hspace{-0.5pt}.\hspace{-0.5pt}r} \in \mathcal{A}^{\prime \prime}} \hspace{-28pt} D_i \, x_{i\hspace{-0.5pt}.\hspace{-0.5pt}n\hspace{-0.5pt}, \nu l\hspace{-0.5pt}.\hspace{-0.5pt}q \rightarrow\hspace{-0.5pt}j\hspace{-0.5pt}.\hspace{-0.5pt}o\hspace{-0.5pt}, \xi m\hspace{-0.5pt}.\hspace{-0.5pt}r} ( \sigma_{\nu l\hspace{-0.5pt}, \mu k} \hspace{-0.75pt} + \hspace{-0.75pt} \pi_{\mu k\hspace{-0.5pt}, \xi m} ) \theta_{i\hspace{-0.5pt}, \nu l\hspace{-0.5pt}.\hspace{-0.5pt}q \rightarrow\hspace{-0.5pt}j\hspace{-0.5pt}, \xi m\hspace{-0.5pt}.\hspace{-0.5pt}r}^{\mu k}\hspace{-0.25pt}\\
    & \hspace{-82.5pt} \leq \hspace{2pt} E_{\mu k}^{\mathrm{bgt}}, \, \forall \, u_{\mu k} \in \mathcal{U}, \, (\mu, k) \neq (\nu, l) \neq (\xi, m).
\end{split}
\end{equation}
The first term in \eqref{eq:fourteen} denotes the computational energy consumption of device $u_{\mu k}$. The next three terms represent its communication energy consumption, considering data transmitted from (second term) and received at (third term) $u_{\mu k}$, as well as the case where $u_{\mu k}$ is used for data transfers between other devices (fourth term).

\paragraph{Non-negativity \& binary constraints} 
The non-negativity \eqref{eq:nineteen}, \eqref{eq:twenty} and binary nature \eqref{eq:fifteen}--\eqref{eq:eighteen} of the continuous and binary variables, respectively, should be ensured:
\begin{equation}
\label{eq:nineteen}
    t_{i\hspace{-0.5pt}.\hspace{-0.5pt}n} \geq 0, \, \forall \, N_{i\hspace{-0.5pt}.\hspace{-0.5pt}n}^{\prime \prime} \in \mathcal{N}^{\prime \prime},
\end{equation}
\begin{equation}
\label{eq:twenty}
     T \geq 0,
\end{equation}
\begin{equation}
\label{eq:fifteen}
     x_{i\hspace{-0.5pt}.\hspace{-0.5pt}n\hspace{-0.5pt}, \mu k\hspace{-0.5pt}.\hspace{-0.5pt}q} \in \{ 0, 1 \}, \, \forall \, N_{i\hspace{-0.5pt}.\hspace{-0.5pt}n\hspace{-0.5pt}, \mu k\hspace{-0.5pt}.\hspace{-0.5pt}q} \in \mathcal{N}^{\prime \prime},
\end{equation}
\begin{equation}
\label{eq:sixteen}
     x_{i\hspace{-0.5pt}.\hspace{-0.5pt}n\hspace{-0.5pt}, \mu k\hspace{-0.5pt}.\hspace{-0.5pt}q \rightarrow\hspace{-0.5pt}j\hspace{-0.5pt}.\hspace{-0.5pt}o\hspace{-0.5pt}, \nu l\hspace{-0.5pt}.\hspace{-0.5pt}r} \in \{ 0, 1 \}, \, \forall \, A_{i\hspace{-0.5pt}.\hspace{-0.5pt}n\hspace{-0.5pt}, \mu k\hspace{-0.5pt}.\hspace{-0.5pt}q \rightarrow\hspace{-0.5pt}j\hspace{-0.5pt}.\hspace{-0.5pt}o\hspace{-0.5pt}, \nu l\hspace{-0.5pt}.\hspace{-0.5pt}r} \in \mathcal{A}^{\prime \prime}, 
\end{equation}
\begin{equation}
\label{eq:sixteen2}
    \hspace{-3.85pt} x_{i\hspace{-0.5pt}, \mu k\hspace{-0.5pt}.\hspace{-0.5pt}q\hspace{-0.5pt}, \nu l\hspace{-0.5pt}.\hspace{-0.5pt}r} \hspace{-0.5pt} \in \hspace{-0.5pt} \{ 0, 1\}\hspace{-0.25pt}, \hspace{0.5pt} \forall \hspace{1pt} N_{i\hspace{-0.5pt}.\hspace{-0.5pt}1\hspace{-0.5pt}, \mu k\hspace{-0.5pt}.\hspace{-0.5pt}q}, N_{i\hspace{-0.5pt}.\hspace{-0.5pt}2\hspace{-0.5pt}, \nu l\hspace{-0.5pt}.\hspace{-0.5pt}r} \hspace{-0.5pt} \in \hspace{-0.5pt} \mathcal{N}^{\prime \prime}\hspace{-0.5pt}, \hspace{0.75pt} \zeta_{i\hspace{-0.5pt}, \mu k\hspace{-0.5pt}.\hspace{-0.5pt}q} \hspace{-1pt} = \hspace{-1pt} 1,
\end{equation}
\begin{equation}
\label{eq:seventeen}
\begin{split}
    x_{i\hspace{-0.5pt}.\hspace{-0.5pt}n\hspace{-0.5pt}, j\hspace{-0.5pt}.\hspace{-0.5pt}o} \in \{ 0, 1 \}, 
    \, \forall \, N_{i\hspace{-0.5pt}.\hspace{-0.5pt}n\hspace{-0.5pt}, \mu k\hspace{-0.5pt}.\hspace{-0.5pt}q}, N_{j\hspace{-0.5pt}.\hspace{-0.5pt}o\hspace{-0.5pt}, \mu k\hspace{-0.5pt}.\hspace{-0.5pt}q} \hspace{-0.5pt} \in \hspace{-0.5pt} \mathcal{N}^{\prime \prime}\hspace{-0.5pt},\\ 
    i \hspace{-1pt} < \hspace{-1pt} j \hspace{-1pt} \lor \hspace{-1pt} (i \hspace{-1pt} = \hspace{-1pt} j \hspace{-1pt} \land \hspace{-1pt} n \hspace{-1pt} < \hspace{-1pt} o)\hspace{-0.25pt}, \tau_{\hspace{-0.25pt}i} \hspace{-1pt} \notin \hspace{-1pt} \mathcal{Q}_j\hspace{-0.25pt}, \tau_{\hspace{-0.25pt}j} \hspace{-1pt} \notin \hspace{-1pt} \mathcal{Q}_i,  
\end{split}
\end{equation}
\begin{equation}
\label{eq:eighteen}
    x_h^{i\hspace{-0.5pt}.\hspace{-0.5pt}n\hspace{-0.5pt}, \mu k\hspace{-0.5pt}.\hspace{-0.5pt}q}, \hspace{0.5pt} \hat{x}_h^{i\hspace{-0.5pt}.\hspace{-0.5pt}n\hspace{-0.5pt}, \mu k\hspace{-0.5pt}.\hspace{-0.5pt}q} \in \{ 0, 1 \}, \, \forall \, s_h \in \mathcal{S}, \, \forall \, N_{i\hspace{-0.5pt}.\hspace{-0.5pt}n\hspace{-0.5pt}, \mu k\hspace{-0.5pt}.\hspace{-0.5pt}q} \in \mathcal{N}^{\prime \prime}.
\end{equation}

The main notations used in this work are summarized in Appendix A in the supplementary material.

\subsubsection{MILP Complexity}
\label{subsubsec:complexity}
The computational complexity of our MILP approach depends on both the problem size and the employed solver. Well-established solvers such as Gurobi \cite{gurobi} typically use undisclosed proprietary algorithms and thus their computational complexity cannot be derived \cite{Mo2023}. 
The problem size (i.e., the number of variables and constraints) depends on the size of the resulting ETAG $G^{\prime \prime}$, whose growth with respect to the initial TG $G$ is analyzed in \cref{subsubsec:etagSize}. 
As relevant applications do not typically involve an excessive number of tasks \cite{Zheng2021, Alam2017, Kashino2019, Savva2021}, the benefits of our approach outweigh the moderate complexity introduced by the size increase of $G^{\prime \prime}$. 
We demonstrate this experimentally in \cref{subsubsec:syntheticResults}.

%% file: 4_heuristic.tex
\section{Multi-Objective \& Multi-Constrained HEFT}
\label{sec:heft}

\subsection{Extension Methodology Overview}
\label{subsec:HEFToverview}

As discussed in \cref{sec:related}, our exact MILP method is the first to provide a holistic and optimal solution to the specific multi-objective and multi-constrained problem in the examined CPS. 
Nevertheless, it remains important to quantify the practical benefit of this optimality by comparing it against a strong and well-established baseline under identical modeling assumptions.
Adapting other exact approaches to accommodate the same objectives, constraints, and architecture would yield formulations essentially equivalent to the one proposed in this work, which is among our key contributions, offering limited additional insight.
Therefore, we consider HEFT, which remains one of the most widely used, effective, and easily adapted scheduling heuristics for workflow applications \cite{Topcuoglu2002, Kuhbacher2019, Aldegheri2020}.
HEFT follows a sequential two-phase procedure: (a) a task prioritization phase, where tasks are prioritized based on their upward rank (i.e., the longest distance to an exit task, in terms of latency), and (b) a core selection phase, where each task, in order of priority, is assigned to the core that minimizes its finish time (i.e., its overall latency), which serves as the objective function at each step.
Due to these properties, HEFT can be systematically extended to accommodate the key aspects of the examined problem. 

The proposed extension preserves the fundamental heuristic nature of HEFT, which remains a greedy list-scheduling heuristic, as task priorities are computed once using upward ranks and tasks are then scheduled sequentially in that order. The extension does not introduce global search, backtracking, or iterative improvement. Once a task (and, when applicable, its replica) is committed, the decision is not revised.
Selective task duplication and the objectives and constraints used in our MILP method are embedded into HEFT in an analogous manner, without changing its algorithmic paradigm. Specifically, selective task duplication is not discovered or optimized dynamically. In contrast, it is pre-encoded in the input ETAG, which is also used by the MILP approach. After rank calculation in the first phase, primary and (when required) replica candidate node sets are constructed, and are then evaluated during the existing second phase of HEFT. Reliability, deadline, capability, memory, storage, and energy constraints are embedded in the second phase, along with the existing precedence and non-overlapping constraints, and are enforced as hard feasibility checks that discard invalid candidates. Among the remaining feasible options for each task, selection is based on a local (myopic) multi-objective score with respect to the current partial schedule, using the same latency, energy, and reliability objectives as in the MILP formulation, while preserving the main selection mechanism of HEFT.

\subsection{Algorithmic Implementation Details}
\label{subsec:HEFTdetails}

Since HEFT is originally designed for latency-driven scheduling under precedence and non-overlapping constraints, we enhance it to incorporate the same objectives, constraints, and selective task duplication technique as the proposed MILP method, under identical modeling assumptions, to ensure a fair and meaningful comparison. 
Without incorporating these into HEFT, any comparison would be inherently biased in favor of the MILP approach. 
Our extended version of HEFT is shown in \cref{alg:heft}. We use as input the ETAG $G^{\prime \prime}$, which integrates selective task duplication.
In the first phase (lines 1--5), we determine the upward rank of each candidate node $N_{i\hspace{-0.5pt}.\hspace{-0.5pt}n\hspace{-0.5pt}, \mu k\hspace{-0.5pt}.\hspace{-0.5pt}q} \hspace{-1pt} \in \hspace{-1pt} G^{\prime \prime}$ (line 2), based on the upward rank of its child nodes \cite{Topcuoglu2002}:
\begin{equation}
\label{eq:rank}
\rho_{i\hspace{-0.6pt}.\hspace{-0.6pt}n\hspace{-0.6pt}, \mu k\hspace{-0.6pt}.\hspace{-0.6pt}q} \hspace{-1.25pt} = \hspace{-1.25pt} L_{i\hspace{-0.6pt}, \mu k\hspace{-0.6pt}.\hspace{-0.6pt}q} + \hspace{-4pt} \max_{A_{i\hspace{-0.75pt}.\hspace{-0.75pt}n\hspace{-0.75pt}, \mu k\hspace{-0.75pt}.\hspace{-0.75pt}q \rightarrow \hspace{-0.75pt} j\hspace{-0.75pt}.\hspace{-0.75pt}o\hspace{-0.75pt}, \nu l\hspace{-0.75pt}.\hspace{-0.75pt}r} \in  \mathcal{A}^{\prime \prime}} \hspace{-1pt} \{ \hspace{-1pt} CL_{i\hspace{-0.6pt}, \mu k\hspace{-0.6pt}.\hspace{-0.6pt}q \rightarrow \hspace{-0.6pt} j\hspace{-0.6pt}, \nu l\hspace{-0.6pt}.\hspace{-0.6pt}r} + \rho_{j\hspace{-0.6pt}.\hspace{-0.6pt}o\hspace{-0.6pt}, \nu l\hspace{-0.6pt}.\hspace{-0.6pt}r} \hspace{-1pt}\}\hspace{-1pt}.
\end{equation}
Given that task duplication is required for certain primary candidate nodes (as in \eqref{eq:one2} of our MILP approach), and to jointly consider the scheduling of both the primary and replica candidate nodes in such cases, we construct a list $\mathit{\Lambda}$ comprising sets of candidate nodes for each task (line 4). 
For primary candidate nodes requiring task duplication, we generate a set for each combination of the primary candidate node with a replica candidate node.
Otherwise, we create a set containing only the specific primary candidate node. The sets in $\mathit{\Lambda}$ are prioritized based on the upward rank of the primary candidate node in each set (line 5).


\begin{algorithm}[!t]

\caption{Extended HEFT.}
\label{alg:heft}

\scriptsize

\SetInd{0.65em}{0.65em} 

\KwIn{ETAG $G^{\prime \prime}=(\mathcal{N}^{\prime \prime}\hspace{-1pt}, \mathcal{A}^{\prime \prime})$.}

\KwOut{$x_{i\hspace{-0.75pt}.\hspace{-0.75pt}n\hspace{-0.75pt}, \mu k\hspace{-0.75pt}.\hspace{-0.75pt}q} \hspace{0.5pt} \forall \hspace{0.5pt} N_{i\hspace{-0.75pt}.\hspace{-0.75pt}n\hspace{-0.75pt}, \mu k\hspace{-0.75pt}.\hspace{-0.75pt}q} \hspace{-1.25pt} \in \hspace{-1.25pt} \mathcal{N}^{\prime \prime}$\hspace{-1pt}, \hspace{-2.5pt} $x_{i\hspace{-0.75pt}.\hspace{-0.75pt}n\hspace{-0.75pt}, \mu k\hspace{-0.75pt}.\hspace{-0.75pt}q \rightarrow\hspace{-0.75pt}j\hspace{-0.75pt}.\hspace{-0.75pt}o\hspace{-0.75pt}, \nu l\hspace{-0.75pt}.\hspace{-0.75pt}r} \hspace{0.5pt} \forall \hspace{0.5pt} A_{i\hspace{-0.75pt}.\hspace{-0.75pt}n\hspace{-0.75pt}, \mu k\hspace{-0.75pt}.\hspace{-0.75pt}q \rightarrow\hspace{-0.75pt}j\hspace{-0.75pt}.\hspace{-0.75pt}o\hspace{-0.75pt}, \nu l\hspace{-0.75pt}.\hspace{-0.75pt}r} \hspace{-1.25pt} \in \hspace{-1.25pt} \mathcal{A}^{\prime \prime}$\hspace{-1.25pt}, and \hspace{0.75pt} $t_{i\hspace{-0.75pt}.\hspace{-0.75pt}n} \hspace{0.75pt} \forall \hspace{0.75pt} N_{i\hspace{-0.75pt}.\hspace{-0.75pt}n}^{\prime \prime} \hspace{-1pt} \in \hspace{-1pt} G^{\prime \prime}$\hspace{-0.5pt}.}

\tcp{Phase A - candidate node prioritization:}

\ForEach{\textup{candidate node \hspace{-4pt} $N_{i\hspace{-0.75pt}.\hspace{-0.75pt}n\hspace{-0.75pt}, \mu k\hspace{-0.75pt}.\hspace{-0.75pt}q} \hspace{-1.25pt} \in \hspace{-1.25pt} \mathcal{N}^{\prime\prime}$ \hspace{-4pt} starting from exit task candidate nodes}}{
Calculate upward rank $\rho_{i\hspace{-0.75pt}.\hspace{-0.75pt}n\hspace{-0.75pt}, \mu k\hspace{-0.75pt}.\hspace{-0.75pt}q}$ using \eqref{eq:rank}\;
}

$\mathit{\Lambda} \gets \{ \{ N_{i\hspace{-0.75pt}.\hspace{-0.75pt}1\hspace{-0.75pt}, \mu k\hspace{-0.75pt}.\hspace{-0.75pt}q}, N_{i\hspace{-0.75pt}.\hspace{-0.75pt}2\hspace{-0.75pt}, \nu l\hspace{-0.75pt}.\hspace{-0.75pt}r} \} \, | \, N_{i\hspace{-0.75pt}.\hspace{-0.75pt}1\hspace{-0.75pt}, \mu k\hspace{-0.75pt}.\hspace{-0.75pt}q}, N_{i\hspace{-0.75pt}.\hspace{-0.75pt}2\hspace{-0.75pt}, \nu l\hspace{-0.75pt}.\hspace{-0.75pt}r} \hspace{-1pt} \in \hspace{-1pt} \mathcal{N}^{\prime\prime}\hspace{-1pt}, \zeta_{i\hspace{-0.75pt}, \mu k\hspace{-0.75pt}.\hspace{-0.75pt}q} \hspace{-1pt} = \hspace{-1pt} 1 \}  \cup \allowbreak \{ \{ N_{i\hspace{-0.75pt}.\hspace{-0.75pt}1\hspace{-0.75pt}, \mu k\hspace{-0.75pt}.\hspace{-0.75pt}q} \} \, | \, N_{i\hspace{-0.75pt}.\hspace{-0.75pt}1\hspace{-0.75pt}, \mu k\hspace{-0.75pt}.\hspace{-0.75pt}q} \hspace{-1pt} \in \hspace{-1pt} \mathcal{N}^{\prime\prime}\hspace{-1pt}, \zeta_{i\hspace{-0.75pt}, \mu k\hspace{-0.75pt}.\hspace{-0.75pt}q} \hspace{-1pt} = \hspace{-1pt} 0 \}$\;

Sort list of sets $\mathit{\Lambda}$ by non-increasing order of $\rho_{i\hspace{-0.75pt}.\hspace{-0.75pt}n\hspace{-0.75pt}, \mu k\hspace{-0.75pt}.\hspace{-0.75pt}q}$ of primary candidate nodes\;

\tcp{Phase B - candidate node selection:}
$isInfeasible \gets 0, \, \mathcal{N}^{\prime\prime}_{\mathrm{sel}} \gets \varnothing, \, \mathcal{A}^{\prime\prime}_{\mathrm{sel}} \gets \varnothing$\;

\While{\textup{$\exists$ unscheduled tasks corresponding to sets of candidate nodes in $\mathit{\Lambda}$}}{
    
    Select first unscheduled task $\tau_i$ based on order of sets in $\mathit{\Lambda}$\;

    \ForEach{\textup{set $\mathit{\Lambda}_v \hspace{-1pt}\in \hspace{-1pt} \mathit{\Lambda}$ corresponding to $\tau_i$}}{
        $isSkipped_v \gets 0$\;
    
        Calculate total reliability $\hat{R}_{i\hspace{-0.75pt}, \mu k\hspace{-0.75pt}.\hspace{-0.75pt}q}$ provided by $\mathit{\Lambda}_v$ using \eqref{eq:totalReliability}\;
        
        \If{\textup{$(\zeta_{i\hspace{-0.75pt}, \mu k\hspace{-0.75pt}.\hspace{-0.75pt}q} \hspace{-1pt} = \hspace{-1pt} \phi_i \hspace{-1pt} = \hspace{-1pt} 1 \hspace{-1pt} \land \hspace{-1pt} (\mu,k) \hspace{-1pt} \neq \hspace{-1pt} (\nu,l)) \lor \hat{R}_{i\hspace{-0.75pt}, \mu k\hspace{-0.75pt}.\hspace{-0.75pt}q} \hspace{-1pt} < \hspace{-1pt} R_i^{\mathrm{thr}}$}}{
            $isSkipped_v \gets 1$; Continue\tcp*[r]{$\mathit{\Lambda}_v$ is not applicable}
        }
    
        $\mathcal{N}^{\prime\prime}_{\mathrm{temp}} \gets \mathcal{N}^{\prime\prime}_{\mathrm{sel}}$\;
        
        \ForEach{\textup{candidate node $N_{i\hspace{-0.75pt}.\hspace{-0.75pt}n\hspace{-0.75pt}, \mu k\hspace{-0.75pt}.\hspace{-0.75pt}q} \hspace{-1pt} \in \hspace{-1pt} \mathit{\Lambda}_v$}}{
            
            $\mathcal{N}^{\prime\prime}_{\hat{\mathrm{temp}}} \gets \mathcal{N}^{\prime\prime}_{\mathrm{temp}} \cup \{ N_{i\hspace{-0.75pt}.\hspace{-0.75pt}n\hspace{-0.75pt}, \mu k\hspace{-0.75pt}.\hspace{-0.75pt}q} \}$\;
    
            \vspace{0.75pt}$\mathcal{A}^{\prime\prime}_{\mathrm{temp}} \gets \mathcal{A}^{\prime\prime}_{\mathrm{sel}} \cup \{ A_{j\hspace{-0.75pt}.\hspace{-0.75pt}o\hspace{-0.75pt}, \nu l\hspace{-0.75pt}.\hspace{-0.75pt}r \rightarrow\hspace{-0.5pt}i\hspace{-0.75pt}.\hspace{-0.75pt}n\hspace{-0.75pt}, \mu k\hspace{-0.75pt}.\hspace{-0.75pt}q} \in \mathcal{A}^{\prime\prime} \,| \, N_{j\hspace{-0.75pt}.\hspace{-0.75pt}o\hspace{-0.75pt}, \nu l\hspace{-0.75pt}.\hspace{-0.75pt}r} \in \mathcal{N}^{\prime\prime}_{\mathrm{sel}}\}$\;
    
            $energyBgtExceeded \gets 0$\;
            \ForEach{\textup{device $u_{\nu l} \in \mathcal{U}$}}{

                \If{\textup{constraint \eqref{eq:fourteen} is violated for $u_{\nu l}$ using $\mathcal{N}^{\prime\prime}_{\hat{\mathrm{temp}}}, \mathcal{A}^{\prime\prime}_{\mathrm{temp}}$ in place of $\mathcal{N}^{\prime\prime}, \mathcal{A}^{\prime\prime}$, respectively}}{
                    $energyBgtExceeded \gets 1$; Break\;           
                }
            }
            \eIf{\textup{$energyBgtExceeded \hspace{0.5pt} \lor \hspace{1pt} M_i > M_{\mu k}^{\mathrm{bgt}} \hspace{0.5pt} \lor \hspace{0.5pt} S_i > S_{\mu k}^{\mathrm{bgt}}$}}{
            $isSkipped_v \gets 1$; Break\tcp*[r]{$\mathit{\Lambda}_v$ is not applicable}
            }
            {
                Calculate earliest finish time $EFT_{i\hspace{-0.75pt}.\hspace{-0.75pt}n\hspace{-0.75pt}, \mu k\hspace{-0.75pt}.\hspace{-0.75pt}q}$ using \cref{alg:heftEFT}\;

                $\mathcal{N}^{\prime\prime}_{\mathrm{temp}} \gets \mathcal{N}^{\prime\prime}_{\mathrm{temp}} \hspace{-1pt} \cup \hspace{-1pt} \{ N_{i\hspace{-0.75pt}.\hspace{-0.75pt}n\hspace{-0.75pt}, \mu k\hspace{-0.75pt}.\hspace{-0.75pt}q} \}$\;
            } 
        }
        \eIf{\textup{$isSkipped_v$}}{Continue\;}
        {
            \tcp{total latency, energy, and reliability of $\mathit{\Lambda}_v$:}
            $L_v \gets \hspace{-4pt} \max\limits_{N_{i\hspace{-0.75pt}.\hspace{-0.75pt}n\hspace{-0.75pt}, \mu k\hspace{-0.75pt}.\hspace{-0.75pt}q} \in \mathit{\Lambda}_v} \{ EFT_{i\hspace{-0.75pt}.\hspace{-0.75pt}n\hspace{-0.75pt}, \mu k\hspace{-0.75pt}.\hspace{-0.75pt}q} \}$\;
            
            \vspace{4pt}$E_v \gets \hspace{-4pt} \sum\limits_{N_{i\hspace{-0.75pt}.\hspace{-0.75pt}n\hspace{-0.75pt}, \mu k\hspace{-0.75pt}.\hspace{-0.75pt}q} \in \mathit{\Lambda}_v} \hspace{-15pt} E_{i\hspace{-0.75pt}, \mu k\hspace{-0.75pt}.\hspace{-0.75pt}q} \hspace{5pt} \allowbreak + \hspace{-18pt} \sum\limits_{\substack{N_{i\hspace{-0.75pt}.\hspace{-0.75pt}n\hspace{-0.75pt}, \mu k\hspace{-0.75pt}.\hspace{-0.75pt}q} \in \mathit{\Lambda}_v \hspace{-1pt},\\ A_{j\hspace{-0.75pt}.\hspace{-0.75pt}o\hspace{-0.75pt}, \nu l\hspace{-0.75pt}.\hspace{-0.75pt}r \rightarrow\hspace{-0.5pt}i\hspace{-0.75pt}.\hspace{-0.75pt}n\hspace{-0.75pt}, \mu k\hspace{-0.75pt}.\hspace{-0.75pt}q} \in \mathcal{A}^{\prime\prime}_{\mathrm{sel}}}} \hspace{-30pt} CE_{j\hspace{-0.75pt}, \nu l\hspace{-0.75pt}.\hspace{-0.75pt}r \rightarrow\hspace{-0.5pt}i\hspace{-0.75pt}, \mu k\hspace{-0.75pt}.\hspace{-0.75pt}q}$\;
            
            \vspace{-2.5pt}$R_v \gets \hat{R}_{i\hspace{-0.75pt}, \mu k\hspace{-0.75pt}.\hspace{-0.75pt}q}$\;
        }
    }
    \eIf{\textup{$\exists$ $\mathit{\Lambda}_v \hspace{-1pt}\in \hspace{-1pt} \mathit{\Lambda}$ corresponding to $\tau_i$ with $isSkipped_v = 0$}}{
        \ForEach{\textup{set $\mathit{\Lambda}_v \hspace{-1pt}\in \hspace{-1pt} \mathit{\Lambda}$ corresponding to $\tau_i$ with $isSkipped_v = 0$}}{

            Normalize $L_v$, $E_v$, $R_v$ in $[0,1]$ to yield $\dot{L}_v$, $\dot{E}_v$, $\dot{R}_v$, respectively\; 
            
            $g_v \gets w_{\mathrm{lat}}\dot{L}_v + w_{\mathrm{en}}\dot{E}_v -  w_{\mathrm{rel}}\dot{R}_v$\tcp*[r]{multi-objective}
        }
        
        Select $\mathit{\Lambda}_v \hspace{-1pt}\in \hspace{-1pt} \mathit{\Lambda}$ corresponding to $\tau_i$ that minimizes $g_v$\;
        
        \eIf{$L_v \leq L_{\mathrm{thr}}$}{
             \ForEach{\textup{candidate node $N_{i\hspace{-0.75pt}.\hspace{-0.75pt}n\hspace{-0.75pt}, \mu k\hspace{-0.75pt}.\hspace{-0.75pt}q} \hspace{-1pt} \in \hspace{-1pt} \mathit{\Lambda}_v$}}{
                $x_{i\hspace{-0.75pt}.\hspace{-0.75pt}n\hspace{-0.75pt}, \mu k\hspace{-0.75pt}.\hspace{-0.75pt}q} \gets 1$,
                $\mathcal{N}^{\prime\prime}_{\mathrm{sel}} \gets \mathcal{N}^{\prime\prime}_{\mathrm{sel}} \cup \{ N_{i\hspace{-0.75pt}.\hspace{-0.75pt}n\hspace{-0.75pt}, \mu k\hspace{-0.75pt}.\hspace{-0.75pt}q}\}$\;
                
                \ForEach{$A_{j\hspace{-0.75pt}.\hspace{-0.75pt}o\hspace{-0.75pt}, \nu l\hspace{-0.75pt}.\hspace{-0.75pt}r \rightarrow\hspace{-0.5pt}i\hspace{-0.75pt}.\hspace{-0.75pt}n\hspace{-0.75pt}, \mu k\hspace{-0.75pt}.\hspace{-0.75pt}q} \in \mathcal{A}^{\prime \prime}$}{
                    \If{$N_{j\hspace{-0.75pt}.\hspace{-0.75pt}o\hspace{-0.75pt}, \nu l\hspace{-0.75pt}.\hspace{-0.75pt}r} \in \mathcal{N}^{\prime\prime}_{\mathrm{sel}}$}{
                        $x_{j\hspace{-0.75pt}.\hspace{-0.75pt}o\hspace{-0.75pt}, \nu l\hspace{-0.75pt}.\hspace{-0.75pt}r \rightarrow\hspace{-0.5pt}i\hspace{-0.75pt}.\hspace{-0.75pt}n\hspace{-0.75pt}, \mu k\hspace{-0.75pt}.\hspace{-0.75pt}q} \gets 1$,
                        $\mathcal{A}^{\prime\prime}_{\mathrm{sel}} \hspace{-1.5pt} \gets \mathcal{A}^{\prime\prime}_{\mathrm{sel}} \hspace{-0.5pt} \cup \hspace{-0.5pt} \{ A_{j\hspace{-0.75pt}.\hspace{-0.75pt}o\hspace{-0.75pt}, \nu l\hspace{-0.75pt}.\hspace{-0.75pt}r \rightarrow\hspace{-0.5pt}i\hspace{-0.75pt}.\hspace{-0.75pt}n\hspace{-0.75pt}, \mu k\hspace{-0.75pt}.\hspace{-0.75pt}q} \}$\;
                    }
                }    
            }
            Mark task $\tau_i$ as scheduled\;
        }
        {
            $isInfeasible \gets 1$; 
            Break\;   
        }
    }
    {
        $isInfeasible \gets 1$; 
        Break\;
    }    
}
\Return{$isInfeasible$}\;
\end{algorithm}

\begin{algorithm}[!t]

\caption{Calculation of EFT in extended HEFT.}
\label{alg:heftEFT}

\scriptsize

\SetInd{0.65em}{0.65em} 

\KwIn{Candidate node $N_{i\hspace{-0.75pt}.\hspace{-0.75pt}n\hspace{-0.75pt}, \mu k\hspace{-0.75pt}.\hspace{-0.75pt}q}$ and sets $\mathcal{N}^{\prime\prime}_{\mathrm{temp}}$ and $\mathcal{A}^{\prime \prime}_{\mathrm{temp}}$ from \cref{alg:heft}\hspace{-0.5pt}.}

\KwOut{$EFT_{i\hspace{-0.75pt}.\hspace{-0.75pt}n\hspace{-0.75pt}, \mu k\hspace{-0.75pt}.\hspace{-0.75pt}q}$ and $t_{i\hspace{-0.75pt}.\hspace{-0.75pt}n}$ of task $\tau_{i\hspace{-0.75pt}.\hspace{-0.75pt}n}$ on core $p_{\mu k\hspace{-0.75pt}.\hspace{-0.75pt}q}$\hspace{-0.5pt}.}

    \vspace {2pt}$t_{i\hspace{-0.75pt}.\hspace{-0.75pt}n} \gets \hspace{-6pt} \max\limits_{A_{j\hspace{-0.75pt}.\hspace{-0.75pt}o\hspace{-0.75pt}, \nu l\hspace{-0.75pt}.\hspace{-0.75pt}r \rightarrow\hspace{-0.5pt}i\hspace{-0.75pt}.\hspace{-0.75pt}n\hspace{-0.75pt}, \mu k\hspace{-0.75pt}.\hspace{-0.75pt}q} \in \mathcal{A}^{\prime \prime}_{\mathrm{temp}}} \hspace{-8pt} \{ t_{j\hspace{-0.75pt}.\hspace{-0.75pt}o} + L_{j\hspace{-0.75pt}, \nu l\hspace{-0.75pt}.\hspace{-0.75pt}r} + CL_{j\hspace{-0.75pt}, \nu l\hspace{-0.75pt}.\hspace{-0.75pt}r \rightarrow \hspace{-0.75pt} i\hspace{-0.75pt}, \mu k\hspace{-0.75pt}.\hspace{-0.75pt}q}\}$\;

    $\mathit{\Lambda}^{\mu k} \gets \{ N_{j\hspace{-0.75pt}.\hspace{-0.75pt}o\hspace{-0.75pt}, \mu k\hspace{-0.75pt}.\hspace{-0.75pt}r} \in \mathcal{N}^{\prime\prime}_{\mathrm{temp}} \, | \, t_{j\hspace{-0.75pt}.\hspace{-0.75pt}o} + L_{j\hspace{-0.75pt}, \mu k\hspace{-0.75pt}.\hspace{-0.75pt}r} > t_{i\hspace{-0.75pt}.\hspace{-0.75pt}n} \}$\;
    
    Sort list $\mathit{\Lambda}^{\mu k}$ by non-decreasing order of $t_{j\hspace{-0.75pt}.\hspace{-0.75pt}o} + L_{j\hspace{-0.75pt}, \mu k\hspace{-0.75pt}.\hspace{-0.75pt}r}$\;
    \For{\textup{$\eta \gets 1$ \KwTo $|\mathit{\Lambda}^{\mu k}|$}}{
        
        $N_{j\hspace{-0.75pt}.\hspace{-0.75pt}o\hspace{-0.75pt}, \mu k\hspace{-0.75pt}.\hspace{-0.75pt}r} \gets \mathit{\Lambda}^{\mu k}_\eta$,      
        $\mathcal{N}^{\prime\prime}_{\hat{\mathrm{temp}}} \gets \{N_{i\hspace{-0.75pt}.\hspace{-0.75pt}n\hspace{-0.75pt}, \mu k\hspace{-0.75pt}.\hspace{-0.75pt}q}\} \cup \{ N_{b\hspace{-0.75pt}.\hspace{-0.75pt}h\hspace{-0.75pt}, \mu k\hspace{-0.75pt}.\hspace{-0.75pt}d} \in \mathit{\Lambda}^{\mu k}\, |\, d \neq q, t_{i\hspace{-0.75pt}.\hspace{-0.75pt}n} < (t_{b\hspace{-0.75pt}.\hspace{-0.75pt}h} + L_{b\hspace{-0.75pt}, \mu k\hspace{-0.75pt}.\hspace{-0.75pt}d}) \land (t_{i\hspace{-0.75pt}.\hspace{-0.75pt}n} + L_{i\hspace{-0.75pt}, \mu k\hspace{-0.75pt}.\hspace{-0.75pt}q}) > t_{b\hspace{-0.75pt}.\hspace{-0.75pt}h} \}$\;
        
        \vspace{1pt}\If{\textup{$(r \hspace{-1pt} = \hspace{-1pt} q \hspace{-1pt} \land \hspace{-1pt} t_{i\hspace{-0.75pt}.\hspace{-0.75pt}n} \hspace{-1pt} < \hspace{-1pt} (t_{j\hspace{-0.75pt}.\hspace{-0.75pt}o} \hspace{-1pt} + \hspace{-1pt} L_{j\hspace{-0.75pt}, \mu k\hspace{-0.75pt}.\hspace{-0.75pt}r}) \hspace{-1pt} \land \hspace{-1pt} (t_{i\hspace{-0.75pt}.\hspace{-0.75pt}n} \hspace{-1pt} + \hspace{-1pt} L_{i\hspace{-0.75pt}, \mu k\hspace{-0.75pt}.\hspace{-0.75pt}q}) \hspace{-1pt} > \hspace{-1pt} t_{j\hspace{-0.75pt}.\hspace{-0.75pt}o})
        \allowbreak \lor \hspace{-2pt} \bigg( \sum\limits_{N_{b\hspace{-0.75pt}.\hspace{-0.75pt}h\hspace{-0.75pt}, \mu k\hspace{-0.75pt}.\hspace{-0.75pt}d} \in \mathcal{N}^{\prime\prime}_{\hat{\mathrm{temp}}}} \hspace{-15pt} y_{\mu k}^{c_a}\,z_b^{c_a} > 1 \land z_i^{c_a}=1 \land a>0 \bigg) \allowbreak 
        \lor \hspace{-5pt} \sum\limits_{N_{b\hspace{-0.75pt}.\hspace{-0.75pt}h\hspace{-0.75pt}, \mu k\hspace{-0.75pt}.\hspace{-0.75pt}d} \in \mathcal{N}^{\prime\prime}_{\hat{\mathrm{temp}}}} \hspace{-15pt} M_b > M_{\mu k}^{\mathrm{bgt}}
        \allowbreak \, \lor \hspace{-5pt} \sum\limits_{N_{b\hspace{-0.75pt}.\hspace{-0.75pt}h\hspace{-0.75pt}, \mu k\hspace{-0.75pt}.\hspace{-0.75pt}d} \in \mathcal{N}^{\prime\prime}_{\hat{\mathrm{temp}}}} \hspace{-15pt} S_b > S_{\mu k}^{\mathrm{bgt}}$ 
        }}{
            $t_{i\hspace{-0.75pt}.\hspace{-0.75pt}n} \gets t_{j\hspace{-0.75pt}.\hspace{-0.75pt}o} + L_{j\hspace{-0.75pt}, \mu k\hspace{-0.75pt}.\hspace{-0.75pt}r}$\;
        }

        $\eta \gets \eta + 1$\;
    }
    $EFT_{i\hspace{-0.75pt}.\hspace{-0.75pt}n\hspace{-0.75pt}, \mu k\hspace{-0.75pt}.\hspace{-0.75pt}q} \gets t_{i\hspace{-0.75pt}.\hspace{-0.75pt}n} \hspace{-0.5pt} + \hspace{-0.5pt} L_{i\hspace{-0.75pt}, \mu k\hspace{-0.75pt}.\hspace{-0.75pt}q}$\;

\end{algorithm}


In the second phase (lines 6--62), for each unscheduled task $\tau_i$ we examine each of its corresponding sets in $\mathit{\Lambda}$ (in order of rank) to select the one that minimizes the employed multi-objective function.
Specifically, for each set $\mathit{\Lambda}_v \hspace{-1.5pt} \in \hspace{-1.5pt} \mathit{\Lambda}$, we first check if its candidate nodes concern the allocation of $\tau_{i\hspace{-0.5pt}.\hspace{-0.5pt}1}$ and $\tau_{i\hspace{-0.5pt}.\hspace{-0.5pt}2}$ on different devices (in case $\tau_i$ is an exit task and task duplication is required), and if the resulting total reliability is below the reliability threshold of $\tau_i$ (line 12), as in MILP constraints \eqref{eq:one3} and \eqref{eq:three4}, respectively. 
If one of these conditions holds, we skip $\mathit{\Lambda}_v$ and continue with the next (according to its rank) set of $\tau_i$ (line 13).

\begin{figure}[!t]
    \centering
    \includegraphics[width=0.75\columnwidth]{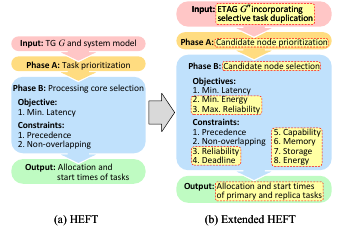}
    \caption{Overview of our extension to HEFT.}
    \label{fig:HEFT}
\end{figure}

Otherwise, for each candidate node $N_{i\hspace{-0.5pt}.\hspace{-0.5pt}n\hspace{-0.5pt}, \mu k\hspace{-0.5pt}.\hspace{-0.5pt}q} \hspace{-1.5pt} \in \hspace{-1.5pt} \mathit{\Lambda}_v$, we examine if it would exceed the energy budget of any device (lines 20--25), using \eqref{eq:fourteen}.
Additionally, we check if the memory or storage requirements of $\tau_i$ exceed the respective budgets of device $u_{\mu k}$ (line 25). If any of these conditions hold, we skip $\mathit{\Lambda}_v$ and continue with the next set of $\tau_i$ (line 26). 
Otherwise, we determine the earliest finish time $EFT_{i\hspace{-0.5pt}.\hspace{-0.5pt}n\hspace{-0.5pt}, \mu k\hspace{-0.5pt}.\hspace{-0.5pt}q}$ of candidate node $N_{i\hspace{-0.5pt}.\hspace{-0.5pt}n\hspace{-0.5pt}, \mu k\hspace{-0.5pt}.\hspace{-0.5pt}q} \hspace{-1.5pt}$ using \cref{alg:heftEFT}.
In \cref{alg:heftEFT}, we ensure that the precedence constraints of primary or replica task  $\tau_{i\hspace{-0.5pt}.\hspace{-0.5pt}n}$ are satisfied (line 1), as in \eqref{eq:four}.
Moreover, we ensure that the execution of $\tau_{i\hspace{-0.5pt}.\hspace{-0.5pt}n}$ does not overlap with other tasks allocated on $p_{\mu k\hspace{-0.5pt}.\hspace{-0.5pt}q}$, and that the capability, memory, and storage constraints of device $u_{\mu k}$ are not violated (lines 2--11), similar to \eqref{eq:eight1}, \eqref{eq:eight2}, and \eqref{eq:eleven}--\eqref{eq:thirteen} in our MILP method, respectively.
Subsequently, in \cref{alg:heft}, we determine the total latency, energy, and reliability of set $\mathit{\Lambda}_v$ (lines 35--37), and we use their normalized values to calculate the multi-objective function for each set of $\tau_i$ (lines 41--44). 
Finally, we select for $\tau_i$ the set of candidate nodes that minimizes the multi-objective function, without exceeding the defined deadline (lines 45--55), as in \eqref{eq:six}.
If there is a task for which no set satisfies all of the above constraints, then the problem is infeasible.
The output of extended HEFT includes the selected candidate nodes and arcs in $G^{\prime \prime}$, i.e., the resulting allocation of primary and replica tasks, and their start times. 
\cref{fig:HEFT} shows an overview of our extension to HEFT, with our modifications highlighted in yellow.

\subsection{Extended HEFT Complexity}
\label{subsec:HEFTtimeComplexity}

In the worst case, each task $\tau_i \hspace{-1.5pt} \in \hspace{-1.5pt} \mathcal{T}$ can be allocated on any processing core $p_{\mu k\hspace{-0.5pt}.\hspace{-0.5pt}q} \hspace{-1.5pt} \in \hspace{-1.5pt} \mathcal{P}$, with each allocation requiring task duplication. 
Hence, there are $|\mathcal{N}^{\prime \prime}| \hspace{-2.5pt} = \hspace{-2.5pt} 2 |\mathcal{T}| |\mathcal{P}|$ candidate nodes in $\mathcal{N}^{\prime \prime}$, and therefore $|\mathcal{T}| |\mathcal{P}|^2$ sets of candidate nodes in list $\mathit{\Lambda}$, where $|\mathcal{T}| \hspace{-2.5pt} > \hspace{-2.5pt} |\mathcal{P}|$.
Moreover, in dense ETAGs the number of arcs $|\mathcal{A}^{\prime \prime}|$ is proportional to $|\mathcal{N}^{\prime \prime}|^2$ \cite{Topcuoglu2002}.
Thus, the worst-case time complexity of the first phase of extended HEFT is dominated by the operations in lines 1--3, resulting in $O(|\mathcal{T}|^3|\mathcal{P}|^3)$.
The worst-case time complexity of the second phase is $O(|\mathcal{U}| |\mathcal{T}|^3 |\mathcal{P}|^4)$, determined by the operations in lines 20--24. 
Hence, the overall worst-case time complexity of extended HEFT is $O(|\mathcal{U}| |\mathcal{T}|^3 |\mathcal{P}|^4)$. 
Regarding ETAG $G^{\prime \prime}$, which is used as input in both extended HEFT and our MILP approach, as mentioned in \cref{subsubsec:etagSize}, 
in the worst case the number of its nodes and arcs increases by linear and quadratic factors, respectively, with respect to the number of available cores $|\mathcal{P}|$, compared to TG $G$.

%% file: 5_results.tex
\section{Evaluation}
\label{sec:evaluation}

We assessed the proposed MILP method in comparison with extended HEFT, using a relevant real-world IoT workflow under different system configurations, while varying the relative importance of latency, energy, and reliability.
We further validated our approach and investigated its scalability across different task graph sizes, using suitable synthetic IoT workflows we developed for this purpose.

\begin{figure}[t]
    \centering
    \begin{minipage}[b]{0.33\columnwidth}
        \includegraphics[width=\columnwidth]{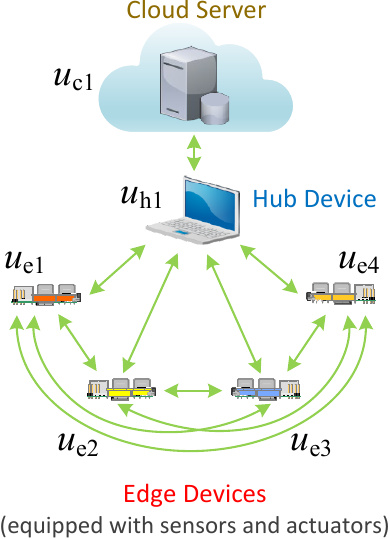}
        \vspace{-15pt}
        \caption{Examined CPS.}
        \label{fig:system}
    \end{minipage}
    \hfill
    \begin{minipage}[b]{0.57\columnwidth}
        \begin{table}[H]
    \setlength{\tabcolsep}{0.5pt}
    \centering
    \caption{Device Capabilities}
    \resizebox{\columnwidth}{!}{
        \begin{tabular}{llcc}
            \toprule
            \multirow{2}{*}{$c_a$} & \multirow{2}{*}{\hspace{1pt}Capability} & Device & Task\\
                                &                                                       & Type   & Type\\
            \hline
            0 & \hspace{1pt}Basic computational capability & All & All\\
            1 & \hspace{1pt}Thermal camera & $\mathrm{e}$ & Entry\\
            2 & \hspace{1pt}LiDAR sensor & $\mathrm{e}$ & Entry\\
            3 & \hspace{1pt}Multispectral camera & $\mathrm{e}$ & Entry\\
            
            4 & \hspace{1pt}High-precision GNSS${}^{1}$ module & $\mathrm{e}$ & Interm.\\
            
            5 & \hspace{1pt}Tag release mechanism & $\mathrm{e}$ & Exit\\
            6 & \hspace{1pt}UAV coordination module & $\mathrm{h}$ & Interm.\\
            7 & \hspace{1pt}Integrated display & $\mathrm{h}$ & Exit\\
            8 & \hspace{1pt}High-performance GPU & $\mathrm{c}$ & Interm.\\
            9 & \hspace{1pt}High-availability storage & $\mathrm{c}$ & Exit\\
            \bottomrule
            \multicolumn{4}{l}{${}^{1}$Global Navigation Satellite System.}\\
            
        \end{tabular}
    }
    \label{tab:capabilities}
    \end{table}
    \end{minipage}
\end{figure}

\begin{table*}[!t]
    \setlength{\tabcolsep}{1.5pt}
    \centering
    \caption{System Devices}
    \resizebox{0.97\textwidth}{!}{
    \begin{tabular}{
    @{\extracolsep{4pt}}
    cccrrccrrcccc}
    \toprule
    
    \multirow{3}{*}{$u_{\mu k}$} & \multirow{3}{*}{Device} & \multicolumn{4}{c}{Hardware Specifications} & \multicolumn{4}{c}{Budgets \& Reserved\,Cores} & \multicolumn{3}{c}{Capability Assignment per Configuration} \\
    
    \cline{3-6}
    \cline{7-10}
    \cline{11-13}
    
     & & \multirow{2}{*}{Processor} & \multicolumn{1}{c}{Memory} & \multicolumn{1}{c}{Storage} & Battery${}^{3}$ & $M_{\mu k}^{\mathrm{bgt}}$ & \multicolumn{1}{c}{$S_{\mu k}^{\mathrm{bgt}}$} & \multicolumn{1}{c}{$E_{\mu k}^{\mathrm{bgt}}$} & \multirow{2}{*}{$\beta_{\mu k}$} & \multirow{2}{*}{C1} & \multirow{2}{*}{C2} & \multirow{2}{*}{C3}\\
                    
     & & & \multicolumn{1}{c}{(GiB)} & \multicolumn{1}{c}{(GiB)}  & (Wh) & (GiB) & \multicolumn{1}{c}{(GiB)} & \multicolumn{1}{c}{(Wh)} & & & &\\
    \hline

    $u_{\mathrm{e}1}$ & Raspberry\,Pi\,3   & Cortex-A53\,@\,1.4\,GHz                              & 1 \hspace{5pt}   & 16 \hspace{5pt}    & 33.3 & 0.95 & 1.0 \hspace{3pt}  & 1 \hspace{3pt}  & 2	& $\{0, 1, 5\}$    & $\{0, 5\}$  & $\{0, 1, 2, 3, 4, 5\}$\\
    $u_{\mathrm{e}2}$ & Odroid\,XU4        & Cortex-A7\,\&\,Cortex-A15${}^{1}$\,@\,2.0\,GHz       & 2 \hspace{5pt}   & 16 \hspace{5pt}    & 33.3 & 1.00 & 1.5 \hspace{3pt}  & 1 \hspace{3pt}  & 2	& $\{0, 1, 5\}$  & $\{0, 1\}$    & $\{0, 1, 2, 3, 4, 5\}$\\
    $u_{\mathrm{e}3}$ & Jetson\,TX2        & NVIDIA\,Denver2\,\&\,Cortex-A57${}^{2}$\,@\,2.0\,GHz & 8 \hspace{5pt}   & 32 \hspace{5pt}    & 33.3 & 2.00 & 2.0 \hspace{3pt}  & 1 \hspace{3pt}  & 2	& $\{0, 2, 3, 4\}$  & $\{0, 3\}$ & $\{0, 1, 2, 3, 4, 5\}$\\ 
    $u_{\mathrm{e}4}$ & Jetson\,Xavier\,NX & NVIDIA\,Carmel ARMv8.2\,@\,1.4\,GHz                  & 8 \hspace{5pt}   & 32 \hspace{5pt}    & 33.3 & 2.00 & 2.5 \hspace{3pt}  & 1 \hspace{3pt}  & 2	& $\{0, 2, 3, 4\}$ & $\{0, 2, 4\}$  & $\{0, 1, 2, 3, 4, 5\}$\\
    $u_{\mathrm{h}1}$ & Mi\,Notebook\,Pro  & Intel\,i5\,8250U\,@\,1.6\,GHz                        & 8 \hspace{5pt}   & 512 \hspace{5pt}   & 60.0 & 3.00 & 5.0 \hspace{3pt}  & 2 \hspace{3pt}  & 4	& $\{0, 6, 7\}$    & $\{0, 6, 7\}$    & $\{0, 6, 7\}$\\
    $u_{\mathrm{c}1}$ & HPE\,DL580\,Gen10  & Intel\,Xeon\,Gold\,6240\,@\,2.6\,GHz                 & 400 \hspace{5pt} & 10\hspace{0.75pt}240 \hspace{5pt} & --   & 4.00 & 10.0 \hspace{3pt} & 10 \hspace{3pt} & 6	& $\{0, 8, 9\}$    & $\{0, 8, 9\}$    & $\{0, 8, 9\}$\\
    \bottomrule
    
    \multicolumn{13}{l}{${}^{1,2}$Without loss of generality, the reserved cores are considered to be located on Cortex-A15 and Cortex-A57, respectively.}\\
    \multicolumn{13}{l}{${}^{3}$A compatible external battery (TalentCell YB1203000-USB) is considered for the edge devices.}
    \end{tabular}
    }
    \label{tab:devices}
\end{table*}

\subsection{Experimental Setup}
\label{subsec:setup}
The examined edge-hub-cloud CPS (\cref{fig:system}) comprised four different edge devices, a hub device, and a cloud server, each modeled after typical real-world counterparts with heterogeneous multicore processors and diverse memory, storage, and energy capacities.
All devices also featured various sensing, actuating, or other specialized capabilities, based on the examined real-world use case. 
These capabilities, denoted by integers 0--9, were assigned to system devices based on their type (edge $\mathrm{e}$, hub $\mathrm{h}$, or cloud $\mathrm{c}$), accommodating the requirements of entry, intermediate, and exit tasks, as shown in \cref{tab:capabilities}. 
Each edge device was assumed to be attached to a UAV. 
We examined three system configurations (C1--C3) with varied capabilities assigned to the edge devices, to reflect different provisioning strategies. 
In configuration C1 (moderately provisioned), each specialized capability in $\{1, \allowbreak 2, \allowbreak 3, \allowbreak 4, \allowbreak 5\}$ was assigned to two edge devices. In C2 (minimally provisioned), each of these capabilities was assigned to only one edge device, whereas in C3 (over-provisioned), each capability was assigned to all four edge devices. 
In all configurations, the hub device and the cloud server were assigned the specialized capabilities $\{6, \allowbreak 7\}$ and $\{8, \allowbreak 9\}$, respectively, while all devices featured the basic computational capability (denoted by 0).

As the computational resources in the targeted CPS are typically limited and shared among different applications, the budgets $M_{\mu k}^{\mathrm{bgt}}$, $S_{\mu k}^{\mathrm{bgt}}$, and $E_{\mu k}^{\mathrm{bgt}}$, as well as the number of reserved cores $\beta_{\mu k}$, were subsets of the respective resources on each device.
\cref{tab:devices} shows the system devices and their hardware specifications, the considered budgets and reserved cores on each device, and the assigned capabilities per configuration.
The core failure rates for each device type (edge $\mathrm{e}$, hub $\mathrm{h}$, and cloud $\mathrm{c}$) are shown in \cref{tab:failureRates}. They were derived from \cite{Cui2021}, reflecting that edge devices are typically less reliable than a hub device, which is in turn less reliable than a cloud server.
\cref{tab:channels} shows the bandwidth and energy parameters for the communication channels between each pair of devices, derived from real-world measurements \cite{Huang2012, Vladan2021}. 
As we adopt an offline optimization approach, all application, system, and network parameters were treated as fixed inputs during optimization. These parameters were instantiated with conservative values derived from real-world measurements and application profiling to ensure feasibility under adverse operating conditions. 
In line with the safety-critical nature of the targeted use cases, the reserved cores across the system devices were assumed to be used exclusively for the considered workflow application.
While parameters do not vary at runtime, the framework can be re-executed at design time with different parameter sets to evaluate alternative operating conditions.
We implemented our MILP approach and extended HEFT in C++. The formulated MILP problem was solved using Gurobi Optimizer 11.0.3 \cite{gurobi}, on a server running CentOS 7.9, equipped with an Intel Xeon Gold 6240 processor\,@\,2.6\,GHz and 400\,GiB of RAM.
%
All experiments were executed using the default Gurobi solver parameters (including feasibility and optimality tolerances), except for the Integrality Focus parameter, which was set from its default value of 0 to 1 to emphasize the early discovery of integer-feasible solutions. No parameter tuning was performed to avoid bias and ensure result generality.

\subsection{Experiments With Real-World IoT Workflow}
\label{subsec:real}

\subsubsection{Overview}
\label{subsubsec:realOverview}
\begin{figure}[t]
    \centering
    \begin{minipage}[b]{0.68\columnwidth}
        \begin{table}[H]
        \setlength{\tabcolsep}{2pt}
        \centering
        \caption{Real-World IoT Workflow Tasks}
        \resizebox{\columnwidth}{!}{
            \begin{tabular}{llcc}
            \toprule
            Task & Description & $c_a$ & $R_i^{\mathrm{thr}}$\\
            \hline
            $N_1$ & Capture multispectral image & 3 & 0.9999\\
            $N_2$ & Multispectral image preprocessing & 0 & 0.9996\\
            $N_3$ & Detect power transmission lines & 0 & 0.9994\\
            $N_4$ & Detect vegetation encroachment & 0 & 0.9994\\
            $N_5$ & Perform LiDAR scan & 2 & 0.9998\\
            $N_6$ & Data preprocessing/get GNSS data & 4 & 0.9997\\
            $N_7$ & Detect power transmission towers & 0 & 0.9996\\
            $N_8$ & Detect structural integrity problems & 0 & 0.9997\\
            $N_9$ & Capture thermal image & 1 & 0.9999\\
            $N_{10}$ & Thermal image preprocessing & 0 & 0.9995\\
            $N_{11}$ & Detect overheated components & 0 & 0.9993\\
            $N_{12}$ & Multi-source data fusion & 8 & 0.9999\\
            $N_{13}$ & Save data on high-availability storage & 9 & 0.9995\\
            $N_{14}$ & Plan path/coordinate tag deployment & 6 & 0.9998\\
            $N_{15}$ & Display final output & 7 & 0.9993\\
            $N_{16}$ & Deploy tags at problematic sections & 5 & 0.9998\\
            \bottomrule
            \end{tabular}
        }
        \label{tab:realTasks}
        \end{table}
    \end{minipage}
     \begin{minipage}[b]{0.25\columnwidth}
        \centering
        \includegraphics[width=0.7\columnwidth]{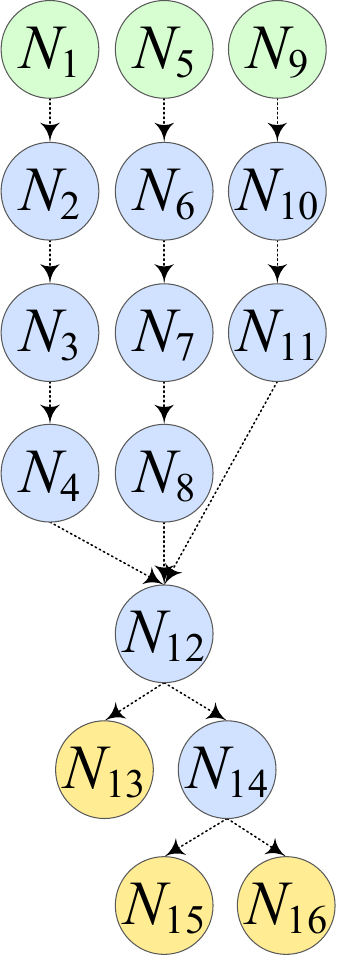}
        \caption{Real-world IoT workflow TG.}
        \label{fig:realWorkflow}
        \vspace{2pt}
    \end{minipage}
\end{figure}

We considered a real-world IoT workflow for the UAV-enabled autonomous inspection of power transmission towers and lines, based on \cite{Savva2021}.
\cref{tab:realTasks} lists its tasks along with their required capabilities and reliability thresholds. \cref{fig:realWorkflow} illustrates its TG. 
Entry and exit tasks are depicted in green and yellow, respectively, while intermediate tasks are shown in blue. 
The workflow involves the collaboration of multiple UAVs, each equipped with an edge device, to capture multispectral, LiDAR, and thermal images for the detection of infrastructure issues, such as vegetation encroachment, structural integrity problems, and overheated components, respectively. 
The collected data are fused to create a visual representation of the infrastructure, necessitating a cloud-based high-performance GPU (as data fusion is computationally demanding) and high-availability storage (to ensure data accessibility). 
The fused data are used by the hub device for tag deployment path planning. The output is displayed on the hub device, while a UAV starts deploying location-transmitting tags at the identified problematic sections of the infrastructure.

\begin{figure}[t]
    \centering
    \begin{minipage}[b]{0.37\columnwidth}
        \begin{table}[H]
        \setlength{\tabcolsep}{1.5pt}
        \centering
        \caption{Processing Core\\ Failure Rates}
        \resizebox{\columnwidth}{!}{
            \begin{tabular}{cc}
                \toprule
                Device & $\lambda_{\mu k\hspace{-0.5pt}.\hspace{-0.5pt}q}$\\
                Type   & (failures/s)\\
                \hline
                $\mathrm{e}$ & $[6 \times 10^{-4}, 8 \times 10^{-4}]$\\
                $\mathrm{h}$ & $[4 \times 10^{-4}, 6 \times 10^{-4}]$\\
                $\mathrm{c}$ & $[2 \times 10^{-4}, 4 \times 10^{-4}]$\\
                \bottomrule
            \end{tabular}
        }
        \label{tab:failureRates}
        \end{table}
    \end{minipage}
    \hfill
    \begin{minipage}[b]{0.52\columnwidth}
        \begin{table}[H]
        \setlength{\tabcolsep}{1.5pt}
        \centering
        \caption{Communication Channels}
        \resizebox{\columnwidth}{!}{
        \begin{tabular}{lccc}
        \toprule
        \multicolumn{1}{c}{Comm.} & $\epsilon_{\mu k, \nu l}$ & $\pi_{\mu k, \nu l}$ & $\sigma_{\mu k, \nu l}$\\
        \multicolumn{1}{c}{Channel} & (Mbit/s) & (\SI{}{\micro \joule}/bit) & (\SI{}{\micro \joule}/bit)\\
        \hline
        \hspace{3pt} $u_{\mathrm{e} k} \leftrightarrow u_{\mathrm{e} l}$ & $[6,9]$ & $[0.6,1.0]$ & $[0.4,0.6]$\\
        \hspace{3pt} $u_{\mathrm{e} k} \rightarrow u_{\mathrm{h} 1}$ & $[9,13]$ & $[0.8,1.2]$ & $[0.6,0.8]$\\
        \hspace{3pt} $u_{\mathrm{h} 1} \rightarrow u_{\mathrm{e} l}$ & $[7,10]$ & $[0.7,1.1]$ & $[0.5,0.7]$\\
        \hspace{3pt} $u_{\mathrm{h} 1} \rightarrow u_{\mathrm{c} 1}$ & $[10,15]$ & $[1.8,2.7]$ & $[0.8,1.2]$\\
        \hspace{3pt} $u_{\mathrm{c} 1} \rightarrow u_{\mathrm{h} 1}$ & $[16,24]$ & $[2.0,3.0]$ & $[1.0,1.5]$\\
        \bottomrule
        \end{tabular}
        }
        \label{tab:channels}
        \end{table}
    \end{minipage}
\end{figure}

We investigated the execution of this workflow under system configurations C1--C3 (\cref{tab:devices}), while varying the weight of each objective ($w_{\mathrm{lat}}$, $w_{\mathrm{en}}$, and $w_{\mathrm{rel}}$) in increments and decrements of $1/3$ within the interval $[0,1]$, to examine a comprehensive range of scenarios.
For each configuration, the workflow TG was transformed into the corresponding ETAG. 
The number of ETAG candidate nodes and arcs in each case is shown in \cref{tab:realETAGs}.
The ETAG parameters $M_i$, $S_i$, $D_i$, $L_{i\hspace{-0.5pt}, \mu k\hspace{-0.5pt}.\hspace{-0.5pt}q}$, and $P_{i\hspace{-0.5pt}, \mu k\hspace{-0.5pt}.\hspace{-0.5pt}q}$ were determined through profiling and power monitoring tools (perf and Powertop) \cite{powertop} across all system devices.
The reliability threshold $R_i^{\mathrm{thr}}$ of each task was set within $[0.9990, 0.9999]$, so that based on \eqref{eq:nodeReliability} it would not always be met by solely executing the primary task \cite{Cui2021}.
The values of the ETAG parameters are listed in \cref{tab:realParams}. 
$E_{i\hspace{-0.5pt}, \mu k\hspace{-0.5pt}.\hspace{-0.5pt}q}$, $R_{i\hspace{-0.5pt}, \mu k\hspace{-0.5pt}.\hspace{-0.5pt}q}$, $\zeta_{i\hspace{-0.5pt}, \mu k\hspace{-0.5pt}.\hspace{-0.5pt}q}$, $CL_{i\hspace{-0.5pt}, \mu k\hspace{-0.5pt}.\hspace{-0.5pt}q \rightarrow\hspace{-0.5pt}j\hspace{-0.5pt}, \nu l\hspace{-0.5pt}.\hspace{-0.5pt}r}$, and $CE_{i\hspace{-0.5pt}, \mu k\hspace{-0.5pt}.\hspace{-0.5pt}q \rightarrow\hspace{-0.5pt}j\hspace{-0.5pt}, \nu l\hspace{-0.5pt}.\hspace{-0.5pt}r} \hspace{-1pt}$ were determined based on \eqref{eq:nodeEnergy}, \eqref{eq:nodeReliability}, \eqref{eq:reqReplica}, \eqref{eq:arcLatency}, and \eqref{eq:arcEnergy}, respectively. $Q_i$ and $\phi_i$ were derived from the TG structure, whereas $\theta_{i\hspace{-0.5pt}, \mu k\hspace{-0.5pt}.\hspace{-0.5pt}q \rightarrow\hspace{-0.5pt}j\hspace{-0.5pt}, \nu l\hspace{-0.5pt}.\hspace{-0.5pt}r}^{\xi m}$ was derived from the corresponding TAG structure. 
For each ETAG, the deadline $L_{\mathrm{thr}}$ was set at 1.5 times the length of its critical path (i.e., the longest path from an entry task candidate node to an exit task candidate node), taking into account the computational and communication latency of the candidate nodes and arcs on the path, respectively, as this was a realistic yet challenging scenario \cite{Bai2021}.

\begin{figure}[t]
    \centering
    \begin{minipage}[b]{0.58\columnwidth}
        \begin{table}[H]
        \setlength{\tabcolsep}{1.5pt}
        \centering
        \caption{Real-World IoT Workflow ETAGs}
        \resizebox{\columnwidth}{!}{
            \begin{tabular}{cccc}
                \toprule
                \multirow{2}{*}{Config.} & \#Nodes/ & \#Variables/ & Avg. Solver\\
                 & Arcs & Constraints & Runtime (s)\\
                \hline
                C1 & 278\,/\,5904 & 23793\,/\,72699 & 249.7\\
                
                C2 & 256\,/\,5440 & 20977\,/\,64359 & 126.3\\
                
                C3 & 314\,/\,6832 & 26869\,/\,83531 & 270.1\\
                \bottomrule
                \vspace{5pt}
            \end{tabular}
        }
        \label{tab:realETAGs}
        \end{table}
    \end{minipage}
    \hfill
    \begin{minipage}[b]{0.386\columnwidth}
        \begin{table}[H]
        \setlength{\tabcolsep}{1.5pt}
        \centering
        \caption{Real-World Workflow ETAG Parameter Ranges}
        \resizebox{\columnwidth}{!}{
            \begin{tabular}{lc}
                \toprule
                Param. & Range\\
                \hline
                $M_i$ & $[12.4, 453.0]$\,MiB\\
                $S_i$ & $[29.3, 448.9]$\,MiB\\
                $D_i$ & $[0.4, 18.1]$\,MiB\\
                $L_{i\hspace{-0.75pt}, \mu k\hspace{-0.75pt}.\hspace{-0.75pt}q}$ & $[2.6, 12648.4]$\,ms\\
                $P_{i\hspace{-0.75pt}, \mu k\hspace{-0.75pt}.\hspace{-0.75pt}q}$ & $[0.3, 23.7]$\,W\\
                $R_i^{\mathrm{thr}}$ & $[0.9990, 0.9999]$\\
                \bottomrule
            \end{tabular}
        }
        \label{tab:realParams}
        \end{table}
    \end{minipage}
\end{figure}

\subsubsection{Results}
\label{subsubsec:realResults}
\begin{figure}[t]
    \centering
    \subfloat[]{\includegraphics[width=0.97\columnwidth]{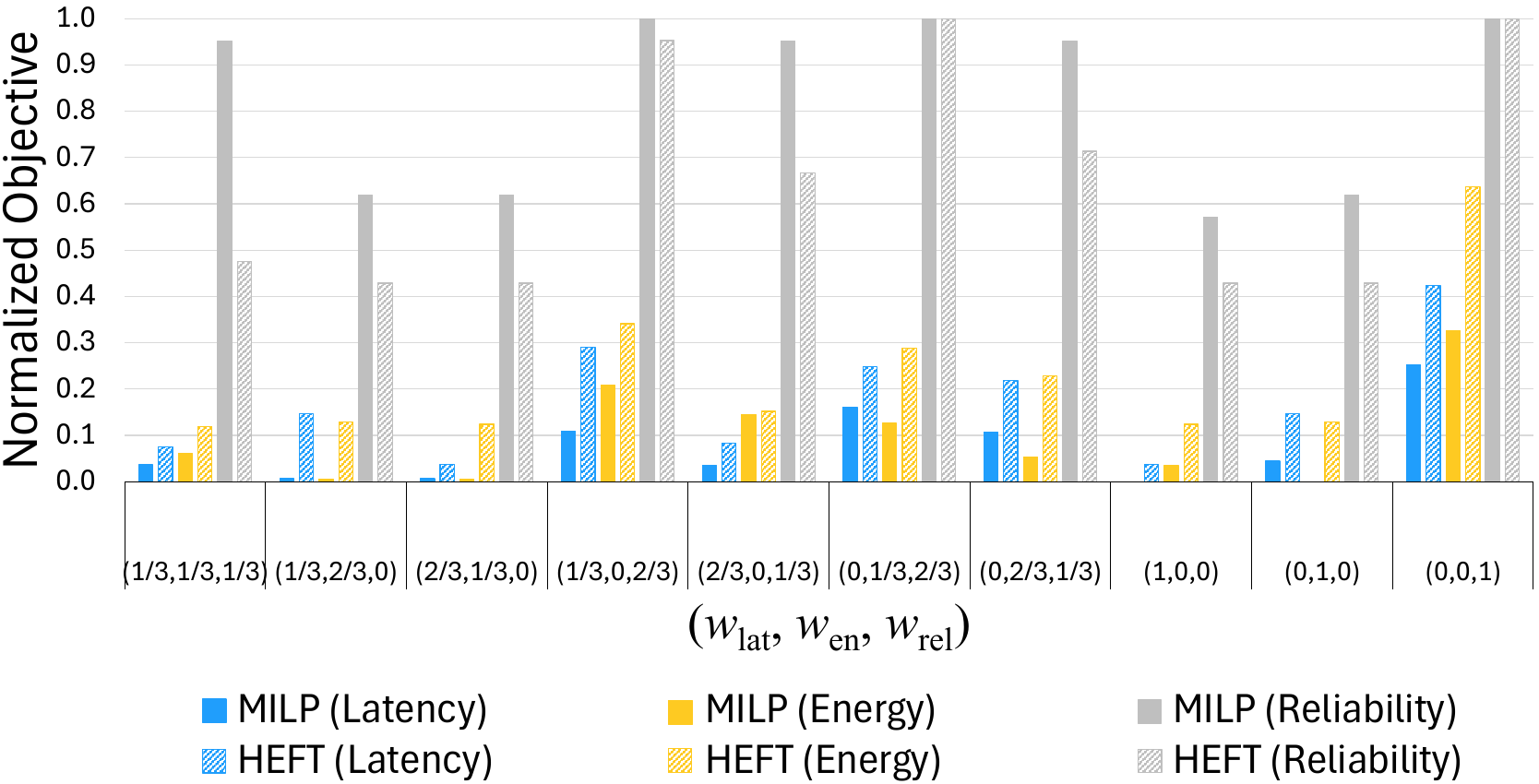}
        \label{fig:normalized}}\\
    \subfloat[]{\includegraphics[width=0.97\columnwidth]{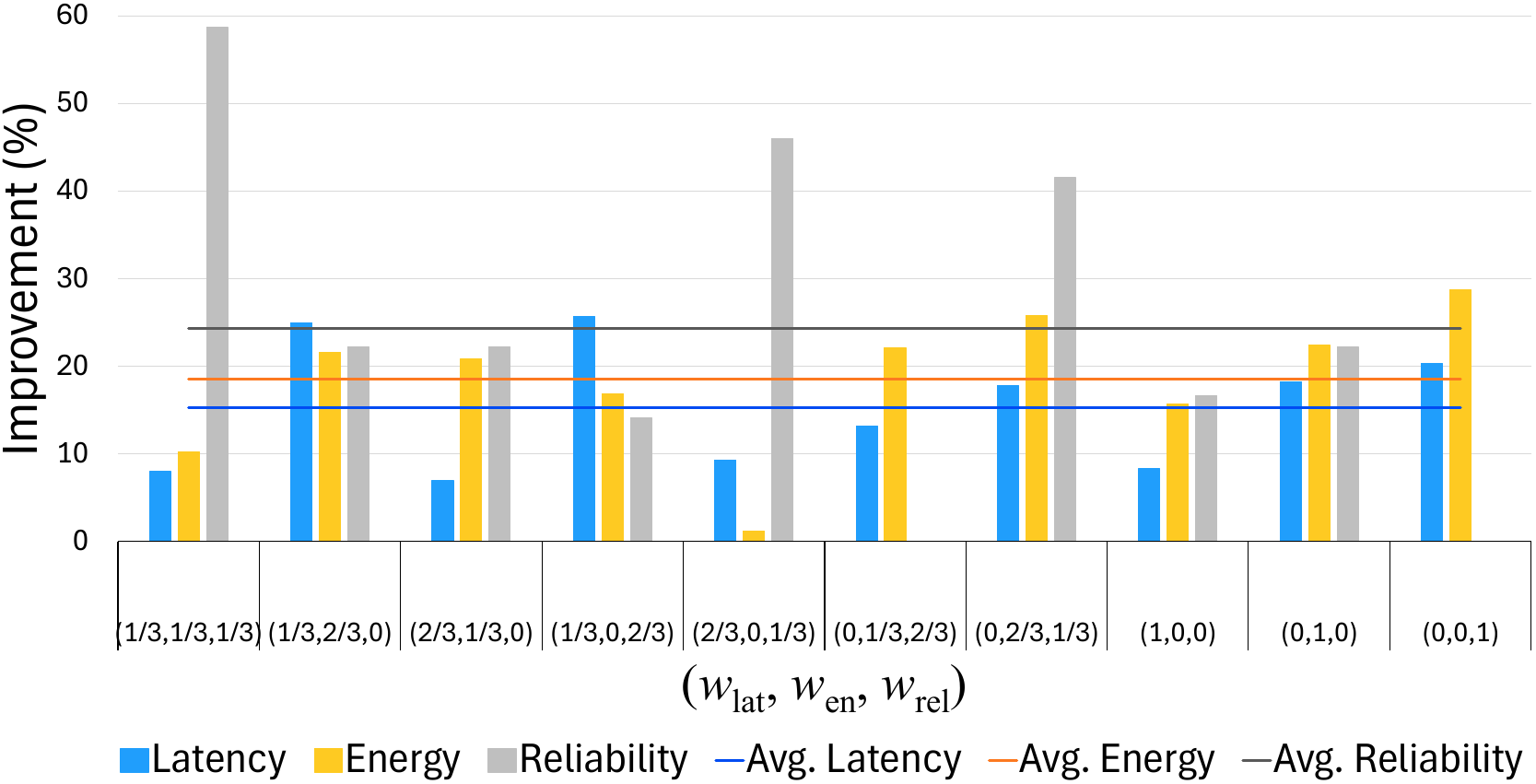}
        \label{fig:improvement}}\\
    \subfloat[]{\includegraphics[width=0.97\columnwidth]{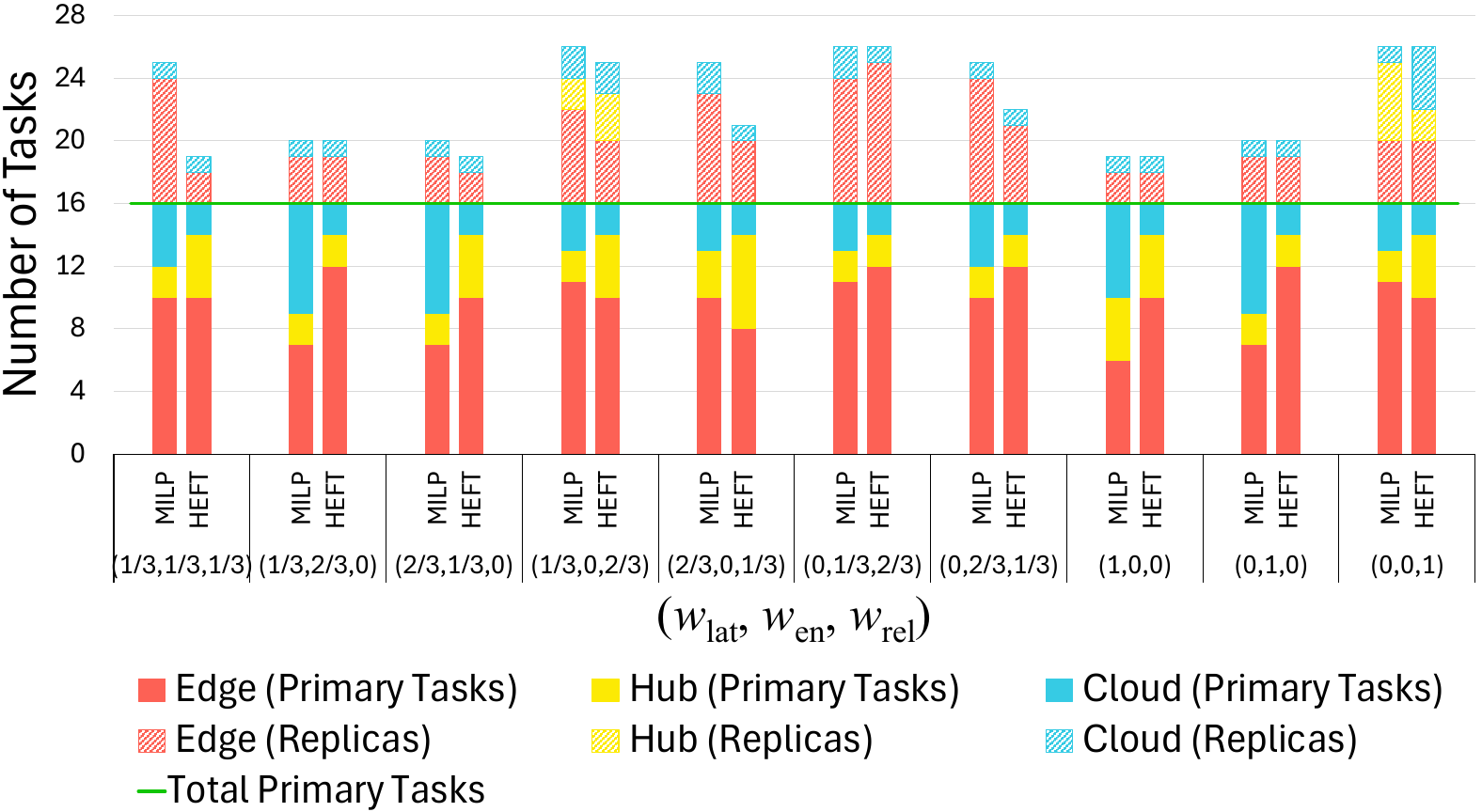}
        \label{fig:allocation}}
    \caption{Comparison between proposed MILP approach and extended HEFT for the real-world IoT workflow under system configuration C1.}
    \label{fig:realResults}
\end{figure}

\begin{figure}[t]
    \centering
    \subfloat[]{\includegraphics[width=0.97\columnwidth]{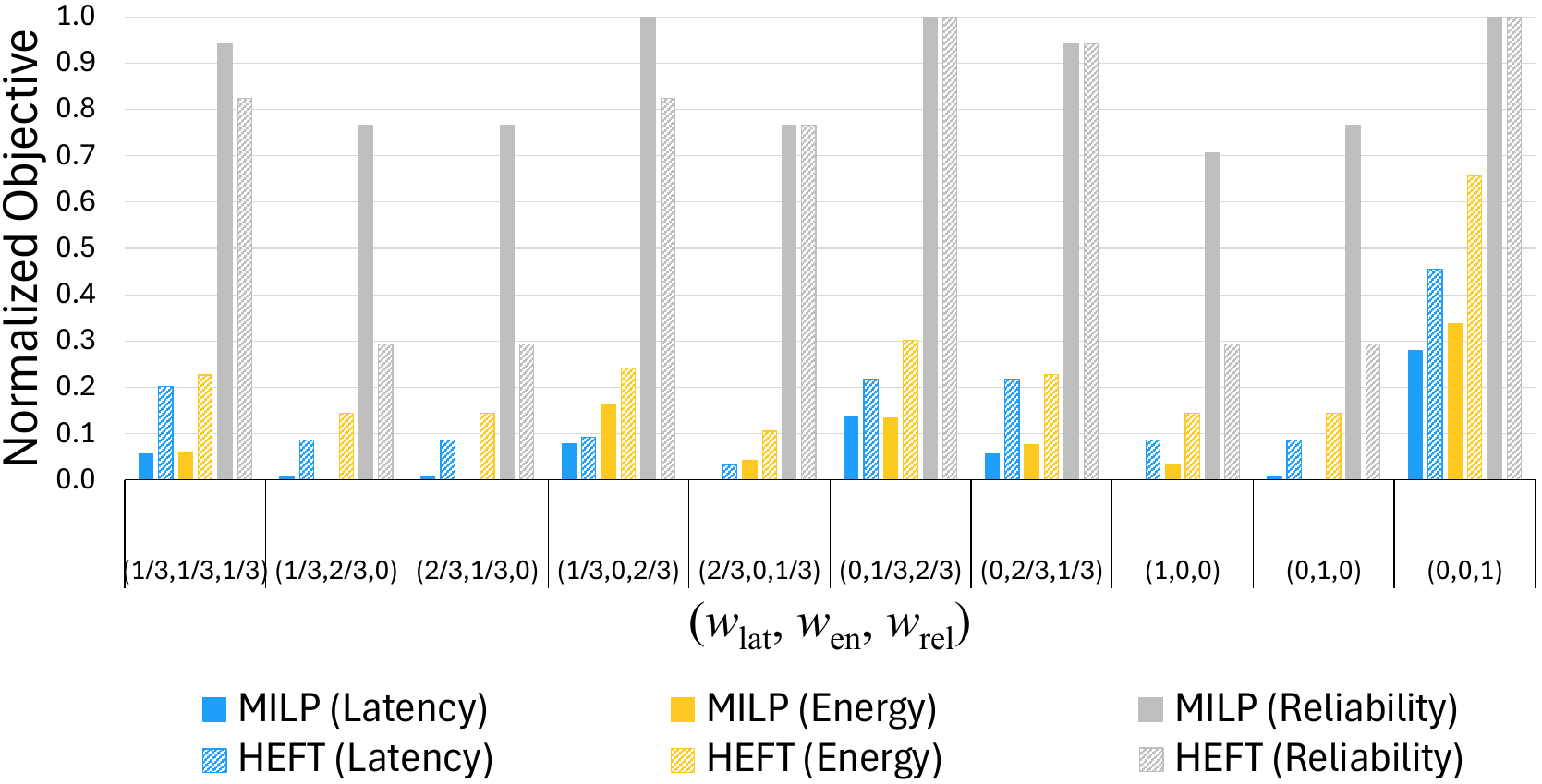}
        \label{fig:normalized2}}\\
    \subfloat[]{\includegraphics[width=0.97\columnwidth]{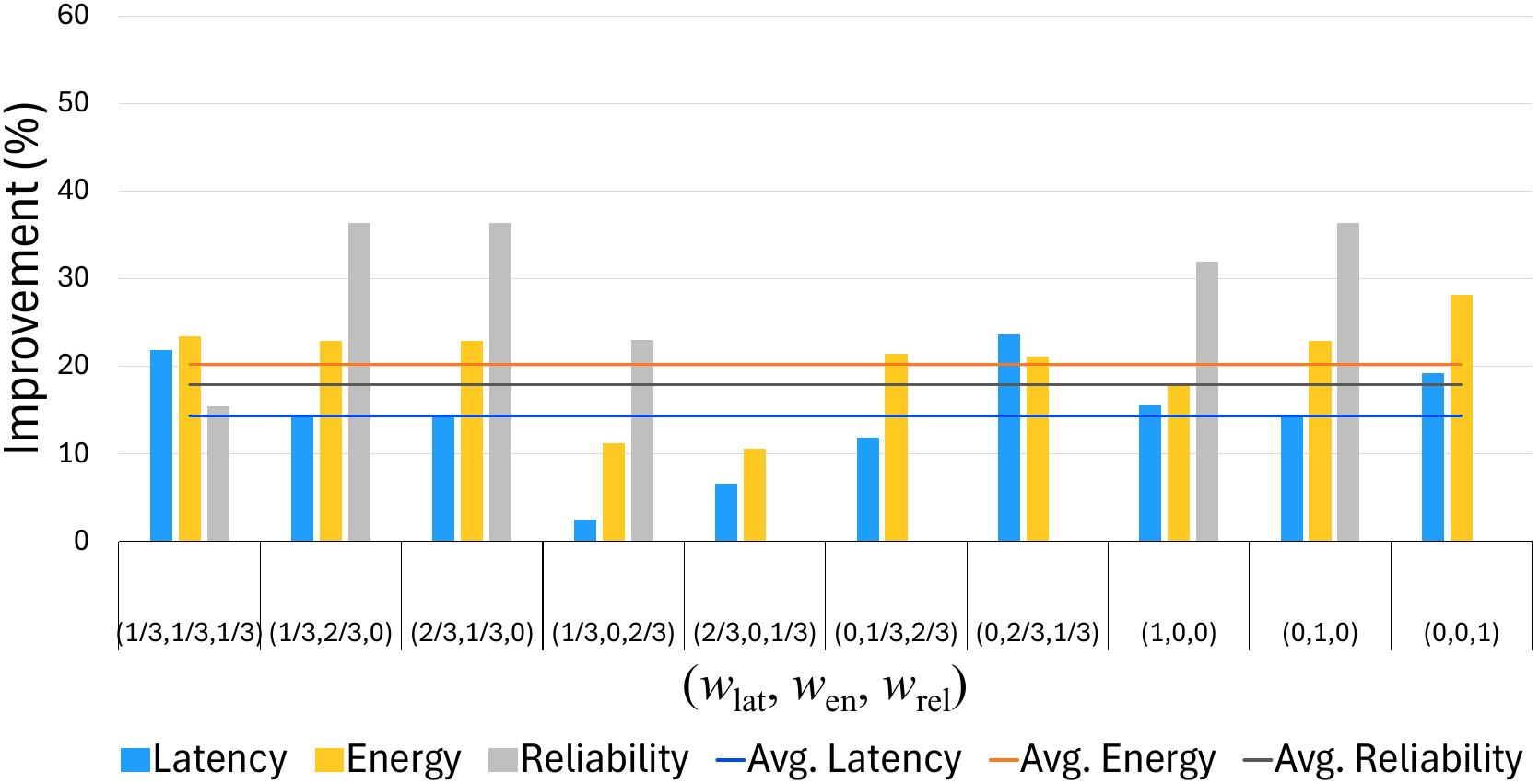}
        \label{fig:improvement2}}\\
    \subfloat[]{\includegraphics[width=0.97\columnwidth]{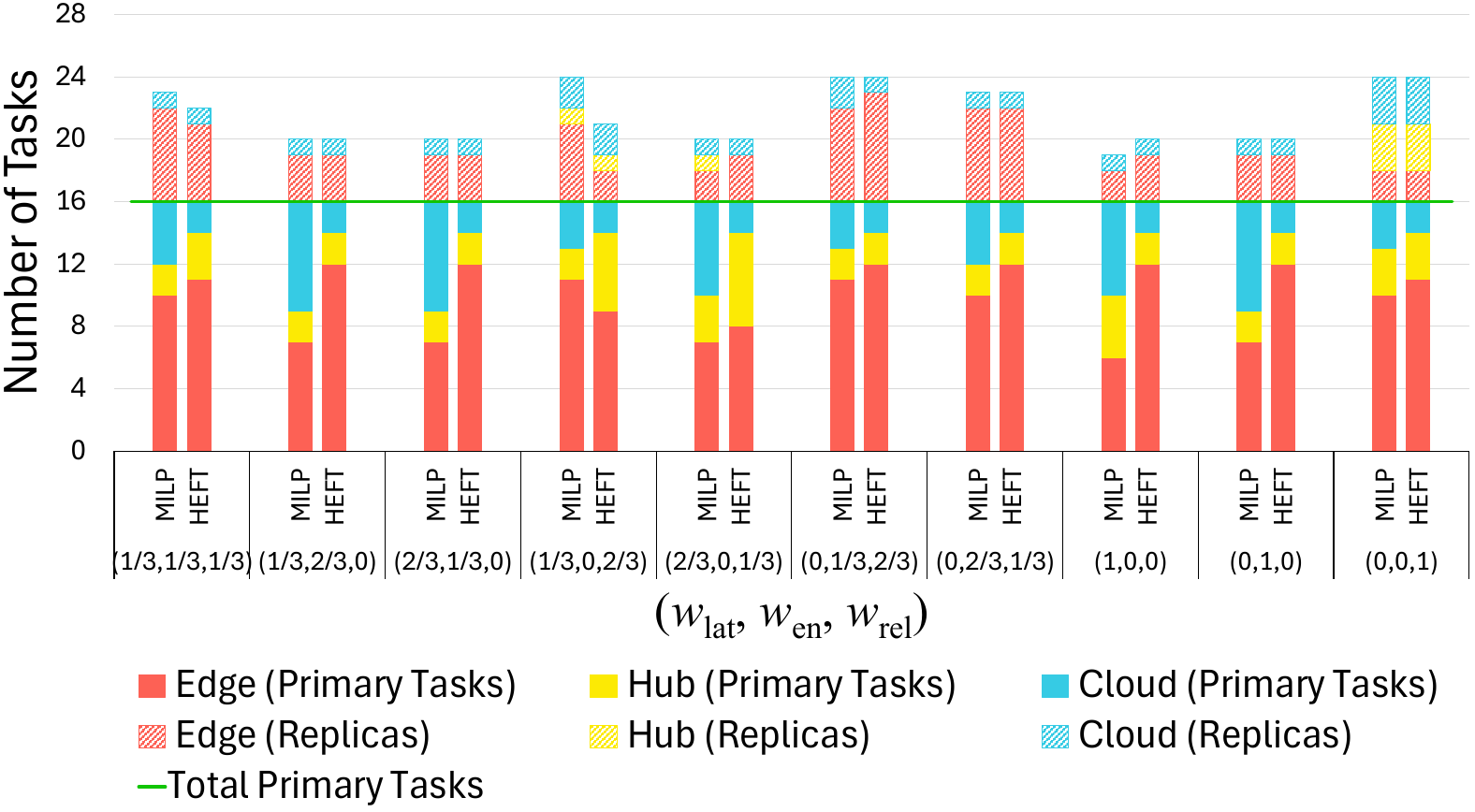}
        \label{fig:allocation2}}
    \caption{Comparison between proposed MILP approach and extended HEFT for the real-world IoT workflow under system configuration C2.}
    \label{fig:realResults2}
\end{figure}

\begin{figure}[!t] 
    \centering
    \subfloat[]{\includegraphics[width=0.97\columnwidth]{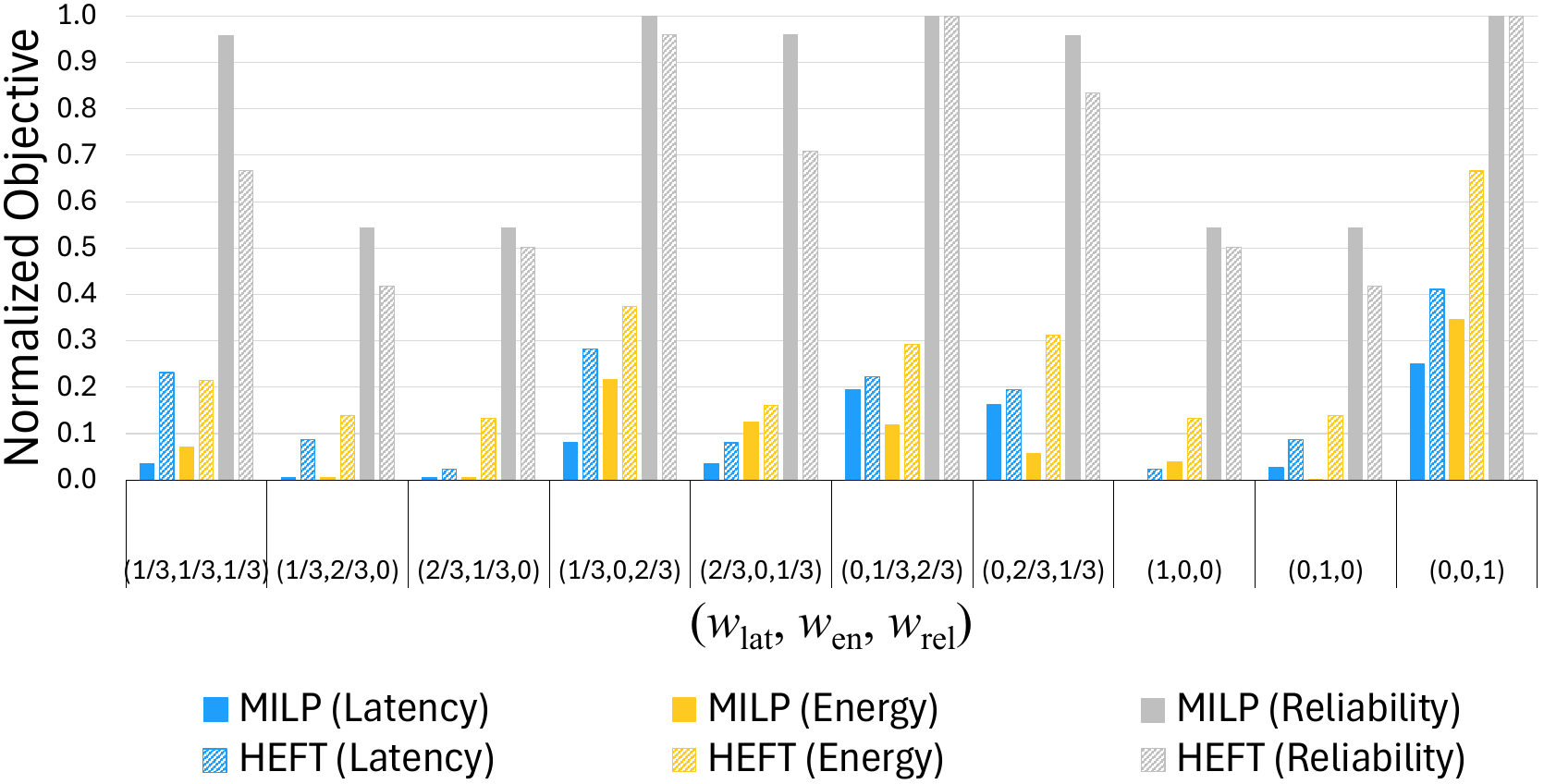}
        \label{fig:normalized3}}\\
    \subfloat[]{\includegraphics[width=0.97\columnwidth]{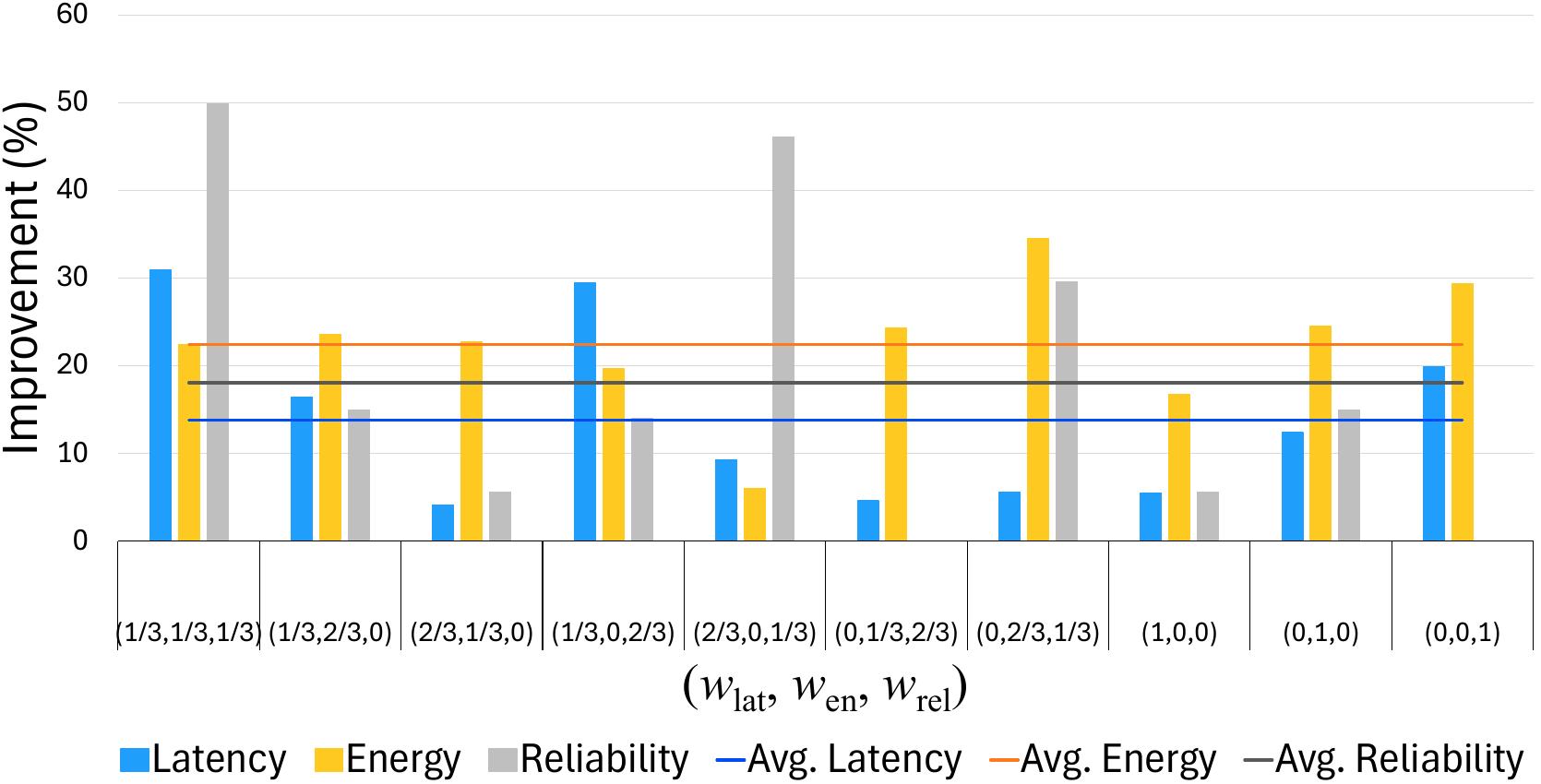}
        \label{fig:improvement3}}\\
    \subfloat[]{\includegraphics[width=0.97\columnwidth]{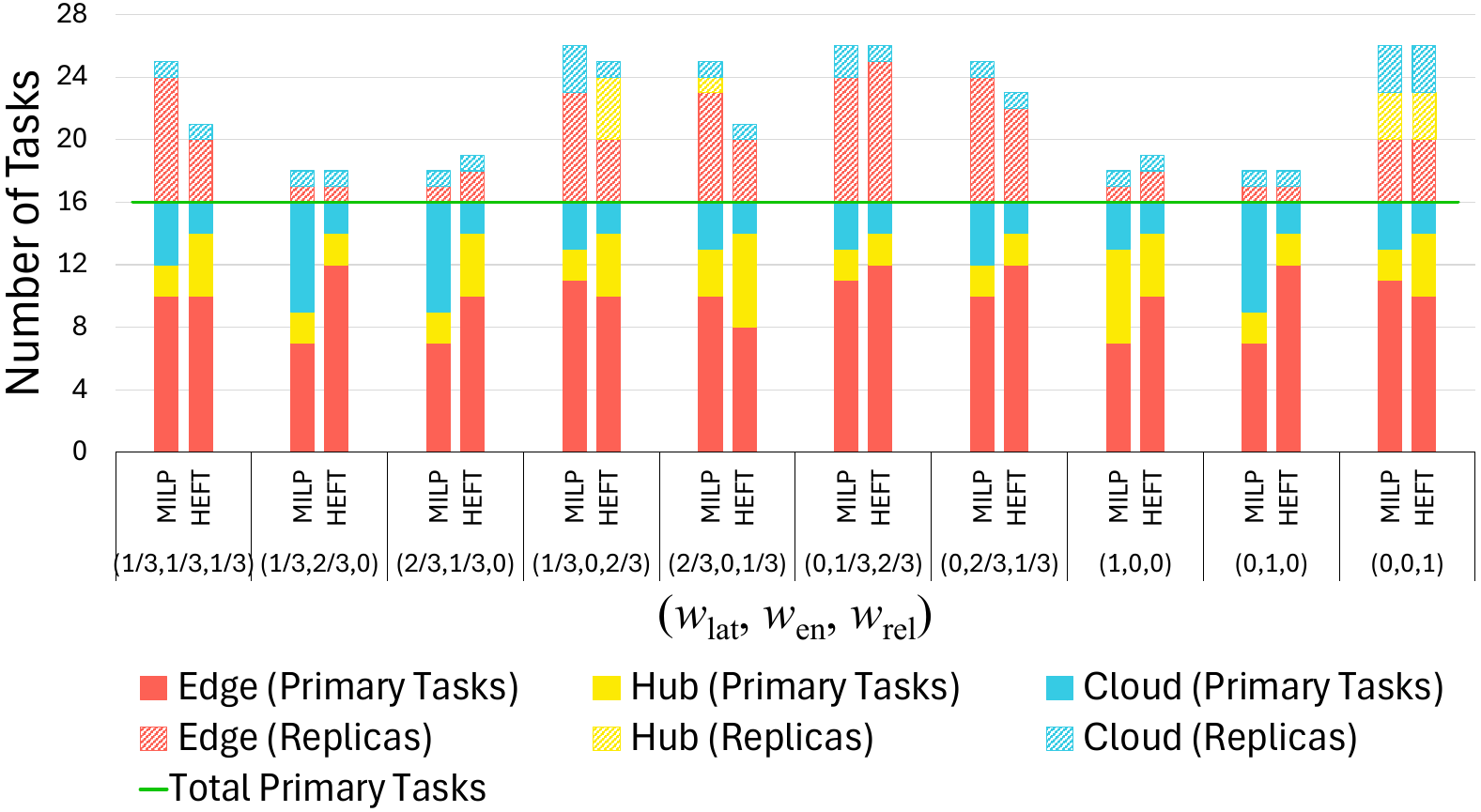}
        \label{fig:allocation3}}
    \caption{Comparison between proposed MILP approach and extended HEFT for the real-world IoT workflow under system configuration C3.}
    \label{fig:realResults3}
\end{figure}

Figs. \ref{fig:realResults}, \ref{fig:realResults2}, and \ref{fig:realResults3} demonstrate the comparison between the proposed MILP approach and extended HEFT for the considered real-world workflow under system configurations C1, C2, and C3, respectively, while varying the weights of the three objectives (latency, energy, and reliability).
Specifically, Figs. \ref{fig:normalized}, \ref{fig:normalized2}, and \ref{fig:normalized3} show the normalized overall latency, energy, and reliability, with respect to $w_{\mathrm{lat}}$, $w_{\mathrm{en}}$, and $w_{\mathrm{rel}}$, yielded by our MILP method (bars in solid color) and extended HEFT (bars in patterned color). 
Lower latency and energy, and higher reliability, indicate better performance.
Figs. \ref{fig:improvement}, \ref{fig:improvement2}, and \ref{fig:improvement3} illustrate the percentage improvement in latency, energy, and reliability provided by the proposed MILP technique over extended HEFT, with respect to the examined weight combinations. 
Figs. \ref{fig:allocation}, \ref{fig:allocation2}, and \ref{fig:allocation3} showcase the allocation of the primary tasks and their replicas (depicted in solid and patterned color, respectively) on the examined CPS for each set of weights.
In all figures, the case where all three objectives were equally important ($w_{\mathrm{lat}} \hspace{-2pt} = \hspace{-2pt} \allowbreak w_{\mathrm{en}} \hspace{-2pt} = \hspace{-2pt} \allowbreak w_{\mathrm{rel}} \hspace{-2pt} = \hspace{-2pt} \allowbreak 1/3$) is shown first, followed by the cases where only two objectives were optimized (in different weight combinations). 
The single-objective cases where only latency ($w_{\mathrm{lat}} \hspace{-2pt} = \hspace{-2pt} \allowbreak 1$), energy ($w_{\mathrm{en}} \hspace{-2pt} = \hspace{-2pt} \allowbreak 1$), or reliability ($w_{\mathrm{rel}} \hspace{-2pt} = \hspace{-2pt} \allowbreak 1$) was optimized, are shown last.

As shown in Figs. \ref{fig:normalized}, \ref{fig:improvement}, \ref{fig:normalized2}, \ref{fig:improvement2}, \ref{fig:normalized3}, and \ref{fig:improvement3}, MILP consistently outperformed HEFT across all weight combinations and system configurations, achieving significant improvements in latency, energy, and reliability.
Specifically, MILP attained an overall average decrease of 14.47\% in latency and 20.37\% in energy, along with an overall average increase of 20.11\% in reliability, compared to HEFT.
In system configurations C1--C3, when optimizing for all three objectives (i.e., $w_{\mathrm{lat}} \hspace{-2pt} = \hspace{-2pt} \allowbreak w_{\mathrm{en}} \hspace{-2pt} = \hspace{-2pt} \allowbreak w_{\mathrm{rel}} \hspace{-2pt} = \hspace{-2pt} \allowbreak 1/3$), MILP outperformed HEFT in latency, energy, and reliability. 
When optimizing for any two of the three objectives, MILP outperformed HEFT in latency and energy, and either matched or exceeded HEFT in reliability. 
For example, for $w_{\mathrm{lat}} \hspace{-2pt} = \hspace{-2pt} \allowbreak 0$, $w_{\mathrm{en}} \hspace{-2pt} = \hspace{-2pt} \allowbreak 1/3$, and $w_{\mathrm{rel}} \hspace{-2pt} = \hspace{-2pt} \allowbreak 2/3$, both methods achieved the same optimal reliability, whereas for $w_{\mathrm{lat}} \hspace{-2pt} = \hspace{-2pt} \allowbreak 1/3$, $w_{\mathrm{en}} \hspace{-2pt} = \hspace{-2pt} \allowbreak 0$, and $w_{\mathrm{rel}} \hspace{-2pt} = \hspace{-2pt} \allowbreak 2/3$, MILP yielded higher reliability than HEFT.
When focusing on a single objective, MILP outperformed HEFT in all three objectives, except for $w_{\mathrm{rel}} \hspace{-2pt} = \hspace{-2pt} \allowbreak 1$, where it yielded the same optimal reliability as HEFT.
%
These observations across all objective weight combinations indicate that, for the specific IoT workflow, reliability was more sensitive to its corresponding weight and the employed system configuration than latency and energy. This led to more pronounced fluctuations in reliability, as well as a few cases where MILP and HEFT achieved the same optimal reliability. 
This increased sensitivity is directly linked to the strong dependence of reliability on the utilized capability provisioning strategy, as primary tasks and their replicas requiring specialized capabilities such as sensors or actuators could only be allocated on specific devices, with configuration C2 being more restrictive than C1 and C3 in that regard.
The average improvement in each objective per configuration is shown in \cref{tab:improvement}.

As demonstrated in Figs. \ref{fig:allocation}, \ref{fig:allocation2}, and \ref{fig:allocation3}, the task allocation yielded by each method was not always intuitive or straightforward. 
For example, in cases where the reliability objective was not taken into account (e.g., $w_{\mathrm{lat}} \hspace{-2pt} = \hspace{-2pt} \allowbreak 2/3$, $w_{\mathrm{en}} \hspace{-2pt} = \hspace{-2pt} \allowbreak 1/3$, and $w_{\mathrm{rel}} \hspace{-2pt} = \hspace{-2pt} \allowbreak 0$), some tasks were still duplicated. This was due to the reliability constraints considered in both MILP and HEFT. Furthermore, MILP generally allocated more primary tasks on the cloud server than HEFT, while utilizing at least as many replicas as HEFT. 
This is particularly evident for $w_{\mathrm{lat}} \hspace{-2pt} = \hspace{-2pt} \allowbreak w_{\mathrm{en}} \hspace{-2pt} = \hspace{-2pt} \allowbreak w_{\mathrm{rel}} \hspace{-2pt} = \hspace{-2pt} \allowbreak 1/3$.
Despite this trend, MILP still outperformed HEFT by optimally allocating the primary tasks and their replicas in the examined CPS under each weight combination, while using a reasonable number of replicas (ranging from 2 to 10 for 16 primary tasks).

Overall, the experimental results in Figs. \ref{fig:realResults}, \ref{fig:realResults2}, and \ref{fig:realResults3} showcase that the proposed MILP approach is more adaptable and effective in jointly optimizing latency, energy, and reliability across different objective weight combinations and system configurations, compared to HEFT.
The inferior performance of HEFT stems from its inherent limitation of scheduling one task at a time without exploring the entire solution space, in contrast to MILP.
The Gurobi solver required between 126.3 and 270.1 seconds to return a solution for our MILP method, as demonstrated in \cref{tab:realETAGs}, which also shows the number of variables and constraints for each system configuration.
Given the pre-programmed nature of the examined workflow, which enables its offline scheduling, the solver runtime is short and practical.

\begin{table}[t]
    \centering
    \caption{Average Improvement in Latency, Energy, \\and Reliability for Real-World IoT Workflow}
    \resizebox{0.58\columnwidth}{!}{
    \begin{tabular}{cccc}
        \toprule
        \multirow{2}{*}{Config.} & \multicolumn{3}{c}{Avg. Improvement}\\
        \cline{2-4}
         & Latency & Energy & Reliability\\
        \hline
        C1 & 15.27\% & 18.56\% & 24.36\%\\
        C2 & 14.35\% & 20.17\% & 17.92\%\\
        C3 & 13.80\% & 22.38\% & 18.05\%\\
        \hline
        Overall & 14.47\% & 20.37\% & 20.11\%\\
        \bottomrule
    \end{tabular}
    }
    \label{tab:improvement}
\end{table}

\subsection{Experiments With Synthetic IoT Workflows}
\label{subsec:scalability}

\subsubsection{Overview}
\label{subsubsec:syntheticWorkflows}

As demonstrated by the real-world IoT workflow in \cref{subsec:real}, the targeted workflows for the examined system architecture typically have a coarse-grained nature, featuring a small-to-moderate number of tasks (10--20 tasks) \cite{Zheng2021, Alam2017, Kashino2019, Savva2021}. However, to further validate our MILP approach and explore its scalability under more demanding workloads, we additionally investigated larger TGs. Specifically, we generated 25 random TGs, divided into five sets of different sizes, using the generator in \cite{Dick1998}. Each set included five TGs with 10, 20, 30, 40, or 50 tasks \cite{Genez2020, Liu2019, Mo2022, Mo2023}, and an average in/out degree (incoming/outgoing arcs per node) of 1.45.
We randomly assigned specialized capabilities to the tasks of each TG, based on the type of each task, considering that entry and exit tasks often require a specific sensor or actuator, respectively.

We transformed the generated TGs into the corresponding ETAGs based on system configuration C1, as it represented the most balanced capability provisioning strategy.
We considered equal importance for all three objectives, i.e., $w_{\mathrm{lat}} \hspace{-2pt} = \hspace{-2pt} \allowbreak w_{\mathrm{en}} \hspace{-2pt} = \hspace{-2pt} \allowbreak w_{\mathrm{rel}} \hspace{-2pt} = \hspace{-2pt} \allowbreak 1/3$. 
\cref{tab:syntheticETGs} shows the average number of nodes and arcs of the TGs and the corresponding ETAGs for the considered synthetic workflows.
It also includes the theoretical upper bounds on the number of ETAG nodes and arcs, derived from the analysis in \cref{subsubsec:etagSize}, for $|\mathcal{P}| = 18$ reserved cores across the system.
We randomly assigned values to the device-independent ETAG candidate node parameters $M_i$, $S_i$, and $D_i$ from those obtained in the real-world application.
The device/core-dependent parameters $L_{i\hspace{-0.5pt}, \mu k\hspace{-0.5pt}.\hspace{-0.5pt}q}$ and $P_{i\hspace{-0.5pt}, \mu k\hspace{-0.5pt}.\hspace{-0.5pt}q}$ were first determined for the candidate nodes involving the least computationally capable device, Raspberry Pi 3 (which served as the reference device), by using values measured for the particular device in the real-world use case.
For candidate nodes involving other devices, $L_{i\hspace{-0.5pt}, \mu k\hspace{-0.5pt}.\hspace{-0.5pt}q}$ and $P_{i\hspace{-0.5pt}, \mu k\hspace{-0.5pt}.\hspace{-0.5pt}q}$ were calculated based on the performance ratio of each device with respect to the reference device, where the performance ratio was determined by running relevant benchmarks \cite{geekbench5, openbenchmarking} on all devices, as shown in \cref{tab:perfRatio}.
The remaining ETAG candidate node and arc parameters were determined as outlined in \cref{subsubsec:realOverview}.
To support reproducibility, all synthetic workflows used in our experiments are publicly available at \cite{ZenodoMOMC}.

\begin{table}[t]
    \setlength{\tabcolsep}{1.5pt}
    \centering
    \caption{Synthetic IoT Workflow TGs \& ETAGs}
    \resizebox{0.85\columnwidth}{!}{
    \begin{tabular}{cccccr}
        \toprule
        TG\,Size   & TG\,Avg.  & ETAG\,Avg.   & ETAG\,Upper\,Bound   & Avg.\,\#Var./ & \multicolumn{1}{c}{Avg.\,Solver}\\
        (\#Nodes) & \#Arcs     & \#Nodes/Arcs & \#Nodes/Arcs          & Constraints       & \multicolumn{1}{c}{Runtime\,(h)}\\
        \hline
        10 & 13 & 202\,/\,5737  & 360\,/\,16848  & 14581\,/\,48067   & 0.01\hspace{10pt} \\
        20 & 28 & 400\,/\,13315 & 720\,/\,36288  & 44012\,/\,137017  & 0.21\hspace{10pt} \\
        30 & 45 & 620\,/\,24277 & 1080\,/\,58320 & 90629\,/\,280992  & 1.67\hspace{10pt} \\
        40 & 64 & 788\,/\,32289 & 1440\,/\,82944 & 138110\,/\,417383 & 5.66\hspace{10pt} \\
        50 & 74 & 974\,/\,32961 & 1800\,/\,95904 & 557406\,/\,243625 & 10.47\hspace{10pt} \\
        \bottomrule
    \end{tabular}
    }
\label{tab:syntheticETGs}
\end{table}

\begin{table}[t]
    \setlength{\tabcolsep}{1.5pt}
    \centering
    \caption{Benchmark Scores \& Performance Ratios of System Devices}
    \resizebox{\columnwidth}{!}{
        \begin{tabular}{ccrrcrrr}
            \toprule
            \multirow{3}{*}{$u_{\mu k}$} & \multirow{3}{*}{Device} & \multicolumn{5}{c}{Benchmark Scores${}^{1}$} & \multicolumn{1}{c}{\multirow{2}{*}{\hspace{1pt} Perf.}}\\
            \cline{3-7}
             & & \multicolumn{1}{c}{Geek} & \multicolumn{1}{c}{Num} & R & \multicolumn{1}{c}{Scikit} & \multicolumn{1}{c}{Tensor} & \multicolumn{1}{c}{\multirow{2}{*}{\hspace{1pt} Ratio${}^{3}$}}\\
             & & \multicolumn{1}{c}{bench\,5} & \multicolumn{1}{c}{Py} & Bench. & \multicolumn{1}{c}{Learn} & \multicolumn{1}{c}{Flow} &\\
            \hline
            $u_{\mathrm{e}1}$ & Raspberry\,Pi\,3${}^{2}$   &   99\,pts \hspace{5pt} &    4.25\,pts & 9.50\,s  & 3433.85\,s \hspace{1pt} & 17391.52\,s & \hspace{1pt}  1.00 \\
            $u_{\mathrm{e}2}$ & Odroid\,XU4                &  149\,pts \hspace{5pt} &	 4.49\,pts & 9.17\,s  & 3145.17\,s \hspace{1pt} & 13078.08\,s & \hspace{1pt}  1.20 \\
            $u_{\mathrm{e}3}$ & Jetson\,TX2                &  187\,pts \hspace{5pt} &	12.68\,pts & 7.21\,s  &  870.94\,s \hspace{1pt} &  4491.59\,s & \hspace{1pt}  2.80 \\
            $u_{\mathrm{e}4}$ & Jetson\,Xavier\,NX         &  216\,pts \hspace{5pt} &	22.17\,pts & 2.62\,s  &  400.84\,s \hspace{1pt} &  1905.90\,s & \hspace{1pt}  5.74 \\
            $u_{\mathrm{h}1}$ & Mi\,Notebook\,Pro          &  589\,pts \hspace{5pt} &	85.40\,pts & 0.67\,s  &  149.97\,s \hspace{1pt} &  1335.72\,s & \hspace{1pt} 15.23 \\
            $u_{\mathrm{c}1}$ & HPE\,DL580\,Gen10          & 1578\,pts \hspace{5pt} &	91.75\,pts & 0.44\,s  &  102.19\,s \hspace{1pt} &  1099.98\,s & \hspace{1pt} 21.70 \\
            \bottomrule
            \multicolumn{8}{l}{${}^{1}$Higher scores in points (pts) and lower scores in seconds (s) indicate better}\\
            \multicolumn{8}{l}{performance (single-core benchmark scores are reported).}\\
            \multicolumn{8}{l}{${}^{2}$Used as reference device (least computationally capable device).}\\
            \multicolumn{8}{l}{${}^{3}$Average of individual score ratios between each device and the reference device.} 

        \end{tabular}
        }
    \label{tab:perfRatio}
\end{table}

\subsubsection{Scalability Analysis}
\label{subsubsec:syntheticResults}

\begin{figure}[!t]
    \centering
    \includegraphics[width=0.66\columnwidth]{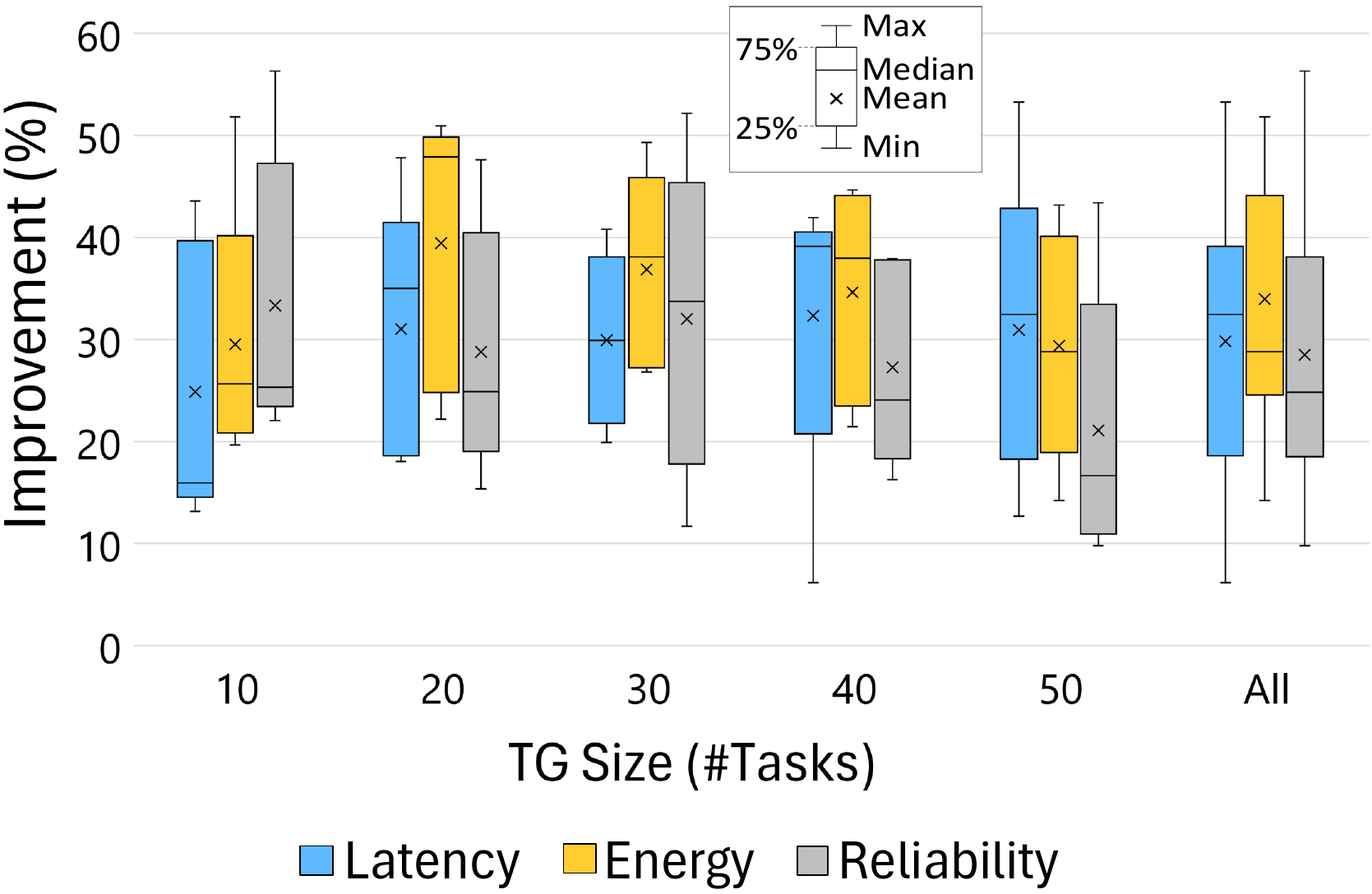}
    \caption{Improvement in latency, energy, and reliability objectives attained by proposed MILP approach over extended HEFT under increasing TG size. Box plots show the distribution of improvement in each objective across ETAGs for each TG size. The rightmost set of box plots aggregates the results across all TG sizes.
    }
    \label{fig:syntheticImprovement}
\end{figure}

\cref{fig:syntheticImprovement} demonstrates the improvement achieved by the proposed MILP method over extended HEFT in all three objectives (latency, energy, and reliability) as we increased the size of the TGs.
Notably, MILP consistently outperformed HEFT in all objectives across all TG sizes. 
Specifically,  
MILP provided a latency decrease over HEFT ranging from 6.15\% to 53.25\%, with a mean and median of 29.83\% and 32.46\%, respectively.
The energy decrease ranged from 14.21\% to 51.79\%, with a mean of 33.96\% and a median of 28.79\%.  
The reliability increase ranged from 9.78\% to 56.31\%, with a mean and median of 28.49\% and 24.81\%, respectively.
It can be observed that the improvement in each objective presented high variability, indicating that the structure and size of the TGs, and by extension the required capabilities of the entry, intermediate, and exit tasks, significantly affect performance in the examined environment.

\begin{figure}[!t]
    \centering
    \includegraphics[width=0.66\columnwidth]{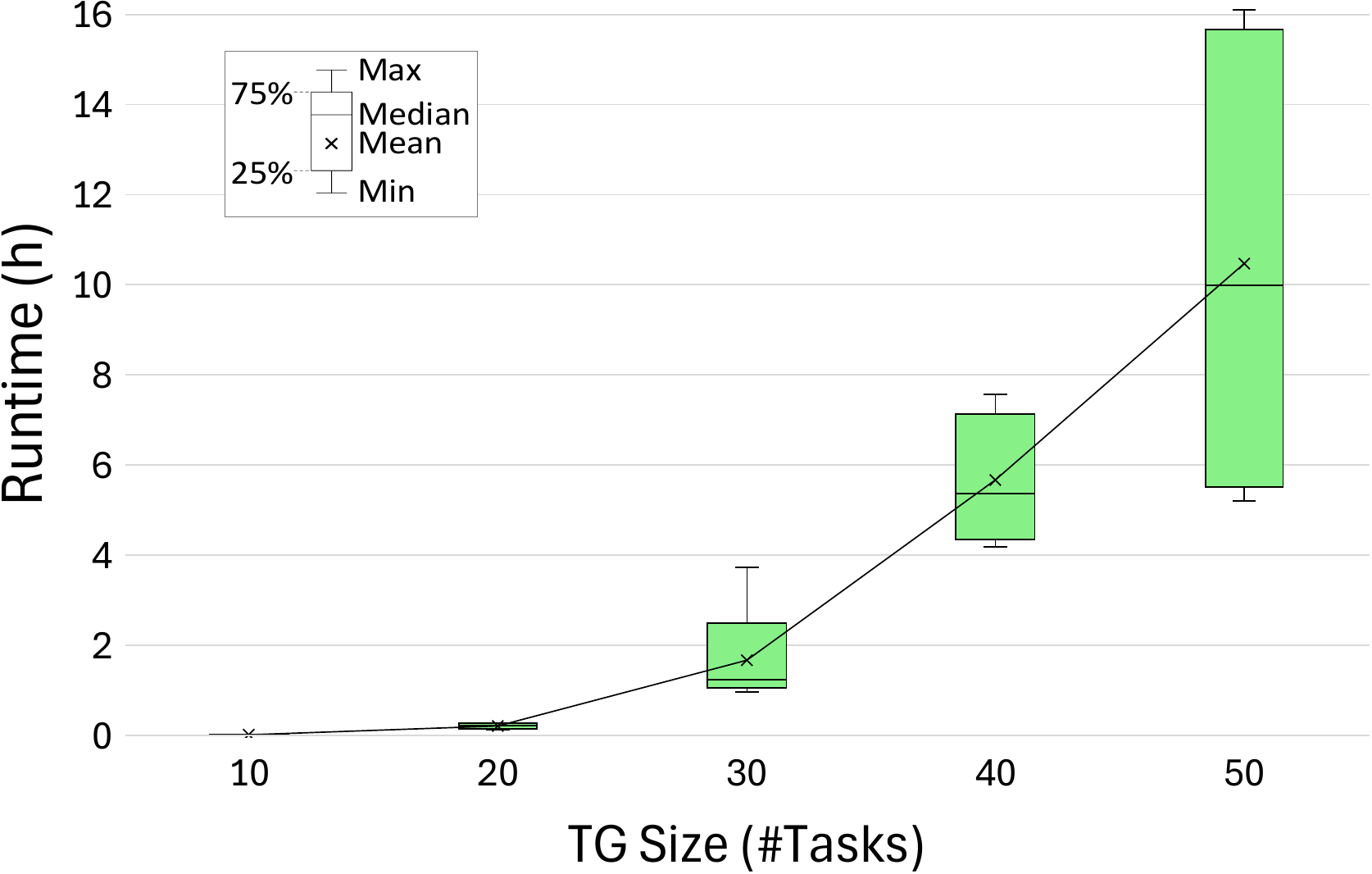}
    \caption{Solver runtime for proposed MILP method under increasing TG size. Box plots show the runtime distribution across ETAGs for each TG size, while the solid line connects the mean runtime values.
    }
    \label{fig:syntheticRuntime}
\end{figure}

The time required by the Gurobi solver to provide a solution for our MILP approach as the TG size increased is shown in \cref{fig:syntheticRuntime}. The average solver runtime is also reported in \cref{tab:syntheticETGs}, along with the average number of variables and constraints for each TG size. 
For reference, the runtime of extended HEFT ranged from 0.1 to 13.6 seconds.
%
Our approach required longer runtimes for larger TGs due to the increase in the number of variables and constraints resulting from the growing size of the corresponding ETAGs.  
However, the resulting ETAGs contained significantly fewer nodes and arcs than the theoretical upper bounds (see \cref{subsubsec:designChoices}). For example, as shown in \cref{tab:syntheticETGs}, for TGs with 50 tasks (which on average comprised 74 arcs), the corresponding ETAGs included on average 974 nodes and 32\,961 arcs, which are substantially fewer than the worst-case bounds of 1800 nodes and 95\,904 arcs, respectively.
The growth of ETAG nodes and arcs under increasing TG size, as well as the comparison between empirical averages and their corresponding theoretical upper bounds, are further illustrated in Appendix B in the supplementary material.
Across all workflow sizes, the empirical ETAG node and arc averages are shown to be well below these bounds, demonstrating the practical tractability of the proposed approach.
Considering the NP-hardness of the problem \cite{Zengen2020}, the exact and offline nature of the proposed MILP method, as well as the significant improvement it consistently provided over HEFT in all three objectives, the time required by the solver is reasonable and practical, even for large TGs.
More importantly, given that relevant applications for the specific architecture typically involve 10--20 tasks (as shown in \cref{subsec:real}), the practicality of our approach is further underscored.
For these TG sizes, the solver yielded optimal solutions in a short time frame, ranging from 20 seconds to 16.6 minutes.

%% file: 6_conclusion.tex
\section{Conclusion \& Future Work}
\label{sec:conclusion}

We proposed a multi-objective and multi-constrained continuous-time MILP approach to optimally schedule an IoT workflow in an edge-hub-cloud CPS. 
Our formulation is enabled by a two-phase task graph transformation technique that incorporates selective task duplication to enhance reliability while avoiding unnecessary task replicas. In contrast to existing approaches, the proposed method comprehensively optimizes latency, energy, and reliability under precedence, deadline, reliability, capability, memory, storage, and energy constraints, while considering heterogeneous multicore processors and multiple sensing, actuating, or other specialized capabilities per device.
We evaluated the proposed approach against the well-established HEFT heuristic using both a representative real-world IoT workflow and synthetic IoT task graphs of varying sizes, under different system configurations and objective trade-offs. To ensure a fair and meaningful comparison, we incorporated into HEFT selective task duplication and the same objectives and constraints as our method.
The experimental results demonstrate the effectiveness of the proposed MILP approach in jointly optimizing the conflicting objectives of the considered problem.
Specifically, MILP consistently outperformed HEFT  in the real-world use case, achieving overall average improvements of 14.47\% in latency, 20.37\% in energy, and 20.11\% in reliability, across all system configurations and objective trade-offs.
For synthetic workflows, it provided average improvements of 29.83\% in latency, 33.96\% in energy, and 28.49\% in reliability, while attaining practical solver runtimes. 
These findings show that the proposed method is effective and scalable for IoT workflows of sizes and characteristics commonly encountered in the targeted applications and system architecture. 
In future work, we will explore hybrid exact and heuristic scheduling approaches that are applicable to different use cases and system architectures. Moreover, we will investigate scenarios involving uncertainty in parameters such as task execution times and network bandwidth.